%% file: ms.tex
\documentclass[fleqn,usenatbib]{mnras}
\pdfoutput=1

\usepackage{newtxtext,newtxmath}

\usepackage[T1]{fontenc}

\DeclareRobustCommand{\VAN}[3]{#2}
\let\VANthebibliography\thebibliography
\def\thebibliography{\DeclareRobustCommand{\VAN}[3]{##3}\VANthebibliography}

\usepackage{graphicx}	

\newcommand{\gcat}{{g_\mathrm{cat}}} 
\newcommand{\Tcat}{{T_\mathrm{cat}}} 
\newcommand{\gLDR}{{g_\mathrm{LDR}}} 
\newcommand{\TLDR}{{T_\mathrm{LDR}}} 
\newcommand{\Teff}{{T_\mathrm{eff}}} 
\newcommand{\FeH}{{\mathrm{[Fe/H]}}} 
\newcommand{\dall}{d_{\rm all}}
\newcommand{\done}{d_{\rm one}}
\newcommand{\dout}{d_{\rm out}}
\newcommand{\IQR}{{\rm IQR}}
\newcommand{\kms}{{\rm km\,s^{-1}}}
\newcommand{\Msun}{{M_{\odot}}}
\newcommand{\Npair}{{N_\mathrm{pair}}}
\newcommand{\epsT}{{\varepsilon_{\,T}}}
\newcommand{\epsG}{{\varepsilon_{G}}}
\newcommand{\delG}{{\delta_{G}}}

\title[LDRs for determination of $\Teff$ and $\log g$]{Line-depth ratios as indicators of effective temperature and surface gravity}

\author[N. Matsunaga et al.]{
Noriyuki Matsunaga,$^{1,2}$\thanks{E-mail: matsunaga@astron.s.u-tokyo.ac.jp}
Mingjie Jian,$^{1}$
Daisuke Taniguchi$^{1}$
and Scarlet S.\ Elgueta$^{1}$
\\
$^{1}$Department of Astronomy, School of Science, The University of Tokyo, 7-3-1, Hongo, Bunkyo-ku, Tokyo 113-0033, Japan \\
$^{2}$Laboratory of Infrared High-resolution spectroscopy (LiH), Koyama Astronomical Observatory, Kyoto Sangyo University, \\
Motoyama, Kamigamo, Kita-ku, Kyoto 603-8555, Japan
}

\date{Accepted 2021 June 16. Received 2021 June 16; in original form 2021 February 25}

\pubyear{2021}

\begin{document}
\label{firstpage}
\pagerange{\pageref{firstpage}--\pageref{lastpage}}
\maketitle

\begin{abstract}
The analysis of stellar spectra depends upon the effective temperature ($\Teff$)
and the surface gravity ($\log g$).
However, the estimation of $\log g$
with high accuracy is challenging. A classical approach is
to search for $\log g$ that satisfies the ionization balance,
i.e., the abundances from neutral and ionized metallic lines
being in agreement.
We propose a method of using empirical relations
between $\Teff$, $\log g$ and line-depth ratios,
for which we meticulously select pairs of \ion{Fe}{i} and
\ion{Fe}{ii} lines and pairs of \ion{Ca}{i} and \ion{Ca}{ii} lines.
Based on $YJ$-band (0.97--1.32\,{$\mu$}m) high-resolution spectra of
42 FGK stars (dwarfs to supergiants),
we selected
five \ion{Fe}{i}--\ion{Fe}{ii} and four \ion{Ca}{i}--\ion{Ca}{ii} line pairs
together with 13 \ion{Fe}{i}--\ion{Fe}{i} pairs (for estimating $\Teff$),
and derived the empirical relations.
Using such relations does not require
complex numerical models and tools for estimating
chemical abundances.
The relations we present allows one to derive $\Teff$ and $\log g$
with a precision of around 50\,K and 0.2\,dex, respectively,
but the achievable accuracy depends on the accuracy of
the calibrators' stellar parameters.
It is essential to revise the calibration by observing stars with
accurate stellar parameters available,
e.g., stars with asteroseismic $\log g$
and stars analyzed with complete stellar models
taking into account the effects of non-local thermodynamic equilibrium
and convection.
In addition, the calibrators we used have a limited metallicity range,
$-0.2 < \FeH < +0.2$\,dex, and our relations need to be tested
and re-calibrated based on a calibrating dataset for a wider range of metallicity.
\end{abstract}




\section{Introduction}
\label{sec:intro}

Surface gravity, $\log g$, is a fundamental stellar parameter
that depends on stellar mass and radius.
In stellar astrophysics, the unit of $g$ is cm\,s$^{-2}$.
The surface gravity of the Sun, $g_{\odot}=27\,400$\,cm\,s$^{-2}$,
corresponds to $\log g_{\odot}=4.44$\,dex.
From the spectroscopic point of view, surface gravity
affects various kinds of absorption features,
such as the damping wings of hydrogen lines.
Several methods have been developed for estimating 
the value of $\log g$ on the basis of such features
\citep[see, e.g.,][]{Catanzaro-2014}.
However, it is not easy to determine $\log g$ with
a high accuracy, and many methods require 
complex calculations with numerical models. 
Systematic differences are often found between
the spectroscopic $\log g$ estimates
and those derived using other approaches
(e.g., asteroseismic analysis; \citealt{Morel-2012}; \citealt{Chaplin-2013};
\citealt{Meszaros-2013}; \citealt{Mortier-2014}).

The main spectral features of stars with the middle spectral types
(i.e., F, G and K types) are the metallic absorption lines 
that characterize their spectra. Among these lines, 
ionized lines are particularly sensitive to $\log g$.  
According to the Saha equation, the ratio of atoms to ions
in the stellar atmosphere is proportional to the electronic pressure, $P_e$,
when the local thermodynamic equilibrium (LTE) holds
and only one ionization stage is available. 
Therefore, the ionization fraction is higher 
with a lower $\log g$, and thus, lower $P_e$
\citep{Gray-2005}. 
Based on this behaviour, it is possible to estimate
$\log g$ by postulating that the iron abundance derived 
from \ion{Fe}{ii} lines ($\log g$-sensitive)
and the counterpart with \ion{Fe}{i} lines ($\log g$-insensitive)
agree with each other.
This method has been used quite often
\citep[][and references therein]{Gray-2019,Tsantaki-2019},
but some authors have questioned the simple assumption of
the balance between \ion{Fe}{i} and \ion{Fe}{ii} lines
under the LTE condition (\citealt{Fuhrmann-1997};
\citealt{Kovtyukh-1999}; \citealt{Korn-2003}).
In addition, previous studies almost always relied on
\ion{Fe}{ii} lines in the optical spectra.
Recently, \citet{Marfil-2020} explored
\ion{Fe}{ii} lines in the infrared range
together with those in the optical range
and detected only two \ion{Fe}{ii} lines that are available
for estimating stellar parameters including $\log g$.

Another commonly-taken approach is to determine
various stellar parameters including $\Teff$, $\log g$ and 
the metallicity (or a bit more detailed chemical abundances) by searching for 
the synthetic spectrum that matches a given observed spectrum best.
The selected synthetic spectrum is required to reproduce
many absorption lines, in a wide spectral range, 
that have different sensitivity to the stellar parameters.
A large set of synthetic spectra and
an efficient tool for the optimization such as {\sc FERRE}
\citep{GarciaPerez-2016} need to be prepared,
and such analysis has been regularly performed,
especially for a large-scale spectroscopic survey
\citep[e.g.][]{AllendePrieto-2006,AllendePrieto-2014,GarciaPerez-2016}.
The resultant precision and accuracy depend
on how well the selected synthetic spectrum
reproduces the observed one. Synthetic spectra for the optical range
are considered to be sufficiently good for stars
in a wide range of stellar parameters, and this is also true for
the spectral band of the APOGEE (a part of the $H$ band)
thanks to their great effort to establish the line list \citep{Shetrone-2015}.
However, as mentioned above, thus-derived $\log g$ tend
to show systematics compared to $\log g$ from other approaches. 
For other wavelength ranges in the infrared, 
the line information tends to have large errors currently
\citep[see, e.g.,][]{Andreasen-2016}.
For example, \citet{Fukue-2021} compiled a list of more than 250 lines
of several elements seen in the $YJ$ band (0.97--1.32\,{$\mu$}m),
including the lines of \ion{Fe}{i} reported in
\citet{Kondo-2019}, that appear relatively isolated in K-type giants.
\citet{Fukue-2021} found that the scatter in the derived abundances are larger
than the results for other wavelengths (see their Section 3.3),
and there appears a systematic offset by 0.1--0.2\,dex
between the results with the oscillator strengths, $\log gf$,
given in two line lists,
the Vienna Atomic Line Database \citep[VALD;][]{Ryabchikova-2015}
and \citet{Melendez-1999}.
Such errors and systematics limit the quality of synthetic spectra 
and thus the precision of the stellar parameters derived 
with the spectral synthesis.

In this study, we propose a new method for determining 
the surface gravity on the basis of neutral and ionized lines
without being affected directly by the uncertainty in the line list.
We focus on the \ion{Fe}{i}, \ion{Fe}{ii}, \ion{Ca}{i} and \ion{Ca}{ii} lines
in the $YJ$ bands.
Moreover, we consider line-depth ratios (LDRs) as
indicators of both temperature and surface gravity
with which the inference of chemical abundances is unnecessary.
An advantage of LDRs is that their use is based on
empirical relations, which do 
not require sophisticated calculations with numerical models.
The accuracy of the $\log g$, derived from the LDR empirical relations, 
relies on the accuracy of the calibrators' $\log g$,
but it is trivial to revise the calibration when the new data set
for better calibrators becomes available. 

LDRs have been often used for estimating $\Teff$.
LDRs of lines
with different excitation potentials (EPs) are
sensitive to $\Teff$, and thus, most previous works
explored the use of LDRs as temperature indicators 
\citep[][and references therein]{Gray-1991,Fukue-2015}.
However, LDRs of some kinds of line pairs depend not only on $\Teff$ but also on
$\log g$ and other stellar parameters.
\citet{Jian-2020} compared the LDRs of
dwarfs, giants and supergiants,
and found that LDRs of some line pairs at each $\Teff$
depend on $\log g$. They examined the
line pairs of neutral atoms, e.g., \ion{Ca}{i}--\ion{Si}{i},
and concluded that the differences in 
ionization potentials of the atoms can explain the different
dependencies of line depth on $\log g$.
LDRs also depend on metallicity and abundance ratios
\citep[see, e.g.,][on the metallicity effect]{Jian-2019}, 
and these effects would complicate the determination of
$\log g$ together with $\Teff$. 
This uncertainty may be avoided if we combine
lines of the same element \citep{Taniguchi-2021}.
Therefore, we consider line pairs of
the same element only.
We use \ion{Fe}{i}--\ion{Fe}{i} pairs
for determining $\Teff$, and 
\ion{Fe}{i}--\ion{Fe}{ii} and \ion{Ca}{i}--\ion{Ca}{ii} pairs
for determining $\log g$.

The main goal of this paper is two fold: (1) to demonstrate
that empirically-calibrated LDR relations can be used for
estimating $\log g$ of stars over a wide range of luminosity class
and (2) to identify the useful line pairs available in the $YJ$ bands.
We use 42 $YJ$-band spectra of FGK stars, dwarfs to supergiants,
adopted from \citet{Jian-2020}. Their sample was not collected
for allowing us the calibration based on 
the most accurate surface gravities, and the lack of ``benchmark stars''
with high-resolution $YJ$-band spectra limits
the accuracy of the LDR relations we can derive in this study.
Nevertheless, we can show that the LDR relations
give useful constraints to $\log g$ together with $\Teff$.
The $YJ$ bands, in particular the $Y$ band, are uniquely important
in determining the surface gravities of reddened stars because
these bands are less affected by interstellar extinction 
($A_Y/A_V \sim 0.4, ~A_J/A_V \sim 0.3$) and host more ionized lines than 
other near-infrared bands \citep{Andreasen-2016,Marfil-2020}.
Many high-resolution spectrographs covering $YJ$ have been
in operation:
e.g., GIANO \citep{Caffau-2019}, HPF \citep{Sneden-2021},
IRD \citep{Kotani-2018} and CRIRES+ \citep{Follert-2014} 
in addition to WINERED with which our data were collected.

\section{Data}
\label{sec:Data}

\begin{figure}
	\includegraphics[width=\columnwidth]{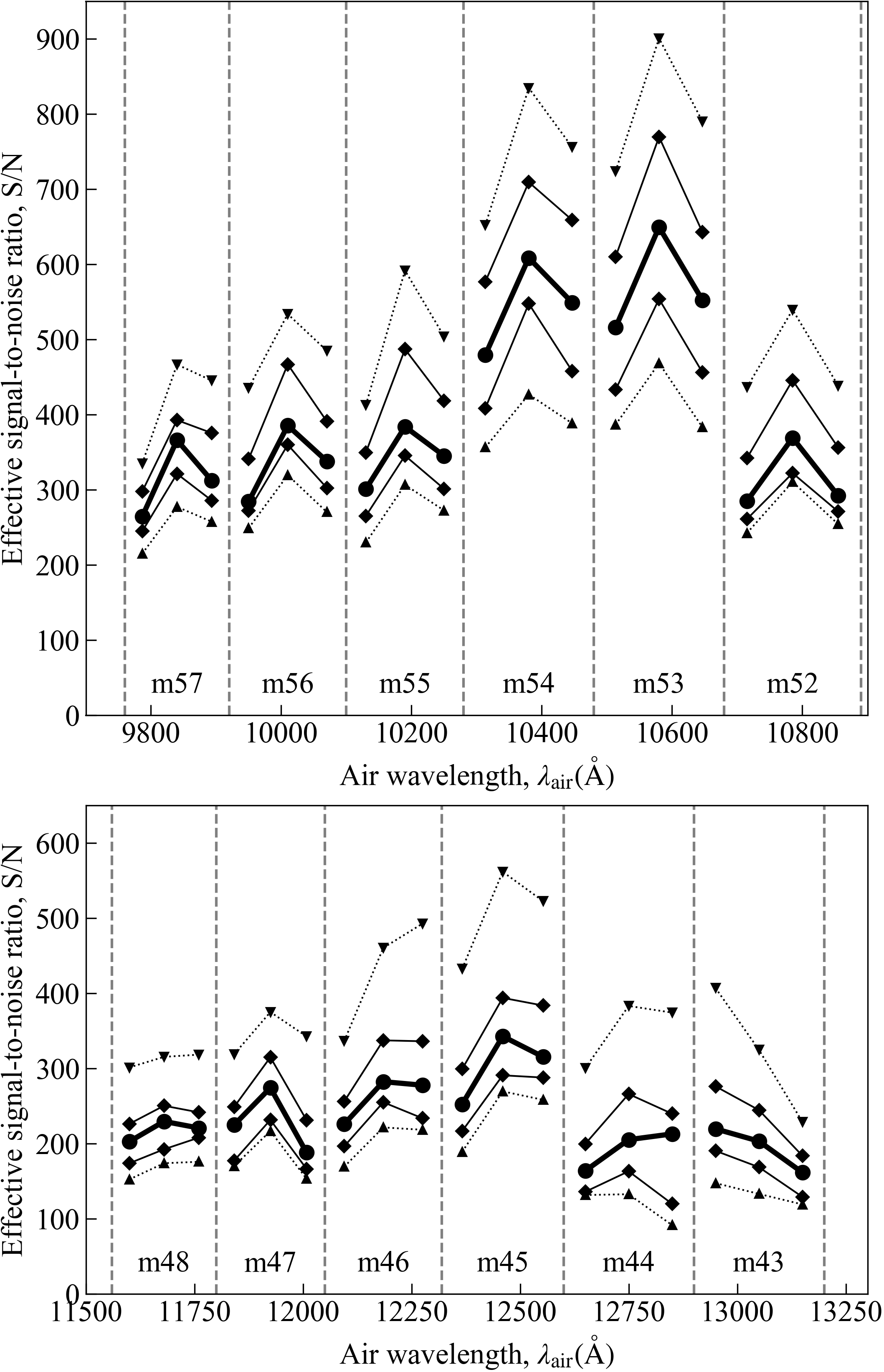}
    \caption{Effective signal-to-noise ratios at three parts of each order (52nd to 57th in $Y$, upper panel, and 43rd to 48th in $J$, lower panel). The S/N values measured by the WINERED pipeline for each target and the corresponding telluric standard are combined to give the effective S/N, except the 53rd and 54th order for which the S/N for the target is used. The S/N values for the spectra of the 42 targets (Table~\ref{tab:targets}) are considered here, and the five points at each wavelength indicate 10, 25, 50, 75 and 90 percentiles, from top to bottom, corresponding to the 4th, 10th, 21st, 31st and 37th values.}
    \label{fig:SNRs}
\end{figure}

We make use of spectra of 42 objects with $\Teff$ between 
4800\,K and 6300\,K (Table~\ref{tab:targets})
adopted from \citet{Jian-2020}.
The WINERED spectrograph was used to collect 
the $YJ$-band spectra with a resolution of around 28\,000
\citep[see the instrumental details on the WINERED in][]{Ikeda-2016}.
The observations were carried out with the 1.3\,m Araki Telescope at
Koyama Observatory, Kyoto Sangyo University in Japan
from October 2015 to May 2016.
The reduction in the spectroscopic data was described in
\citet{Jian-2020} and \citet{Matsunaga-2020}. 
The initial steps of the reduction were performed using
the WINERED pipeline (Hamano et al.\ in preparation).
Then, the telluric absorption lines were removed using 
the observed spectra of telluric standard stars 
with the method developed in \citet{Sameshima-2018},
except for the 53rd--54th orders (10280--10680\,{\AA})
in which significant telluric lines were rather scarce.
The continuum level in every spectrum was normalized to the unity.
In addition to the reduced spectra, the pipeline gives
estimates of the signal-to-noise ratios (S/N) for
three parts of each order by comparing spectra of multiple
exposures with each other. 
We use these S/N values for evaluating the errors of line depths,
discussed below.
Fig.~\ref{fig:SNRs} presents the effective S/N for
which the S/N values of objects and those of telluric standards
are combined except the 53rd--54th orders
(see also Section~\ref{subsec:measure-depth}).

\begin{figure}
	\includegraphics[width=\columnwidth]{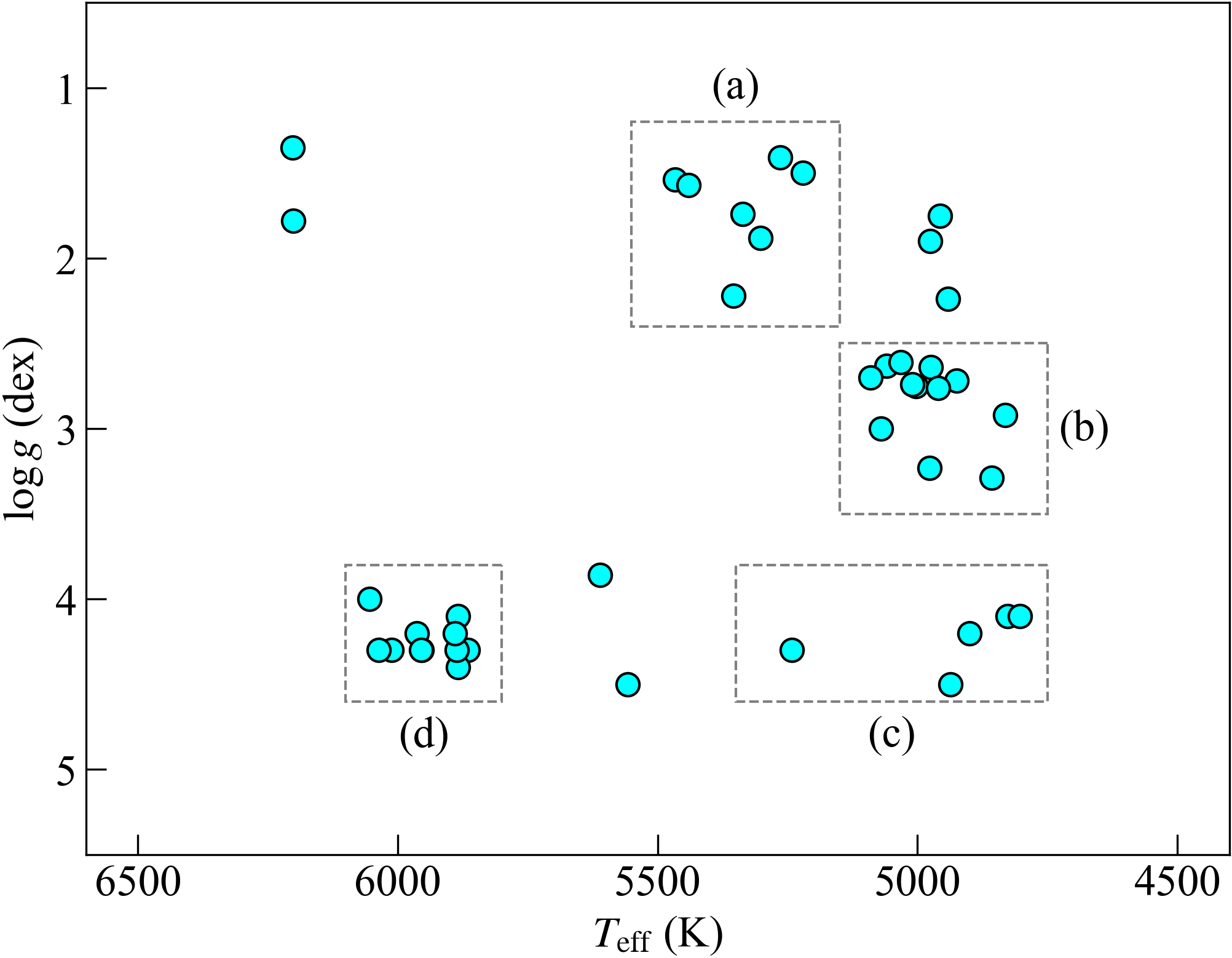}
    \caption{Distribution of the objects (Table~\ref{tab:targets}) on the $(\Teff,~\log g)$ plane. The groups, (a)--(d), are used for selecting line pairs that are useful over a wide range of this plane in Section~\ref{subsec:line-pair-selection}.}
    \label{fig:HRD}
\end{figure}

Our sample is composed of late F to early K-type stars,
and the surface gravity ranges from 1.35 
to 4.5\,dex. The distribution on the $(\log g, ~\Teff)$ plane is,
however, not homogeneous (Fig.~\ref{fig:HRD}).
\citet{Jian-2020} selected these objects, together with stars
in a broader $\Teff$ range,
because the optical LDRs were investigated by
\citet{Kovtyukh-2003,Kovtyukh-2006} and \citet{Kovtyukh-2007}.
In the following analysis, we use the stellar parameters
that \citet{Jian-2020} compiled from a few studies
\citep{Kovtyukh-2004,Prugniel-2011,Park-2013,Liu-2014,Luck-2014,daSilva-2015}.
Although the measurements in these studies were all based on spectroscopic data,
the compiled $\log g$ values are not homogeneous because
of multiple differences, such as those in the spectral resolution and
in the absorption features that had major impacts on the estimates. 
The selection of the objects in \citet{Jian-2020}
was not done for securing
the homogeneous $\log g$ values.
However, it is difficult to find a set of $YJ$-band high-resolution spectra
of stars, extending from dwarfs to supergiants,
with homogeneous and accurate $\log g$ available.
In Section~\ref{subsec:logg-scale}, we consider
a subset of our objects with $\log g$ taken from different studies
to discuss the accuracy of the gravity scale. 

\input{tab1.tex}

HD\,137107 
($\eta$~CrB~A, with $\Teff=6037$\,K and $\log g=4.30$\,dex)
is a double-lined spectroscopic binary
\citep{Pourbaix-2000}. 
The lines of the two components are not split from each other
in our spectrum \citep[see, also,][]{Duquennoy-1991},
but the lines look slightly broader than
those in other objects (Section~\ref{subsec:measure-depth}). 
Because the two components of this binary system are
similar to each other, both being early G-type dwarfs
with ${\sim}1.2\,\Msun$ \citep{Muterspaugh-2010}, 
we include this object in our analysis.
There are six spectroscopic binaries among our sample,
according to \citet{Pourbaix-2004}\footnote{HD\,11559, HD\,26630, HD\,27697, HD\,39587, HD\,137107 and HD\,202109 based on the version 2014-02-26 available at CDS}, 
and HD\,137107 is the only double-lined system among them.

\section{Analysis}
\label{sec:analysis}

\subsection{Line selection based on synthetic spectra}
\label{subsec:select-line-synth}

To perform the line selection, we searched for the lines of the four elements (\ion{Fe}{i}, \ion{Fe}{ii}, \ion{Ca}{i} and \ion{Ca}{ii})
that may be included in line pairs showing tight LDR relations. 
We consider the lines within 9760--10890 and 11600--13200\,{\AA},
corresponding to the echelle orders of 
52--57th ($Y$ band) and 43--48th ($J$ band) of WINERED. 
Not many elements show both neutral and ionized lines.
Other than Fe and Ca, some elements such as Si and Ti show
one or more lines of the neutral and ionized states together for each element.
Looking at the spectra for the lines of those elements, however, we decided not
to include them in our analysis because it would be impossible to
construct multiple pairs of lines that demonstrate useful LDR relations
for our purpose. We focus on Fe and Ca in this study,
although other elements may be useful for different $\Teff$ ranges.

For each of the atoms and ions, we selected the candidates of lines to use
on the basis of the synthetic spectra created with MOOG
\citep[the version released in 2017 February,][]{Sneden-2012}. 
We used the VALD
and the 1D plane-parallel atmosphere models compiled by
R.~Kurucz\footnote{We adopted the files named amXXk2.dat or apXXk2.dat, where XX is to be replaced by two digits indicating the abundances from http://kurucz.harvard.edu/grids/.}~\ 
for synthesizing the following three kinds of spectra:
\begin{description}
\item[(i)] ~the spectra with all the atomic and molecular lines included,
\item[(ii)] ~the spectra with only one of the four species
(\ion{Fe}{i}, \ion{Fe}{ii}, \ion{Ca}{i} and \ion{Ca}{ii}) included, 
\item[(iii)] ~the spectra with all the atomic and molecular lines included
except one of the four species.
\end{description}
The spectra were synthesized for a grid of $(\Teff, ~\log g)$ covering the range of Fig.~\ref{fig:HRD}.
For each wavelength, the three kinds of synthetic spectra give
three theoretical depths---(i)~$\dall$, (ii)~$\done$ and (iii)~$\dout$.
We then listed up the line candidates that were expected to appear
deep enough ($\done \gtrsim 0.03$) without heavy blends
($\dout/\dall \lesssim 0.2$) in a reasonably wide range on the $(\Teff,~\log g)$ diagram. 
The numbers of the line candidates selected at this stage are
78 for \ion{Fe}{i}, 7 for \ion{Fe}{ii}, 12 for \ion{Ca}{i}
and 5 for \ion{Ca}{ii}, giving 102 lines in total.

\subsection{Depth measurements}
\label{subsec:measure-depth}

For the expected wavelength $\lambda_c$ of each candidate line,
we measured the depth in every observed spectrum by fitting a parabola, $f=a+b\lambda+c\lambda^2$,
to four or five pixels around $\lambda_c$.
We then obtained the position of the minimum ($\lambda_0$),
the depth ($d$; the distance at $\lambda_0$ from the unity, i.e., $1-f(\lambda_0)$),
and the full-width at half-maximum (FWHM).
If the parabola was not found to be convex downward, we rejected
the fitting and ignored its measurement for the subsequent analysis.
In addition, to take the error of each line depth into account, 
we considered the S/N values, which the WINERED pipeline gave
for three parts of each order (Fig.~\ref{fig:SNRs}).
The S/N at $\lambda_c$ was estimated
by interpolating the S/N values 
for the two consecutive parts around $\lambda_c$.
The errors, indicated by the reciprocal of the S/N, of
the spectra of an object and the telluric standard were combined
in the quadrature to give the depth error,
$e = \sqrt{e_{\rm obj}^2+e_{\rm tel}^2}$,
except the 53rd and 54th orders for which
$e = e_{\rm obj}$ because we did not make the telluric correction. 

For each object, we measured the depths and the accompanying values
for 102 or fewer lines that were not rejected.
Then, we judged whether the measured absorptions were
disturbed or not by examining the depths, wavelengths
and widths that we obtained for each object.
Suppose that we have $N_{\rm line}$ accepted measurements, i.e., 
the measured depth $d_j$, the depth error ($e_j$),
the offset in wavelength ($V_j=c (\lambda_0-\lambda_c)/\lambda_c$) 
and the FWHM ($W_j$) where $j$ varies from 1 to $N_{\rm line}$
for different lines.
The wavelength offsets ($V_j$) and the FWHMs ($W_j$)
are given in the velocity scale ($\kms$).
Some measurements have to be discarded due to some reasons;
e.g., weak lines may be under the detection limit, or there may be
strong contaminating lines that disturb the target lines.
We require the depth of each line to be
more than 0.01, and then we can use the sufficiently deep lines
to calculate the quartiles of the wavelength offset
($V_{1/4}, ~V_{2/4}, ~V_{3/4}$)
and those of FWHM ($W_{1/4}, ~W_{2/4}, ~W_{3/4}$). 
The quartiles were used to detect outliers according to
the interquartile range (IQR) rule, where
the IQRs were given by $\IQR(V)=V_{3/4}-V_{1/4}$
and $\IQR(W)=W_{3/4}-W_{1/4}$.
Every accepted measurement should account for
the wavelength offset $V_j$
within $[V_{1/4}-3\,\IQR(V):V_{3/4}+3\,\IQR(V)]$
and 
the FWHM $W_j$
within $[W_{1/4}-3\,\IQR(W):W_{3/4}+3\,\IQR(W)]$.
The acceptable range of $V$ was replaced with
$[V_{2/4}-5:V_{2/4}+5]$
if $\IQR(V)$ is larger than $5\,\kms$,
while the range of $W$ was replaced with 
$[W_{2/4}-10:W_{2/4}+10]$
if $\IQR(W)$ is larger than $10\,\kms$.
In addition, the fractional errors of depth, $e_j/d_j$,
needs to be smaller than 0.5. 
We performed the validation of measurements based on
these conditions for every object. 
The FWHM, $W$, ranges from 15 to 23.5\,$\kms$ in 
the chosen 42 objects. 
HD\,137107, the double-lined spectroscopic binary (Section~\ref{sec:Data}),
has the largest width, but the difference from
the other objects is not very high
(e.g., four others have $W$ larger than $20\,\kms$). 

\subsection{Line selection based on observed spectra}
\label{subsec:select-line-obs}

In Section~\ref{subsec:measure-depth},
we measured the depths of the lines
selected by using synthetic spectra (Section~\ref{subsec:select-line-synth}),
and validated the measurements.
The next step is to select the lines that give depths validated
for a significant fraction of objects over a wide range
in the $(\Teff, ~\log g)$ plane. 
The selection of each line was made by considering 
the number and the ranges of $\Teff$ and $\log g$ of the objects
for which the measurements of the line were validated:
\begin{description}
\item[(i)] the number of the validated measurements should be 20 or more for each neutral line and 12 or more for each ionized line (among the 42 objects in total),
\item[(ii)] the $\Teff$ range should be 1000\,K or wider,
\item[(iii)] the $\log g$ range should be 2.5\,dex or wider.
\end{description}
Table~\ref{tab:lines} lists the 97 lines confirmed
(76 \ion{Fe}{i},
5 \ion{Fe}{ii}, 11 \ion{Ca}{i} and 5 \ion{Ca}{ii} lines). 
Among these lines, \ion{Ca}{ii}~9890.6280 includes
two transitions with the oscillator strengths of $\log {gf}=0.900$ and $1.013$ at
the same EP listed in VALD. The electric table provided in Supporting Information lists 
the combined strength ($\log {gf} = 1.261$) with the ``*'' mark. 

\input{tab2.tex}

\subsection{LDR relations of individual line pairs}
\label{subsec:line-pair-selection}

Using the lines in Table~\ref{tab:lines},
we searched for line pairs that show tight LDR relations.
The two lines in each line pair should be located in
the same band, i.e.~$Y$ (9760--10890\,{\AA}) or $J$ (11600--13200\,{\AA}).
We ensured that a $Y$-band line was not combined with any $J$-band line.
In the following analysis, we consider the LDR $r=d_1/d_2$, and 
its logarithm to the base 10, $\log r = \log(d_1/d_2)$,
where $d_1$ and $d_2$ indicate the depths of two lines.
The following approximation formulae were used to obtain
the errors in LDR ($y=r$ or $y=\log r$)
on the basis of the errors in depth ($e_1$ and $e_2$, see Section~\ref{subsec:measure-depth}).
\begin{equation}
e_{y} = \left\{\begin{array}{ll}
\left(\frac{d_1}{d_2}\right) \sqrt{\left(\frac{e_1}{d_1}\right)^2+\left(\frac{e_2}{d_2}\right)^2}
& ~({\rm if} ~y=r) \\
\left(\frac{1}{\ln 10}\right) \sqrt{\left(\frac{e_1}{d_1}\right)^2+\left(\frac{e_2}{d_2}\right)^2} 
& ~({\rm if} ~y=\log r) 
\end{array}\right.
\end{equation}

\subsubsection{\ion{Fe}{i}--\ion{Fe}{i} pairs}
\label{subsubsec:FeI-FeI}

We use \ion{Fe}{i}--\ion{Fe}{i} pairs for determining $\Teff$.
The temperature dependency of line depth is characterized by
the EP, and thus a pair of
low-EP and high-EP lines is expected to be sensitive to $\Teff$.
For \ion{Fe}{i}--\ion{Fe}{i} pairs,
every LDR is a ratio of the depth of a low-EP line to
that of a high-EP line, $r=d_{\rm low}/d_{\rm high}$.
For all the pairs of two lines whose EPs differ by 1\,eV or more, 
we fitted the LDR--$\Teff$ relations with four different forms:
\begin{description}
\item[(T1)] ~$r = \alpha + \beta \,\Teff$,
\item[(T2)] ~$r = \alpha + \beta \,\log \Teff$,
\item[(T3)] ~$\log r = \alpha + \beta \,\Teff$,
\item[(T4)] ~$\log r = \alpha + \beta \,\log \Teff$.
\end{description}

We made the least-square fitting with the weights 
$w=1/e_y^2$ given to each point,
where $y$ is $r$ or $\log r$. 
If the two line depths of a given pair were accepted
in all the objects, we would have 42 points to
fit, but this is not the case in most line pairs.
We ignored the line pairs for which the number of validated
measurements, $N$, were smaller than 20 
for the subsequent analysis. 
We also rejected relations with $\beta \geq 0$ because
the depth ratio of a low-EP line to a high-EP line
is expected to increase with decreasing $\Teff$
in our target range.
The weighted residual around each relation, $y=f(x)$, is given by
\begin{equation}
\sigma_y = \sqrt{\left. \sum_{i=1}^{N} w_i \left[ y_i - f(x_i) \right]^2 \right/ \sum_{i=1}^{N} w_i } , 
\label{eq:sigma-y1} 
\end{equation}
where $x$ stands for $\Teff$ or $\log \Teff$ 
according to the form among the four mentioned above. 
This $\sigma_y$ is converted to the scatter in $\Teff$ around the fitted relation by
\begin{equation} 
\sigma_{\Teff} = \left\{\begin{array}{ll}
\left|\sigma_{y}/\beta \right| 
& ~({\rm if} ~x=\Teff) \\
(\ln 10)\,5400 \, \left| \sigma_{y}/\beta \right|
& ~({\rm if} ~x=\log \Teff) 
\end{array}\right. .
\label{eq:sigma-teff}
\end{equation}
We then rejected relations with $\sigma_{\Teff}$ larger than 200\,K.

\begin{figure}
	\includegraphics[clip,width=\columnwidth]{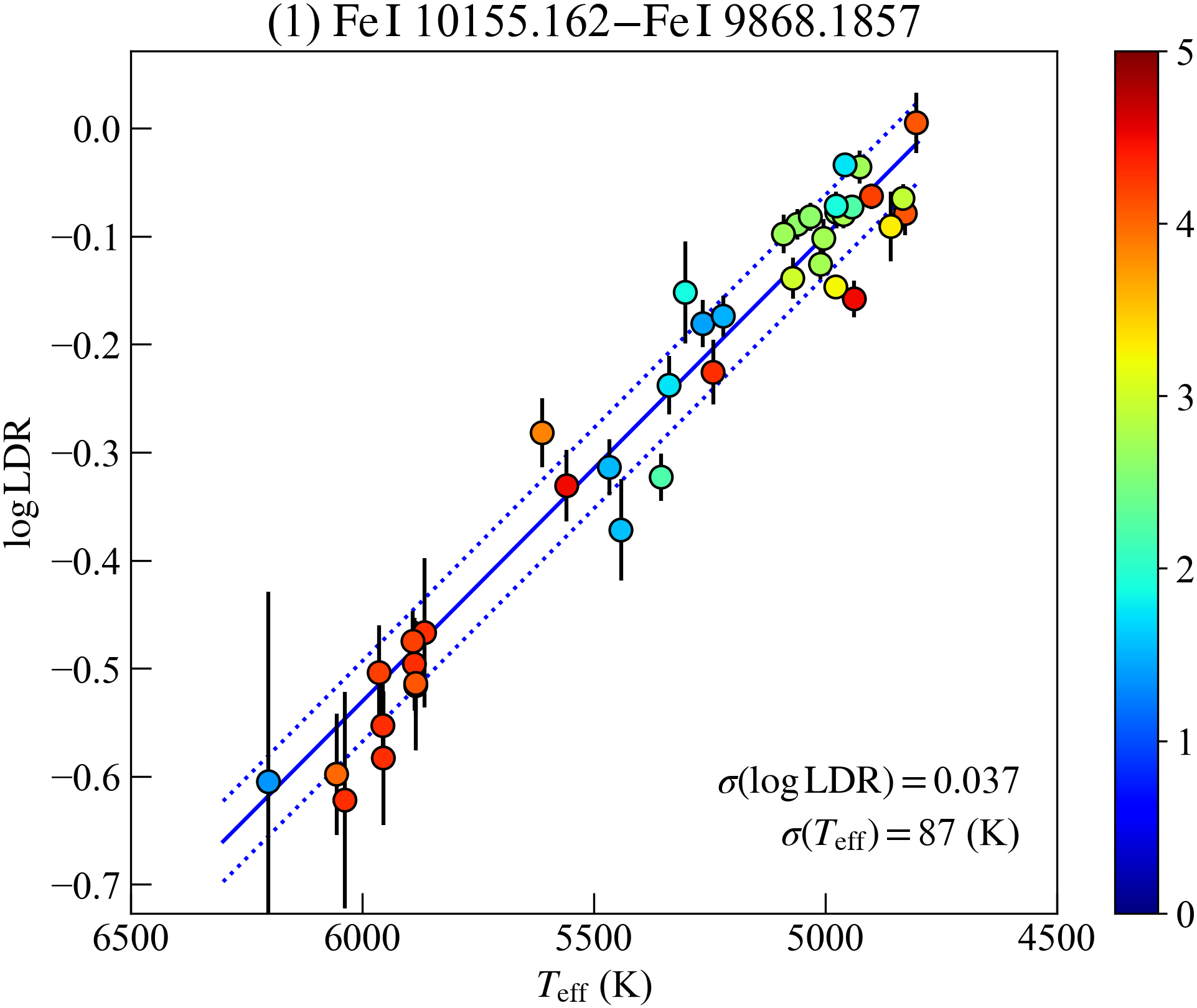}
    \caption{Relations between LDR ($r$ or $\log r$) and effective temperature ($\Teff$ or $\log \Teff$). The points' colours indicate $\log g$ according to the colour scale given on the right.  The solid line indicates the LDR relation (Table~\ref{tab:pairs}) and the dotted lines indicate the width of $\pm \sigma_y$. Plots of all the 13 \ion{Fe}{i}--\ion{Fe}{i} pairs in Table~\ref{tab:pairs} are given in Supporting Information. }
    \label{fig:pairs1}
\end{figure}

In addition, in order to ensure that each relation
can predict the LDRs of stars over a wide $(\Teff, \log g)$ range
of our targets suitably, we consider four groups of stars,
\begin{description}
\item[(a)] ~$5150 < \Teff < 5550$ and $1.2 < \log g < 2.4$ (7 supergiants),
\item[(b)] ~$4750 < \Teff < 5150$ and $2.5<\log g<3.5$ (12 giants),
\item[(c)] ~$4750 < \Teff < 5350$ and $3.8<\log g<4.6$ (5 cool dwarfs),
\item[(d)] ~$5800 < \Teff < 6100$ and $3.8<\log g<4.6$ (11 warm dwarfs),
\end{description}
which are illustrated in Fig.~\ref{fig:HRD}.
First, we rejected the line pairs for which
the number of validated measurements for each group, $N_{\rm grp}$, 
were smaller than 3. 
Then, we compared the LDR-based temperatures ($T_{i}$ from $x_{i}=(y_i-\alpha)/\beta$) 
with the catalogue values ($T_{\mathrm{cat},\,i}$) and calculated
the weighted mean of the deviation for each group,
\begin{equation}
D_{\rm grp} = \left. \sum_{i=1}^{N_{\rm grp}} w_i \left(T_{i}-T_{\mathrm{cat},\,i}\right) \right/ \sum_{i=1}^{N_{\rm grp}} w_i ,
\end{equation}
where the weight is given to each object
according to the measurement error ($w_i=1/e_y^2$).
If any of the $D_{\rm grp}$ values is larger than 120\,K
or smaller than $-120$\,K, we rejected the given relation.
Among the forms that were not rejected due to the conditions of $N_{\rm grp}$ and $D_{\rm grp}$,
we selected the ones with the smallest $\sigma_{\Teff}$ values for each line pair. 
We found 77 \ion{Fe}{i}--\ion{Fe}{i} accepted line pairs,
where each absorption line may be used in many pairs.
The relation between LDR ($r$ or $\log r$) and
effective temperature ($\Teff$ or $\log \Teff$) is illustrated
in Fig.~\ref{fig:pairs1} for every line pair that we finally select
(Section~\ref{subsec:select-unique}).

\subsubsection{\ion{Fe}{i}--\ion{Fe}{ii} and \ion{Ca}{i}--\ion{Ca}{ii} pairs}
\label{subsubsec:FeI-FeII}

We use \ion{Fe}{i}--\ion{Fe}{ii} and \ion{Ca}{i}--\ion{Ca}{ii} pairs for determining $\log g$.
For these pairs, 
every LDR is a ratio of the depth of a neutral line to
that of an ionized line, i.e., $r=d_{\ion{Fe}{i}}/d_{\ion{Fe}{ii}}$
or $d_{\ion{Ca}{i}}/d_{\ion{Ca}{ii}}$.
We searched for the LDR relations in the following forms:
\begin{description}
\item[(TG1)] ~$r = \alpha + \beta\,\Teff + \gamma\,\log g$,
\item[(TG2)] ~$r = \alpha + \beta\,\log \Teff + \gamma\,\log g$,
\item[(TG3)] ~$\log r = \alpha + \beta\,\Teff + \gamma\,\log g$,
\item[(TG4)] ~$\log r = \alpha + \beta\,\log \Teff + \gamma\,\log g$.
\end{description}
We made the least-square fitting 
again with the weights $w=1/e_y^2$.
The subsequent steps for selecting the line pairs 
and the forms of the LDR relations are similar 
to those for \ion{Fe}{i}--\ion{Fe}{i} pairs (Section~\ref{subsubsec:FeI-FeI}),
but the conditions are slightly different.

The minimum number of validated measurements required for
each line pair is 12, instead of 20, considering that
objects showing significant absorption in
both neutral and ionized lines tend to be limited.
The weighted residual around each relation, $y=f(x, \log g)$, is given by
\begin{equation}
\sigma_y = \sqrt{\left. \sum_{i=1}^{N} w_i \left[ y_i - f(x_i, ~\log g_i) \right]^2 \right/ \sum_{i=1}^{N} w_i } , 
\end{equation}
where $x$ and $y$ stand for the temperature ($\Teff$ or $\log \Teff$)
and the LDR ($r$ or $\log r$), respectively.
This $\sigma_y$ is converted to the scatter in $\log g$ around the fitting by
\begin{equation} 
\sigma_{\log g} = \left|\sigma_y / \gamma \right| ,
\label{eq:sigma-logg}
\end{equation}
in addition to $\sigma_\Teff$ given by equation~(\ref{eq:sigma-teff}),
and we rejected relations with $\sigma_{\log g}$ larger than 0.5\,dex.

We also considered the groups of objects in different parts of the $(\Teff, \log g)$ plane,
and calculated the weighted mean of the deviation in $\log g$,
\begin{equation}
D_{\rm grp} = \left. \sum_{i=1}^{N_{\rm grp}} w_i \left(\log g_{i}-\log g_{\mathrm{cat},i}\right) \right/ \sum_{i=1}^{N_{\rm grp}} w_i ,
\end{equation}
where $\log g_{i}$ equals
$(y_i - \alpha - \beta x_i)/\gamma$
in which $x_i$ is provided by the literature $\Teff$, not the LDR-based one.
We stipulated that $N_{\rm grp}$ be 3 or larger and $D_{\rm grp}$ be
smaller than 0.5\,dex for the groups (a) and (d),
i.e.~supergiants and warm dwarfs, but we did not consider 
the groups (b) and (c). Many of the \ion{Fe}{i}--\ion{Fe}{ii}
and \ion{Ca}{i}--\ion{Ca}{ii} pairs lack validated measurements
for the latter two groups because the ionized lines become too weak. 

\begin{figure}
	\includegraphics[clip,width=\columnwidth]{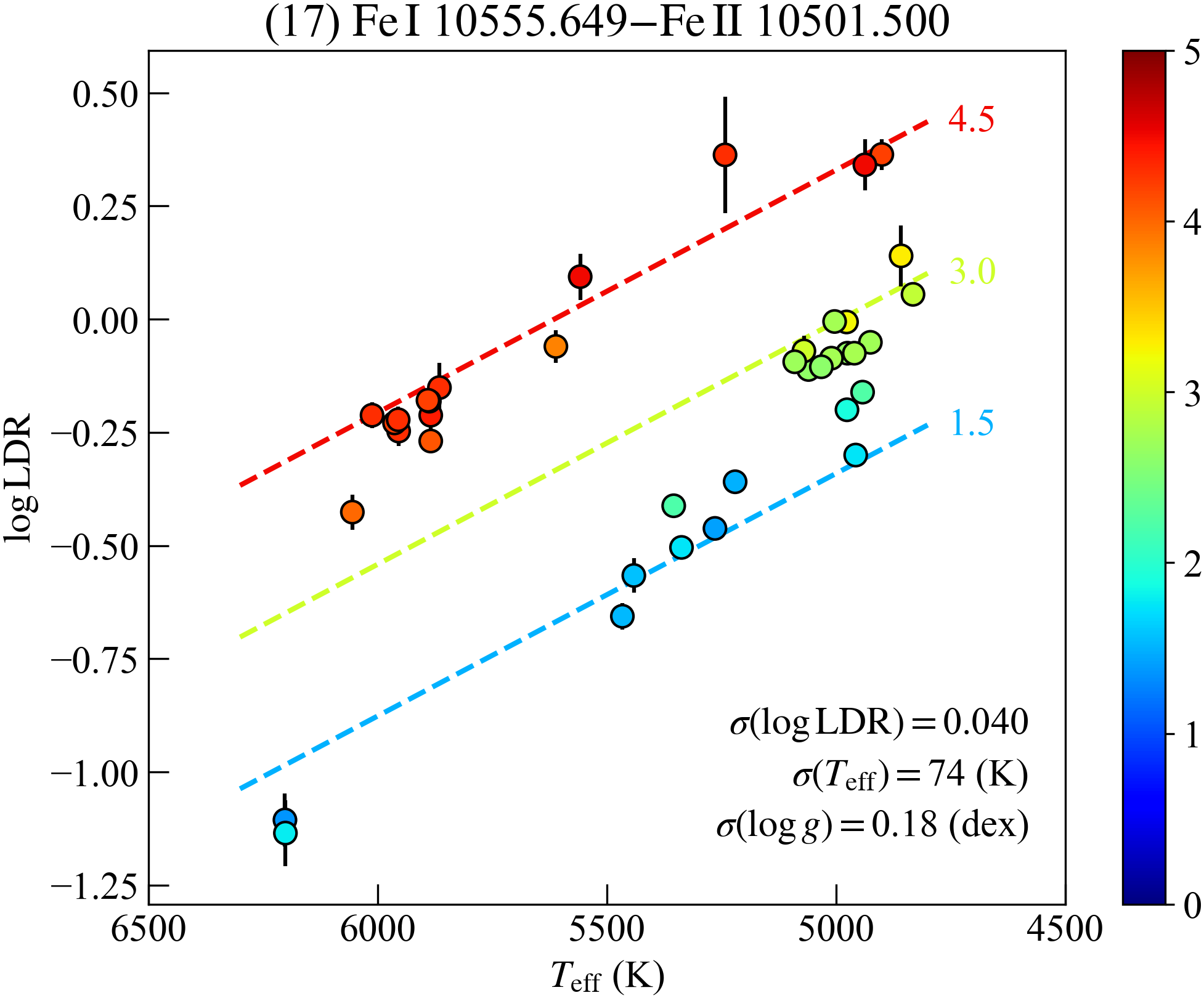}
    \caption{Relations between LDR ($r$ or $\log r$), effective temperature ($\Teff$ or $\log \Teff$) and the surface gravity ($\log g$). The points' colours indicate $\log g$ according to the colour scale given on the right. The three dashed lines indicate the LDR relations at $\log g=1.5$, 3.0 and 4.5\,dex. Plots of all the 5 \ion{Fe}{i}--\ion{Fe}{ii} and 4 \ion{Ca}{i}--\ion{Ca}{ii} pairs in Table~\ref{tab:pairs} are given in Supporting Information. }
    \label{fig:pairs2}
\end{figure}

Finally, we found 175 \ion{Fe}{i}--\ion{Fe}{ii} pairs and 7 \ion{Ca}{i}--\ion{Ca}{ii} pairs.
The relations between LDR ($r$ or $\log r$) and
effective temperature ($\Teff$ or $\log \Teff$) are illustrated
in Fig.~\ref{fig:pairs2} for the line pairs that we finally select
(Section~\ref{subsec:select-unique}).
There are only 5 \ion{Fe}{ii} lines (all the 5 in $Y$),
but each of them is combined
with many \ion{Fe}{i} lines giving a large number of accepted pairs.
The two lines used in \citet{Marfil-2020}, \ion{Fe}{ii}~9997.598 and \ion{Fe}{ii}~10501.500, are included.
In contrast, both \ion{Ca}{i} and \ion{Ca}{ii} lines are
limited in number (4 in $Y$ and 7 in $J$ of \ion{Ca}{i}, and 
3 in $Y$ and 2 in $J$ of \ion{Ca}{ii}).
While all the \ion{Fe}{ii} lines are included in the line pairs accepted,
any pair with \ion{Ca}{ii}~9890.6280 was not accepted
and only four \ion{Ca}{ii} lines are included.

\subsection{Selection of unique line pairs}
\label{subsec:select-unique}

The next step is to select a set of unique line pairs
for which each line is not included in more than one pairs.
The selection was made using a method similar to the one used
by \citet{Taniguchi-2018} but with different algorithms and formulae. 

First, we give the scores of individual
\ion{Fe}{i}--\ion{Fe}{i} line pairs,
\begin{equation}
E_{j} = \sqrt{\left(\sigma_{\Teff, j}\right)^2 + e^2\,\left(\Delta\lambda_j\right) ^2} ,
\label{eq:E1}
\end{equation}
where $j$ stands for the temporary pair ID.
Whereas $\sigma_{\Teff, j}$ is the residual around
each relation (equation~\ref{eq:sigma-teff}),
$\Delta\lambda_j$ is the wavelength separation of 
the paired lines. Pairs with smaller $E_{j}$ values are preferred.
The second term within the square root works as a penalty term that prevents
the selection of pairs of lines with unnecessarily large wavelength separations,
and we used the coefficient $e=0.2$\,K\,{\AA}$^{-1}$.
Then, the score of a set of line pairs is given by
\begin{equation}
E = \frac{1}{\Npair} \sqrt{\sum_{j=1}^{\Npair} E_{j}^2} = \frac{1}{N_{\rm pair}} \sqrt{\sum_{j=1}^{N_{\rm pair}} \left(\sigma_{\Teff, j}\right)^2 + e^2\,\sum_{j=1}^{N_{\rm pair}} \left(\Delta\lambda_j\right) ^2} ,
\label{eq:E}
\end{equation}
where $N_{\rm pair}$ is the number
of line pairs included in the set.
In the case of the \ion{Fe}{i}--\ion{Fe}{ii} and
\ion{Ca}{i}--\ion{Ca}{ii} pairs, we replace
$\sigma_{\Teff}$ in equations~(\ref{eq:E1}) and (\ref{eq:E})
with $\sigma_{\log g}$ and use 
a different coefficient $e$, $5\times 10^{-4}$\,dex\,{\AA}$^{-1}$.

Our purpose is to construct a set of line pairs
whose $E$ is as small as possible. 
On the other hand, having many line pairs is advantageous.
Adding an extra line pair is considered to be useful,
even if it increases $E$ a little.
We calculate the weighted means to estimate
$\TLDR$ and $\log\,\gLDR$, preventing the final precision
from degrading because of the extra pair. We have already selected 
line pairs with relatively small residuals
(Section~\ref{subsec:line-pair-selection}).
Therefore, we try to find the set with the largest number
of line pairs that has the smallest $E$ among
the sets with the same number of line pairs. 

It is simple to find such a set for 
the \ion{Fe}{i}--\ion{Fe}{ii} pairs.
There are five \ion{Fe}{ii} lines.
For each of them, the pair given in Table~\ref{tab:pairs} 
has the smallest $E_{1}$. We selected these five pairs,
leading to the smallest $E$ as a set, 
without including each \ion{Fe}{i} line more than once.
In contrast, among the four \ion{Ca}{ii} lines,
\ion{Ca}{ii}~9854.7588 and \ion{Ca}{ii}~9931.3741 
would give the smallest $E_1$ when they are paired with
\ion{Ca}{i}~10343.819. 
In the case of \ion{Fe}{i}--\ion{Fe}{i}, 
there would be more conflicts if we simply used the pairs
with the smallest $E_1$ for each line.
Therefore,
for the \ion{Ca}{i}--\ion{Ca}{ii} and \ion{Fe}{i}--\ion{Fe}{i} combinations, 
we search for the sets of pairs 
without the conflicts using the following method.

Let us consider the four \ion{Ca}{ii} lines,
which are included in one or more accepted line pairs. 
First, we shuffled the line IDs (1 to 4) to make a queue of
the \ion{Ca}{ii} lines.
Then, starting from the beginning of the queue,
we added to the line-pair set
the pair having a given \ion{Ca}{ii} line 
with the smallest $E_1$ among the pairs that include 
no \ion{Ca}{i} line already used in the set. 
If none of the pairs with a given \ion{Ca}{ii} line
could be selected because of the conflict
(i.e., the \ion{Ca}{i} lines in the pairs had already been used for the set),
we would not include the \ion{Ca}{ii} line, 
However, we were able to include all of the four \ion{Ca}{ii} lines.
By repeating the shuffling of the \ion{Ca}{ii} lines,
we can try different sets of the line pairs and
find the set with the smallest $E$. 
Because the conflicts present between
the \ion{Ca}{i}--\ion{Ca}{ii} pairs and the total number
of those pair itself are limited, we can find the best-achievable
set after just a few trials. 
We used the same algorithm that involved the random shuffling 
to search for the best \ion{Fe}{i}--\ion{Fe}{i} set.
Because there are significantly more \ion{Fe}{i} lines
(14 low-EP lines and 20 high-EP lines) 
and more conflicts involved, 
we had to repeat the procedures from the shuffling
the line IDs (1 to 14 for low-EP lines) to 
the assessment of $E$ of temporary sets
many times. Nevertheless, we obtained the same set
of 13 line pairs after a few thousands
trials, at most {$\sim$}20\,000. 
Among the 14 low-EP \ion{Fe}{i} lines,
both \ion{Fe}{i}~10114.014 and \ion{Fe}{i}~10332.327 
have accepted LDR relations for which the high-EP line is \ion{Fe}{i}~10674.070, and
the finally selected set includes the pair with
\ion{Fe}{i}~10332.327 rather than the pair with \ion{Fe}{i}~10114.014.

\begin{figure}
	\includegraphics[clip,width=\columnwidth]{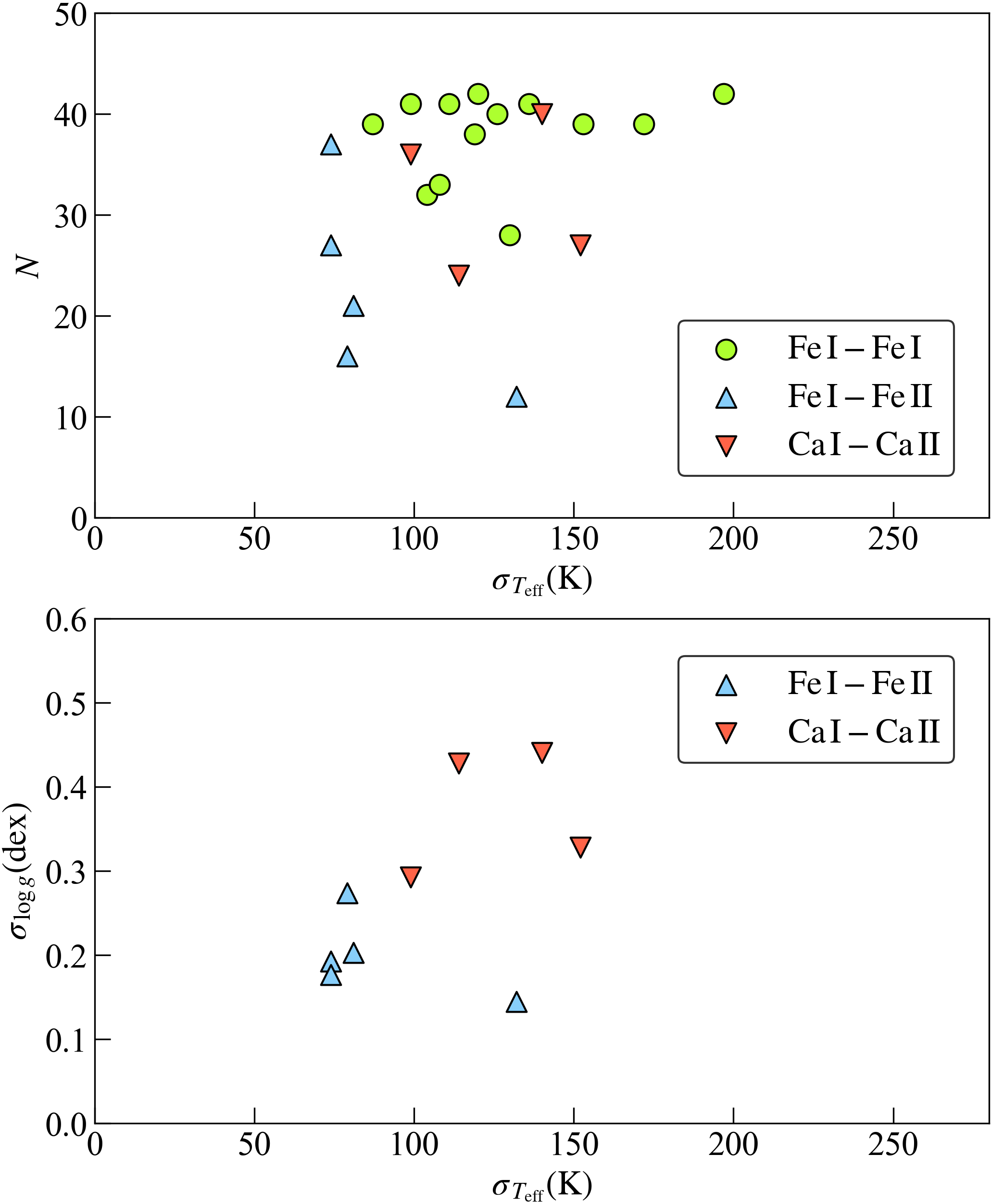}
    \caption{(Upper)~The numbers, $N$, of the objects used for deriving the LDR relations (Table~\ref{tab:pairs}) are plotted against the residuals in $\Teff$ ($\sigma_{\Teff}$ in equation~\ref{eq:sigma-teff}). (Lower)~The residuals in $\log g$ ($\sigma_{\log g}$ in equation~\ref{eq:sigma-logg}) are plotted against $\sigma_{\Teff}$.}
    \label{fig:pair_sigmas}
\end{figure}

Table~\ref{tab:pairs} lists the sets of line pairs selected. 
The numbers of line pairs included in the selected sets are
13 for \ion{Fe}{i}--\ion{Fe}{i}, 5 for \ion{Fe}{i}--\ion{Fe}{ii} and 4 for \ion{Ca}{i}--\ion{Ca}{ii}.
Fig.~\ref{fig:pair_sigmas} plots the numbers, $N$, of 
the objects used for deriving the LDR relations against
$\sigma_{\Teff}$ for all the line pairs in the upper panel
and $\sigma_{\log g}$ against $\sigma_{\Teff}$ for
the \ion{Fe}{i}--\ion{Fe}{ii} and \ion{Ca}{i}--\ion{Ca}{ii} pairs
in the lower panel.
Ionized lines tend to be weak or invisible in
low-$\Teff$ stars, and $N$ for \ion{Fe}{i}--\ion{Fe}{ii} 
and \ion{Ca}{i}--\ion{Ca}{ii} pairs tend to be small.
We discuss this point again in Section~\ref{subsec:expectation}.
There is a correlation seen in the lower panel.
For the \ion{Fe}{i}--\ion{Fe}{ii} and \ion{Ca}{i}--\ion{Ca}{ii}
pairs, $\sigma_{\Teff}$ and $\sigma_{\log g}$ were obtained
with conversions of the same $\sigma_y$ where $y$
is $r$ or $\log r$.
Therefore, the correlation indicates that
the ratio of the $\Teff$ dependency (represented by $\beta$)
to the $\log g$ dependency (represented by $\gamma$)
is similar to each other for most line pairs.
The exception is \ion{Fe}{i}\,10145.561--\ion{Fe}{ii}\,10173.515,
with $\sigma_{\Teff}\sim 130$\,K, 
whose LDRs show relatively strong sensitivity to $\log g$
but limited sensitivity to $\Teff$.
This means that the error in $\Teff$ has only minor impact on 
the estimate of $\log g$.
However, \ion{Fe}{ii}\,10173.515 is rather weak
($N=12$ for its pair),
and thus the usefulness of this line pair
as a surface-gravity indicator may be relatively limited. 

\input{tab3.tex}

In our relations with the $\log g$ term (TG1--TG4),
$\gamma$ indicates the LDR's sensitivity to the surface gravity.
This can be compared with the $\log g$ dependency
detected by \citet{Jian-2020}. They measured 
the offsets in $\log r$, measured at a certain $\Teff$,
between dwarfs, giants and supergiants for dozens of
relations of $YJ$-band LDRs given by \citet{Taniguchi-2018}.
They found non-zero offsets in many line pairs,
indicating significant impacts of surface gravity.
In particular, the offsets between dwarfs and supergiants
(corresponding to the difference of {$\sim$}3\,dex in $\log g$)
tend to be large, up to {$\sim$}0.6\,dex,
depending on the ionization potentials of
the two elements in each line pair. 
Such dependency corresponds to
$0 \leq |\gamma| \lesssim 0.2$. 
Most of the six \ion{Fe}{i}--\ion{Fe}{ii} and \ion{Ca}{i}--\ion{Ca}{ii} pairs with the $y=\log r$ forms,
common with the form in \citet{Jian-2020}, have
$\gamma \sim 0.2$ or higher (the two smallest being
0.1588 and 0.1827).
Among {$\sim$}40 line pairs for which 
the $\log g$ sensitivity was measured in \citet{Jian-2020},
only four\footnote{The pairs, (T10), (T30), (T41) and (T54), in Table~4 of \citet{Jian-2020} have $\Delta \log\,{\rm LDR_{ds}} \geq 0.4$, corresponding to $\gamma \gtrsim 0.14$.}
show such high sensitivity. 
This comparison suggests that the pairs of
neutral and ionized lines are indeed sensitive to $\log g$.

\section{Discussion}
\label{sec:discussion}

\subsection{Application of the LDR relations}
\label{subsec:derivation}

In this section, we describe an application of the LDR relations
to derive $\Teff$ and $\log g$, and verify its performance.
Due to the limited sample of available spectra, we apply the relations
between the calibrators themselves in the subsequent analysis.
This helps us check the self-consistency
and provides an estimate of typical uncertainty. It is necessary to
test the relations and evaluate the systematics on derived parameters, $\Teff$ and $\log g$,
by using a new dataset for different objects in the future. 

The first step of the application procedure we propose is to
use the \ion{Fe}{i}--\ion{Fe}{i} pairs for estimating $\Teff$.
We only consider the pairs of which the LDR ($y=r$ or $\log r$)
is measured and validated for a given object.
Let us suppose $N_T$ LDR values and their errors,
$y_j \pm e_j$ ($1\leq j\leq N_T$), are available
among the 13 \ion{Fe}{i}--\ion{Fe}{i} pairs in Table~\ref{tab:pairs}.
For each LDR relation,
$y_j = \alpha_j + \beta_j\,x$
(where $x=\Teff$ or $\log \Teff$ according to the form selected for each line pair), 
we combine the scatter around the relation ($\sigma_y$ in equation~\ref{eq:sigma-y1}) with the error in the measured LDR
to give the error in $y_j$, i.e., $\varepsilon_j = \sqrt{e^2 + \sigma_y^2}$.
Then, we can estimate the effective temperature and its error with
each individual LDR by
\begin{equation}
T_j \pm \Delta_j = \left\{ \begin{array}{ll}
(y_j-\alpha_j)/\beta_j \pm \varepsilon_j/\beta_j & ~({\rm if}~x=\Teff) \\
10^{(y_j-\alpha_j)/\beta_j} \, \left[ 1 \pm (\ln 10) \varepsilon_j/\beta_j \right] & ~({\rm if}~x=\log \Teff) .
\end{array} \right. 
\label{eq:T_j}
\end{equation}
Then, we calculate the LDR-based temperature by
taking the weighted mean of $T_j$:
\begin{equation}
\TLDR = \left. \sum_{j=1}^{N_T} \left(w_j T_j\right) \right/ \sum_{j=1}^{N_T} w_j ,
\end{equation}
where $w_j$ is the weight given by the error in equation~(\ref{eq:T_j}).
For the error in $\TLDR$, we compare two kinds of errors:
\begin{eqnarray}
\epsT({\rm wav}) &=& \left(\frac{1}{N_T-1} \frac{\sum_{j=1}^{N_T} w_j \left(T_j-\TLDR\right)^2}{\sum_{j=1}^{N_T} w_j} \right)^{1/2} , \label{eq:epsTwav} \\
\epsT({\rm prop}) &=& \left(\sum_{j=1}^{N_T} w_j \right)^{-1/2} \label{eq:epsTprop} .
\end{eqnarray}
The former is the weighted standard error, which 
reflects the dispersion of $T_j$, while the latter is
the error propagated from the errors of $T_j$. 

The second step is to estimate $\log g$ using the \ion{Fe}{i}--\ion{Fe}{ii} and \ion{Ca}{i}--\ion{Ca}{ii} pairs.
$N_G$ denotes the number of these line pairs with LDR available.
Some objects have $N_{G}$ less than 3, and we did not estimate
their $\log\,\gLDR$ (see Section~\ref{subsec:expectation}). 
Each LDR relation in Table~\ref{tab:pairs}, 
$y_j = \alpha_j + \beta_j x_j + \gamma_j \log g$
where the assigned form determines $x_j=\Teff$ or $\log \Teff$, 
leads to an estimate of $\log g$ for a given set of 
$x_j$ and $y_j$ ($1\leq j \leq N_G$). We again
combine the measurement error and the residual around the relation
to give the error in $y_j$, i.e., $\varepsilon_j = \sqrt{e^2+\sigma_y^2}$. 
If we assume the $\Teff$ (leading to $x_j$) to be precise,
the surface gravity and its error
based on each line pair is given by
\begin{equation}
\log g_j = (y_j - \alpha_j - \beta_j x_j ) / \gamma_j \pm \varepsilon_j/\gamma_j .
\label{eq:G_j}
\end{equation}
We can calculate the weighted mean and 
its standard error ($\log \gLDR \pm \epsG$), for the latter of which
we considered the two kinds of error estimates
like equations (\ref{eq:epsTwav}) and (\ref{eq:epsTprop}).
We rejected estimates that deviated significantly from
our parameter range, i.e., $\log g_j < 0$ or $\log g_j > 6$,
and did not include them in calculating the weighted mean and
the subsequent analysis.
We rejected one or two $\log g_j$ for each of six objects (all having
$\log g > 3.5$\,dex in Table~\ref{tab:targets}).
We could have done a similar rejection of the temperature estimates
by equation~(\ref{eq:T_j})
if $T_j < 4000$ or $T_j > 7500$, but none of the $T_j$ was rejected. 
In estimating $\log g_j$, ionized lines (\ion{Fe}{ii} and \ion{Ca}{ii})
tend to be rather weak especially in dwarfs, and 
very large errors may be observed in the estimation.

The error in the temperature, if not negligible, has 
a systematic effect on the LDR-based surface gravity.
$\beta$ is negative in all the relations, and 
an overestimation of $\Teff$ would result in a positive shift
of $\log g_j$ from every line pair, and vice versa.
The error in the temperature, $\epsT$, translates to
the error in $x$,
\begin{equation}
\varepsilon_{x} = \left\{ \begin{array}{ll}
\epsT & ~({\rm if}~x=\Teff) \\
(\ln 10)^{-1} \,\epsT/\Teff & ~({\rm if}~x=\log \Teff) 
\end{array} \right. .
\end{equation}
The systematic error in the LDR-based $\log g$ 
due to the temperature error is then given by
\begin{equation}
\delG = \left. \sum_{j=1}^{N_G} \left(-w_j \beta_j \varepsilon_{x}/\gamma_j\right) \right/ \sum_{j=1}^{N_G} w_j
\end{equation}
where the weights, $w_j=1/\varepsilon_j^2$,
of individual line pairs are the ones used for deriving
$\log \gLDR$ (equation~\ref{eq:G_j}). 

\begin{figure*}
	\includegraphics[clip,width=\textwidth]{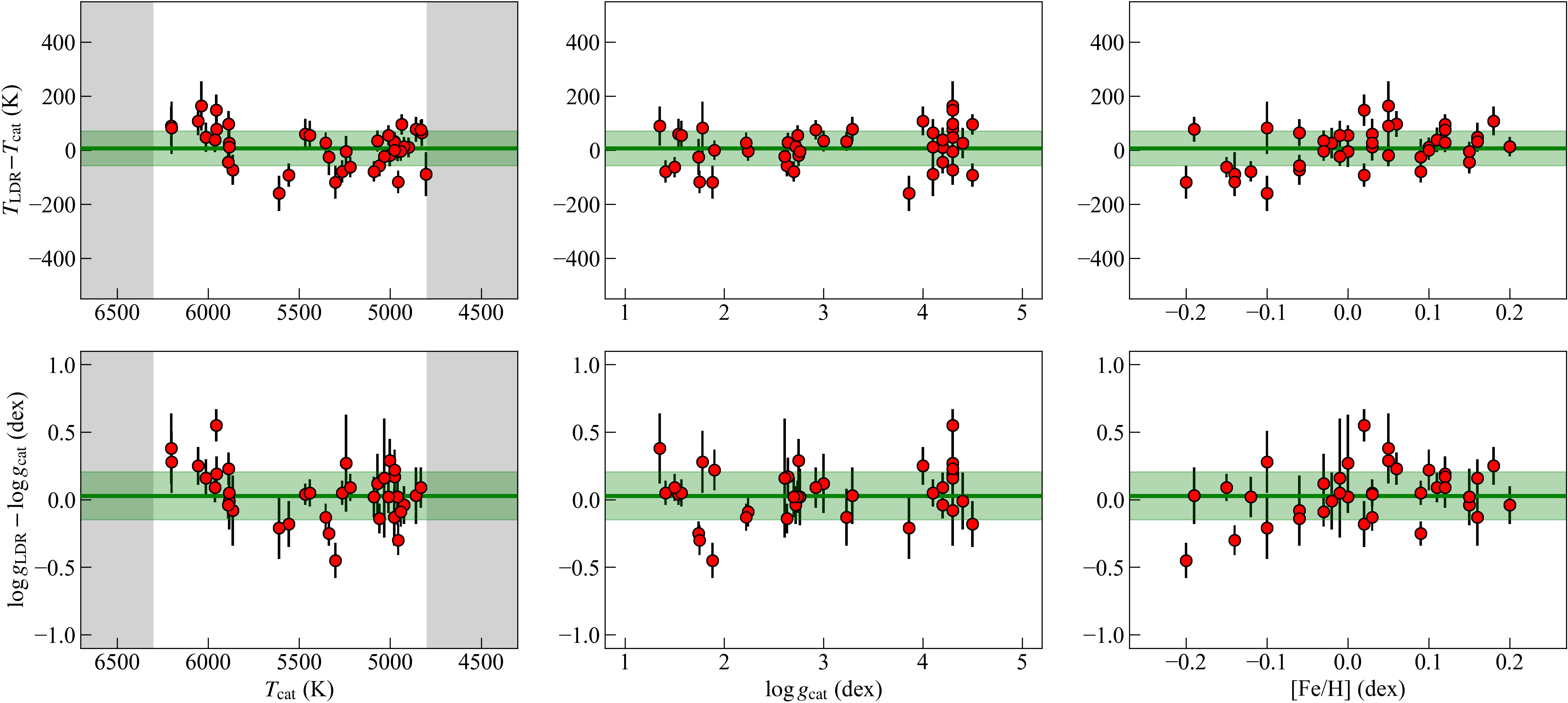}
    \caption{Plotted against three stellar parameters are $\Delta_{T}$ (upper panels) and $\Delta_{G}$ (lower panels). The horizontal strips indicate the weighted standard deviations, {$\pm$}42\,K in $\Delta_{T}$ and {$\pm$}0.17\,dex in $\Delta_{G}$.}
    \label{fig:calcTG}
\end{figure*}

Table~\ref{tab:calcTGs} lists the derived $\TLDR\,(\pm \epsT)$ 
and $\log\,\gLDR\,(\pm \epsG \pm \delG)$
and their differences from the catalogue values,
$\Tcat$ and $\log \gcat$,
\begin{eqnarray}
\Delta_{T} &=& \TLDR - \Tcat , \\
\Delta_{G} &=& \log\,\gLDR - \log\,\gcat .
\end{eqnarray}
Fig.~\ref{fig:calcTG} plots $\Delta_T$ and $\Delta_G$ against
three stellar parameters, i.e., $\Teff$, $\log g$ and $\FeH$.
The weighted standard deviations, 64\,K in $\Delta_{T}$ and 0.18\,dex in $\Delta_{G}$, correspond to the half widths
of the horizontal strips.
There is no clear correlation between the deviations
from catalogue values and the stellar parameters. 

\input{tab4.tex}

\begin{figure}
	\includegraphics[clip,width=\columnwidth]{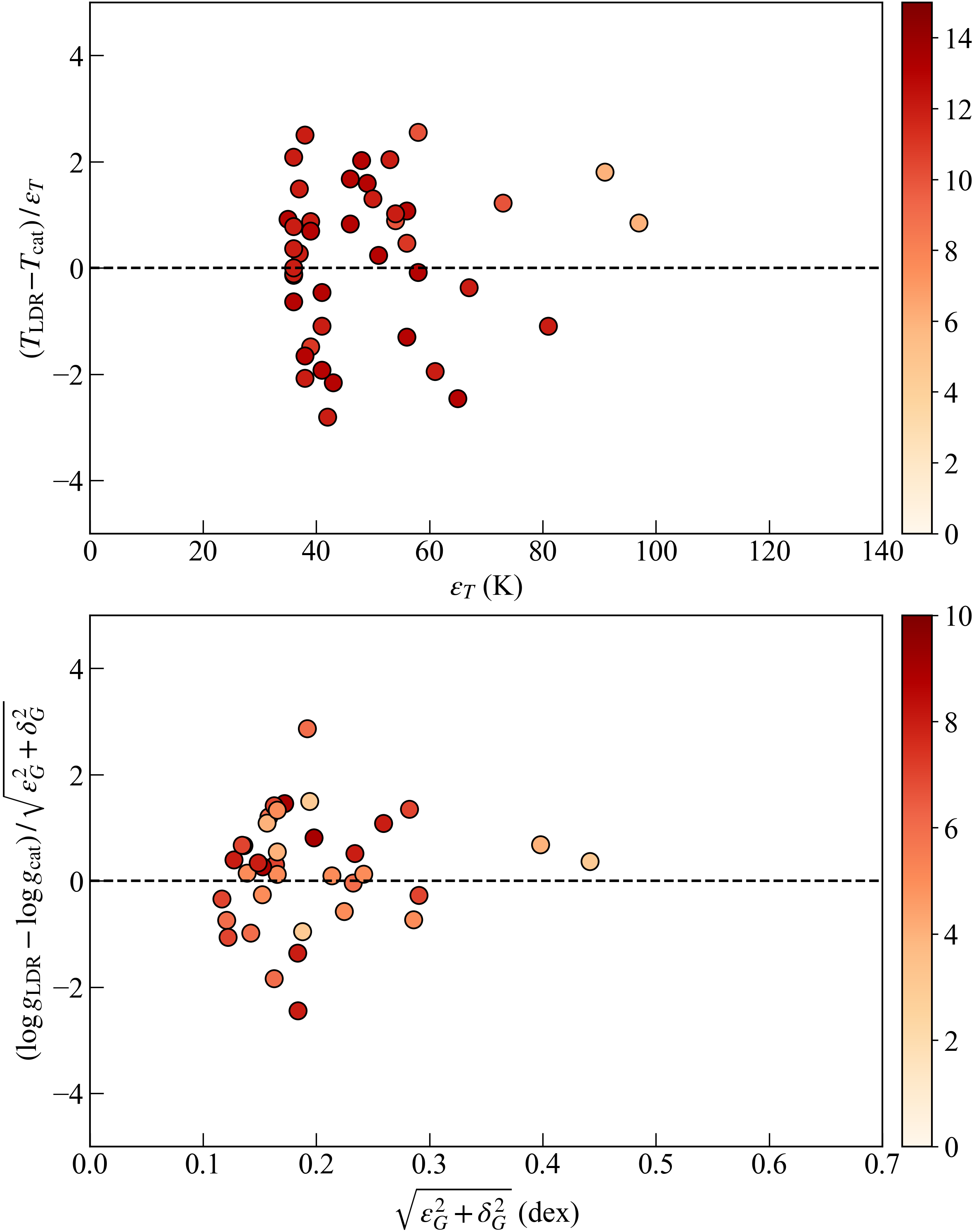}
    \caption{The differences between the LDR-based values and the catalogue values are compared with the estimated errors in the LDR-based values. The points' colours indicate the numbers of line pairs used for the estimation.}
    \label{fig:TGerrs}
\end{figure}

Fig.~\ref{fig:TGerrs} plots $\Delta_{T}/\epsT$
and $\Delta_{G}/\sqrt{\epsG^2+\delG^2}$
against the errors estimated with the LDR method. 
Roughly speaking, the deviations from the catalogue values are 
consistent with the scatter expected with the estimation errors.
In the upper panel, 69\,\% and 88\,\% of the points are
located within $\pm 1$ and $\pm 2$, respectively, while 75\,\% and
97\,\% are within $\pm 1$ and $\pm 2$ in the lower panel. 
The Gaussian error distribution would give 68\,\% and 95\,\%
within $\pm 1$ and $\pm 2$. 
The estimated errors, i.e., $\epsT$ and
$\sqrt{\epsG^2+\delG^2}$,
seem to represent realistic uncertainties. 
The median of $\epsT$ in Table~\ref{tab:calcTGs} is {$\sim$}45\,K,
while the median of 
$\sqrt{\epsG^2+\delG^2}$,
is {$\sim$}0.18\,dex.

\subsection{Comparison with other gravity scales}
\label{subsec:logg-scale}

In the previous section, we have examined the consistency and
the internal precision of the method. 
However, the accuracy of the resultant $\log g$ scale depends entirely on
the $\log g$ used for the calibration. 
The literature values we used were obtained with spectroscopic methods,
and the balance between the neutral and ionized lines has
an essential role in giving constraints on $\log g$. 
The previous works assumed that the approximations of
LTE and hydrostatic equilibrium hold. 
These kinds of equilibrium have been questioned, and
including the non-LTE effects has often modified and improved 
results of spectroscopic analysis (e.g., \citealt{Asplund-2005}; 
\citealt{Bergemann-2019}; \citealt{Amarsi-2020}; \citealt{Bialek-2020}).
Moreover, it has been demonstrated that significant non-LTE effects appear 
in some Fe lines \citep{Bergemann-2012} and Ca lines \citep{Osorio-2019}
that are located within the wavelength range of our interest.
Such non-canonical effects may be important for the LDR method if
the stellar parameters for the calibration are affected by those effects.
In Section \ref{subsec:select-line-synth}, we used synthetic spectra with
the equilibrium assumed for selecting the candidates of lines to use,
but such an assumption has no impact on the steps after this selection
because the calibration of the LDR relations is entirely empirical.
However, if the literature values we used had systematic errors caused by
the classical assumption of equilibrium, the current relations would
lead to results affected by the same errors. 

Here, we compare the $\log\,\gLDR$ values with
the surface gravities obtained using other studies for a sub-sample of the calibrators:
22 objects from \citet{AllendePrieto-1999a}, 
14 objects from \citet{Anders-2019} 
and five objects from \citet{Gray-2019}.
\citet{AllendePrieto-1999a} used
the trigonometric parallaxes from the {\it Hipparcos} mission
to put nearby stars within 100\,pc from the Sun
on the $(B-V)$--$M_V$ plane and compared their positions
with the evolutionary models of \citet{Bertelli-1994}.
Although these observational and theoretical data can be replaced
with more up-to-date sets, their results give the largest
sample for the comparison.
\citet{Anders-2019} obtained the stellar parameters
together with the distance and interstellar extinction
by comparing the broad-band magnitudes from {\it Gaia} DR2
and other large photometric surveys with the theoretical values based on 
the PAdova and TRieste Stellar Evolution Code \citep[PARSEC;][]{Bressan-2012}
and the synthetic stellar spectra \citep{Kurucz-1993}.
These two works, i.e.~\citet{AllendePrieto-1999a} and \citet{Anders-2019}, 
are photometry-based.
In contrast,
\citet{Gray-2019} took a spectroscopic approach
to estimate $\Teff$ and $\log g$, together,
of 26 G and early K-type stars (dwarfs and giants)
by comparing the observed and theoretical equivalent widths
of \ion{Ni}{i}, \ion{V}{i}, \ion{Fe}{i} and \ion{Fe}{ii} lines
(10 lines in total).
Their lines are well-characterized and
situated within a narrow range of the optical regime
6224--6266\,{\AA}.
As far as $\log g$ is concerned, the solutions in their analysis
were obtained by postulating that the stellar abundances
derived with \ion{Fe}{ii} agree with 
those derived with the neutral lines.

\begin{figure}
	\includegraphics[clip,width=\columnwidth]{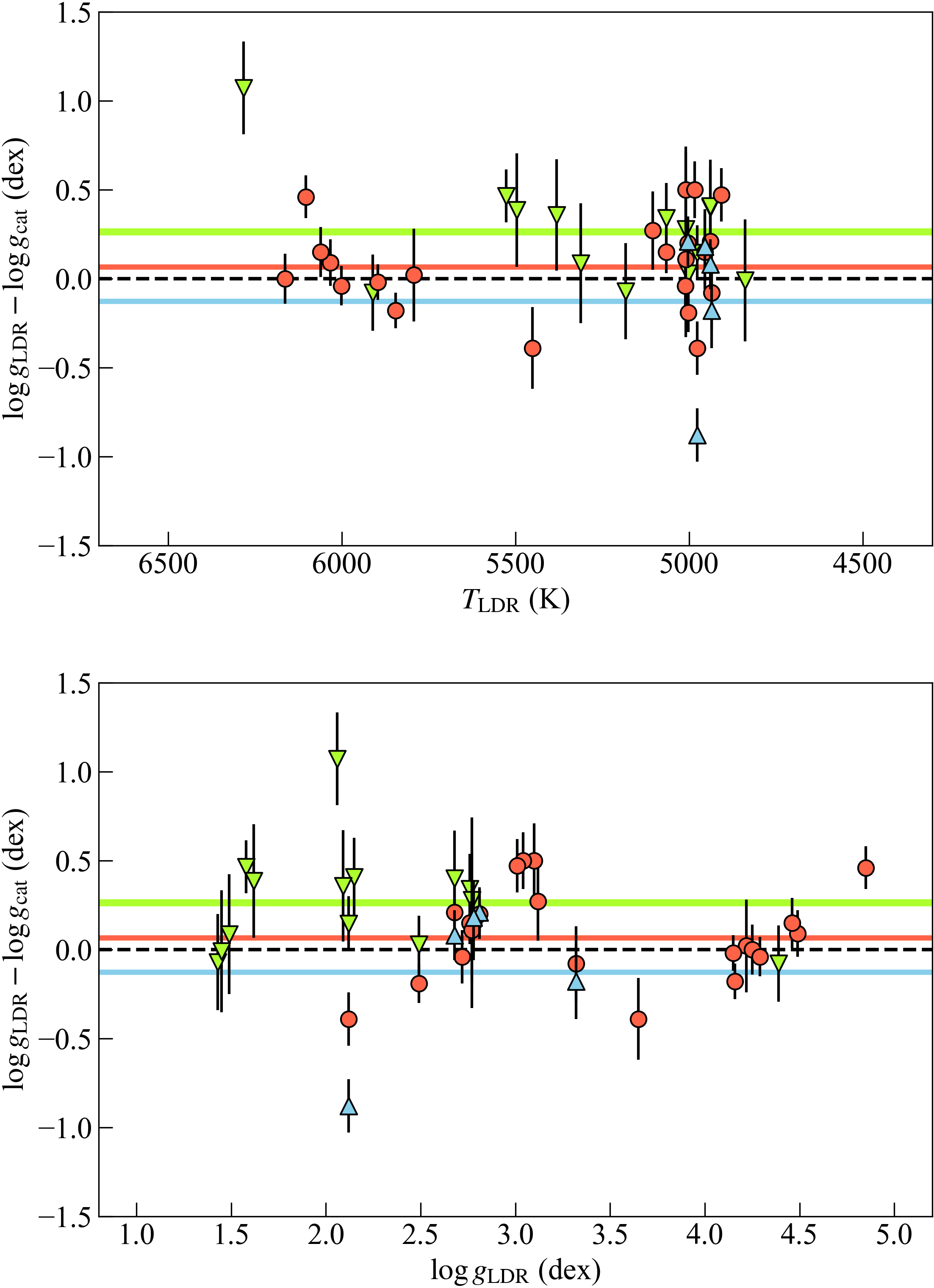}
    \caption{Comparison of the LDR-based surface gravities with $\log g$ obtained by \citet[][indicated by blue upright triangles]{AllendePrieto-1999a}, \citet[][green upside-down triangles]{Anders-2019} and \citet[][red circles]{Gray-2019}. The differences from the literature values are plotted against $\TLDR$ in the upper panel and against $\log\,\gLDR$ in the lower panel. Horizontal solid lines indicate the weighted means ($\mu_\Delta$) of the differences from the literature values (given in Section~\ref{subsec:logg-scale}), and the dashed line merely indicates the zero difference.}
    \label{fig:calcTGs2}
\end{figure}

Fig.~\ref{fig:calcTGs2} presents the comparison of
$\log\,\gLDR$ with the $\log\,\gcat$ from
\citet{AllendePrieto-1999a}, \citet{Anders-2019} and \citet{Gray-2019}.
Using our errors for the weights, $w=(\epsG^2+\delG^2)^{-1}$,
the weighted means ($\mu_\Delta$) and standard deviations ($\sigma_\Delta$)
of the differences
from the literature values are as follows:
$\mu_\Delta=0.07$\,dex and $\sigma_\Delta=0.24$\,dex
in comparison with \citet{AllendePrieto-1999a}, and 
$\mu_\Delta=-0.13$\,dex and $\sigma_\Delta=0.43$\,dex
in comparison with \citet{Gray-2019}.
In the case of \citet{Anders-2019},
their catalogue lists errors in $\log g$,
some of which are as large as our errors.
Therefore, we combined these errors in quadrature
and obtained 
$\mu_\Delta=0.26$\,dex and $\sigma_\Delta=0.27$\,dex.
Although we cannot rule out the systematic trends
between the sets of $\log g$ obtained from different studies, 
our $\log\,\gLDR$ values are found to be similar to
the surface gravity scales in the literature, within {$\sim$}0.2\,dex,
across a wide range over 3\,dex in $\log g$.
The reasonable agreement of our values with
both photometry- and spectroscopy-based results in the literatures
indicates that the systematic error in our $\log g$ scale is not enormous.

Two objects show relatively large offsets.
HD\,202109 ($\zeta$~Cyg, with $\Teff=4976$\,K and $\log g=1.90$\,dex in Table~\ref{tab:targets})
is a spectroscopic binary with
a white dwarf companion \citep{Dominy-1983,Griffin-1992}.
We obtained $\log\,\gLDR=2.12$\,dex. It is slightly higher
than the value in Table~\ref{tab:targets} taken from \citet{Luck-2014}
and 1.98\,dex by \citet{Anders-2019}, 
but lower than 2.51\,dex by \citet{AllendePrieto-1999a} and 3.0\,dex in \citet{Gray-2019}.
HD\,171635 (d~Dra or 45~Dra) has $\Teff=6201$\,K and $\log g=1.78$\,dex in Table~\ref{tab:targets}.
Our value, 2.06\,dex, is higher than the value in Table~\ref{tab:targets} taken from \citet{Luck-2014}
and 0.99\,dex by \citet{Anders-2019}. 
We used five and eight line pairs
(among the nine \ion{Fe}{i}--\ion{Fe}{ii} and \ion{Ca}{i}--\ion{Ca}{ii} pairs in Table~\ref{tab:pairs})
for HD\,202109 and HD\,171635, respectively,
and the estimates with individual line pairs agree with each other
for each object.
It is unclear how the large deviations took place,
and their true $\log g$ values need to be established.

\subsection{Expected targets}
\label{subsec:expectation}

\begin{figure*}
	\includegraphics[clip,width=\textwidth]{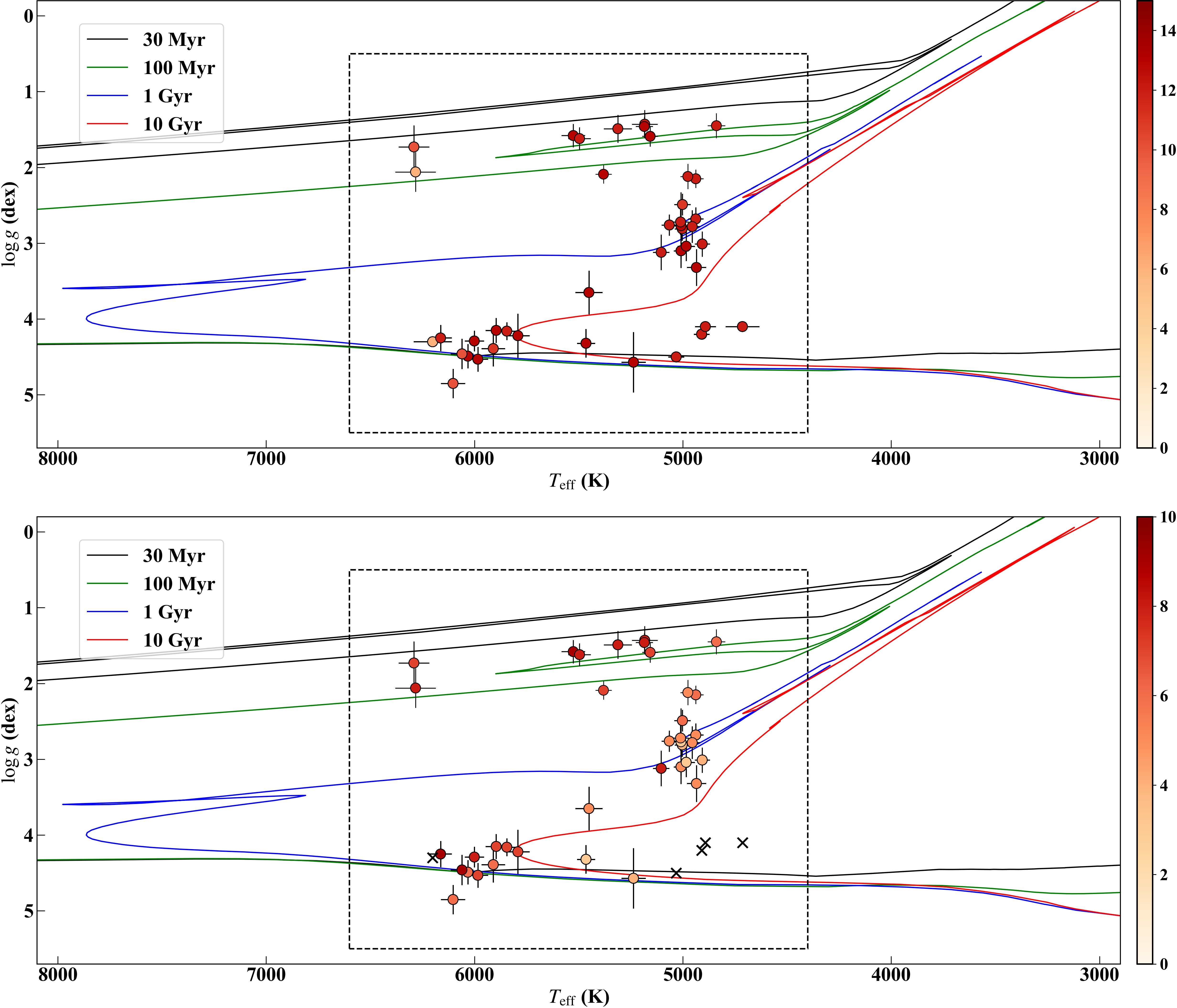}
	\caption{Locations of $\Teff$ and $\log g$ derived with the LDR relations. The colour of circles indicates the number of used line pairs, i.e., \ion{Fe}{i}--\ion{Fe}{i} pairs ($N_T$) in the upper panel and \ion{Fe}{i}--\ion{Fe}{ii} plus \ion{Ca}{I}--\ion{Ca}{ii} pairs ($N_G$) in the lower panel. As presented in the upper panel, $\Teff$ were derived with $N_T\geq 5$ for all the objects. In contrast, $\log g$ was not derived if $N_G$ is smaller than 3, and such objects are indicated by crosses in the lower panel. The isochrones for four different ages are drawn (see more details in text), and the dashed rectangle indicates the range covered in Fig.~\ref{fig:HRD}.}
    \label{fig:HRD2}
\end{figure*}

Our method requires both neutral and ionized lines of
each element to be visible in a given star.
This requirement is satisfied only in stars with
the spectral types around G.
Fig.~\ref{fig:HRD2} represents the numbers of the line pairs
we could use for deriving $\TLDR$ (upper panel)
and $\log\,\gLDR$ (lower panel).
The isochrones for four different ages (30\,Myr, 100\,Myr, 1\,Gyr and 10\,Gyr)
are included to illustrate what kinds of stars can be the targets of 
our relations. We generated these isochrones for the solar metallicity
using the CMD~3.4 tool\footnote{http://stev.oapd.inaf.it/cmd} for which
the PARSEC models are used \citep{Bressan-2012}.
We include evolutionary stages up to early asymptotic giant branch (E-AGB),
i.e.~the thermally pulsing and post-AGB stages are not included.
The upper panels of Fig.~\ref{fig:HRD2} shows that
$T_{\rm LDR}$ was derived for all the 42 objects.
As mentioned in Section~\ref{subsubsec:FeI-FeI}, we required that
every line pair gives 
the LDRs measured for the four groups,
the groups (a) to (d) in Fig.~\ref{fig:HRD}, across the HRD.
In the highest temperature range ($\Teff \gtrsim 6000$\,K),
the numbers of the LDRs tend to be small because
\ion{Fe}{i} lines get weak.
In contrast, 
we could not derive $\log\,\gLDR$ of five objects
indicated by the crosses in the lower panel of Fig.~\ref{fig:HRD2}, 
because the available LDRs are less than 3 for them. 
Towards the low $\Teff$ end ($\Teff \lesssim 5250$\,K),
\ion{Fe}{ii} lines become weak, especially among the higher-$\log g$ objects.
Three of them\footnote{HD\,79555, HD\,82106 and HD\,219134 at $\log g \simeq 4.2$ and $\Teff < 5000$} are located above the main sequence,
where no star is expected. Their $\log g$ are taken from
\citet{Kovtyukh-2004}, while some recent references give 
higher values, $\log g \simeq 4.6$, consistent with the isochrones
\citep[e.g.,][]{Soubiran-2016}. Most of the ionized lines
(\ion{Fe}{ii} and \ion{Ca}{ii}) were not measured well for these objects,
and they have no significant impact on the calibration of our relations.
On the other hand, the warm dwarf
at $\Teff > 6000$\,K lacking $\log\,\gLDR$ is
HD\,137107, the double-lined binary
mentioned in Section~\ref{sec:Data}.
In the case of this object, the large FWHM
(Section~\ref{subsec:measure-depth})
in combination with its relatively high $\Teff$
prevents us from measuring some lines
among both neutral and ionized ones. 

Unfortunately, the $\Teff$ range in which
both neutral and ionized lines appear is narrow.
For \ion{Fe}{i}--\ion{Fe}{ii} and \ion{Ca}{i}--\ion{Ca}{ii} pairs,
our examination of $YJ$-band spectra including synthetic ones
indicates that the LDRs of these pairs are useful at around
4800--6300\,K only.
This range corresponds to the spectral types of
late F to early K, which roughly covers
the Cepheid instability strip. 

Pulsating stars in the instability strip, classical Cepheids
and RR Lyrs in particular, are useful tracers of the Galactic
chemical evolution \citep{Beaton-2018,Matsunaga-2018}.
Our sample does not include any of these pulsating stars,
and it is necessary to test how well our method works for them.
For example, classical Cepheids have been used to
measure the metallicity gradient of the Galactic disk
\citep{Genovali-2014,Genovali-2015,daSilva-2016}.
A vast majority of the previous spectroscopic analyses
on Cepheids were done with optical spectra,
but the application of infrared spectra has been initiated
\citep{Inno-2019,Minniti-2020}. 
In the pioneering two papers, Cepheids' stellar parameters were determined by
searching for the best-matched synthetic spectra (Section~\ref{sec:intro}).
However, the infrared spectra of Cepheids are poorly reproduced
compared to the optical spectra.
In terms of the determined metallicities, for example, 
\citet{Inno-2019} and \citet{Minniti-2020} estimated 
the errors to be 0.15\,dex and 0.20\,dex, respectively, 
while the errors {$\sim$}0.1\,dex or lower have been achieved
with high-quality optical spectra
\citep[e.g.,][]{Luck-2011,Genovali-2013,Takeda-2013}.
The analysis with infrared spectra of Cepheids need to be developed
and tested more extensively. 
Recently, a large number of classical Cepheids
have been discovered across a very large area of the disk \citep{Chen-2019,Dekany-2019,Skowron-2019}.
Many distant Cepheids at ${\gtrsim}4$\,kpc are 
heavily reddened by interstellar dust. 
They tend to be rather faint in the optical range,
but many of them are still bright in the infrared range.
Therefore, developing methods of analysing the
Cepheids' infrared spectra is essential
for future investigations on the emerging sample
of dust-obscured Cepheids. 
We can also find other interesting objects in this temperature range,
e.g., yellow supergiants \citep{Neugent-2010,Neugent-2012}
and yellow hypergiants \citep{deJager-1998}, 
planet-hosting dwarfs with $\Teff$ not far from that of the Sun
\citep{Ramirez-2009,Adibekyan-2013}, etc.

\section{Conclusions}

In near-infrared high-resolution spectra (0.97--1.32\,{$\mu$}m with $\lambda/\Delta\lambda = 28\,000$) of 42 FGK stars,
we have selected line pairs of \ion{Fe}{i}--\ion{Fe}{i},
\ion{Fe}{i}--\ion{Fe}{ii} and \ion{Ca}{i}--\ion{Ca}{ii} 
combinations whose LDRs show tight relations with 
$\Teff$ and $\log g$.
Their simple linear relations allow us to predict 
$\Teff$ and $\log g$ with a moderate precision
(50\,K and 0.2\,dex)
for late F to early K type stars (mainly G-type ones)
in a wide range of luminosity class (from supergiants to dwarfs).
This method does not require complex calculations with numerical models
like the inference of chemical abundances. The current analysis
suggests that the metallicity effect on the derived $\Teff$
and $\log g$ is negligible, at least, within
the metallicity range, $-0.2 < \FeH < 0.2$\,dex,
covered by our sample,
although this independency and the $\FeH$ range of application
need to be tested with larger datasets.
We used simple linear formulae given in
Section~\ref{subsubsec:FeI-FeI} and \ref{subsubsec:FeI-FeII}
considering the limited number and the non-uniform distribution 
of calibrators. Using a larger sample for calibration would allow us
to try higher-order or more complex formulae, and this may reduce
the fitting residuals.
Moreover, it will be essential to improve the calibration
by observing stars with accurate and precise $\log g$ known.
In particular, objects with asteroseismic $\log g$ are
very nice calibrators (e.g., \citealt{Morel-2012};
\citealt{Hekker-2013}; \citealt{Pinsonneault-2018}), 
and empirical LDR relations would enable us to obtain
surface gravities consistent with the asteroseismic estimates 
without introducing the systematic errors caused in
the spectroscopic analysis such as the LTE/non-LTE difference.

\section*{Acknowledgements}

We acknowledge useful comments from the referee, Maria Bergemann.
We are grateful to the staff of the
Koyama Astronomical Observatory for their support during our
observation. This study is supported by JSPS KAKENHI (grant
Nos.~26287028 and 18H01248). The WINERED was developed
by the University of Tokyo and the Laboratory of Infrared High-resolution
spectroscopy (LiH), Kyoto Sangyo University under the
financial supports of KAKENHI (Nos.~16684001, 20340042, and
21840052) and the MEXT Supported Program for the Strategic
Research Foundation at Private Universities (Nos. S0801061 and
S1411028). 
Two of us are financially supported by JSPS Research Fellowship for 
Young Scientists and accompanying Grant-in-Aid for JSPS Fellows,
MJ (No.~21J11301) and DT (No.~21J11555).
DT also acknowledges financial support from Masason Foundation.
SSE is supported by CONICYT BECAS CHILE DOCTORADO EN EL EXTRANJERO 72170404.
This research has made use of the SIMBAD database 
and the VizieR catalogue access tool,
provided by CDS, Strasbourg, France (DOI:~10.26093/cds/vizier).
The original description of the VizieR service was published in 2000 (A\&AS 143, 23).
We also acknowledge the VALD database, operated at
Uppsala University, the Institute of Astronomy RAS in Moscow,
and the University of Vienna.
 
\section*{Data Availability}

The data underlying this article will be shared on a reasonable request with the corresponding author. 


\section*{Supporting Information}

Supplementary data are available at MNRAS online.

{\bf Table~\ref{tab:lines}.}
List of the 97 confirmed lines that are used in search of line-depth ratio pairs (see Section~\ref{subsec:select-line-obs}). The last column lists the number of objects for which the depth measurements were validated. 
Among these lines, \ion{Ca}{ii}~9890.6280 includes
two transitions with $\log {gf}=0.900$ and $1.013$ at
the same EP in VALD. Our electric table lists 
the combined strength ($\log {gf} = 1.261$) with the ``*'' mark. 

{\bf Figure~\ref{fig:pairs1}.}
The relation between LDR ($r$ or $\log r$) and effective temperature ($\Teff$ or $\log \Teff$) for the 13 \ion{Fe}{i}--\ion{Fe}{i} pairs in Table~\ref{tab:pairs}.

{\bf Figure~\ref{fig:pairs2}.}
The relation between LDR ($r$ or $\log r$), effective temperature ($\Teff$ or $\log \Teff$) and the surface gravity ($\log g$) for the 5 \ion{Fe}{i}--\ion{Fe}{ii} and 4 \ion{Ca}{i}--\ion{Ca}{ii} pairs in Table~\ref{tab:pairs}.

\newpage

\setcounter{table}{1}
\begin{table}
\centering
   \caption{
     {\bf The entire part of Table~2 available in Supporting Information.} List of the 97 confirmed lines that are used to search line-depth ratio pairs (see Section~\ref{subsec:select-line-obs}). The last column lists the number of objects for which the depth measurements were validated. \ion{Fe}{i}, \ion{Fe}{ii}, \ion{Ca}{i} and \ion{Ca}{ii} lines are combined together. The first 10 lines are presented here, and the full table is available in Supporting Information.
   } 
\begin{tabular}{crrrrr}
\hline
Species & \multicolumn{1}{c}{$\lambda_\mathrm{air}$} & \multicolumn{1}{c}{EP} & \multicolumn{2}{c}{$\log {gf}$ (dex)} & $N_\mathrm{obj}$ \\
 & \multicolumn{1}{c}{(\AA)} & \multicolumn{1}{c}{(eV)} & VALD & MB99 & \\
\hline
      \ion{Fe}{i} & 9800.3075 &  5.086 & $-0.453$ & \multicolumn{1}{c}{---} & 37 \\ 
      \ion{Fe}{i} & 9811.5041 &  5.012 & $-1.362$ & \multicolumn{1}{c}{---} & 41 \\ 
      \ion{Fe}{i} & 9861.7337 &  5.064 & $-0.142$ & \multicolumn{1}{c}{---} & 42 \\ 
      \ion{Fe}{i} & 9868.1857 &  5.086 & $-0.979$ & \multicolumn{1}{c}{---} & 42 \\ 
      \ion{Fe}{i} & 9889.0351 &  5.033 & $-0.446$ & \multicolumn{1}{c}{---} & 41 \\ 
      \ion{Fe}{i} & 9944.2065 &  5.012 & $-1.338$ & \multicolumn{1}{c}{---} & 40 \\ 
      \ion{Fe}{i} & 9980.4629 &  5.033 & $-1.379$ & \multicolumn{1}{c}{---} & 42 \\ 
      \ion{Fe}{i} & 10041.472 &  5.012 & $-1.772$ & $-1.84$ & 37 \\ 
      \ion{Fe}{i} & 10065.045 &  4.835 & $-0.289$ & $-0.57$ & 40 \\ 
      \ion{Fe}{i} & 10081.393 &  2.424 & $-4.537$ & $-4.53$ & 33 \\ 
      \ion{Fe}{i} & 10114.014 &  2.759 & $-3.692$ & $-3.76$ & 39 \\ 
      \ion{Fe}{i} & 10145.561 &  4.795 & $-0.177$ & $-0.41$ & 42 \\ 
      \ion{Fe}{i} & 10155.162 &  2.176 & $-4.226$ & $-4.36$ & 40 \\ 
      \ion{Fe}{i} & 10167.468 &  2.198 & $-4.117$ & $-4.26$ & 41 \\ 
      \ion{Fe}{i} & 10195.105 &  2.728 & $-3.580$ & $-3.63$ & 42 \\ 
      \ion{Fe}{i} & 10216.313 &  4.733 & $-0.063$ & $-0.29$ & 42 \\ 
      \ion{Fe}{i} & 10218.408 &  3.071 & $-2.760$ & $-2.93$ & 42 \\ 
      \ion{Fe}{i} & 10227.994 &  6.119 & $-0.354$ & \multicolumn{1}{c}{---} & 21 \\ 
      \ion{Fe}{i} & 10265.217 &  2.223 & $-4.537$ & $-4.67$ & 34 \\ 
      \ion{Fe}{i} & 10332.327 &  3.635 & $-2.938$ & $-3.15$ & 35 \\ 
      \ion{Fe}{i} & 10340.885 &  2.198 & $-3.577$ & $-3.65$ & 42 \\ 
      \ion{Fe}{i} & 10347.965 &  5.393 & $-0.551$ & $-0.82$ & 42 \\ 
      \ion{Fe}{i} & 10353.804 &  5.393 & $-0.819$ & $-1.09$ & 42 \\ 
      \ion{Fe}{i} & 10364.062 &  5.446 & $-0.960$ & $-1.19$ & 42 \\ 
      \ion{Fe}{i} & 10395.794 &  2.176 & $-3.393$ & $-3.42$ & 42 \\ 
      \ion{Fe}{i} & 10423.743 &  3.071 & $-2.918$ & $-3.13$ & 41 \\ 
      \ion{Fe}{i} & 10435.355 &  4.733 & $-1.945$ & $-2.11$ & 41 \\ 
      \ion{Fe}{i} & 10469.652 &  3.884 & $-1.184$ & $-1.37$ & 42 \\ 
      \ion{Fe}{i} & 10532.234 &  3.929 & $-1.480$ & $-1.76$ & 42 \\ 
      \ion{Fe}{i} & 10535.709 &  6.206 & $-0.108$ & \multicolumn{1}{c}{---} & 42 \\ 
      \ion{Fe}{i} & 10555.649 &  5.446 & $-1.108$ & $-1.39$ & 39 \\ 
      \ion{Fe}{i} & 10577.139 &  3.301 & $-3.136$ & $-3.28$ & 41 \\ 
      \ion{Fe}{i} & 10611.686 &  6.169 & $0.021$ & $-0.09$ & 42 \\ 
      \ion{Fe}{i} & 10616.721 &  3.267 & $-3.127$ & $-3.34$ & 41 \\ 
      \ion{Fe}{i} & 10674.070 &  6.169 & $-0.466$ & \multicolumn{1}{c}{---} & 39 \\ 
      \ion{Fe}{i} & 10717.806 &  5.539 & $-0.436$ & $-1.68$ & 22 \\ 
      \ion{Fe}{i} & 10725.185 &  3.640 & $-2.763$ & $-2.98$ & 41 \\ 
      \ion{Fe}{i} & 10753.004 &  3.960 & $-1.845$ & $-2.14$ & 40 \\ 
      \ion{Fe}{i} & 10780.694 &  3.237 & $-3.289$ & $-3.59$ & 39 \\ 
      \ion{Fe}{i} & 10783.050 &  3.111 & $-2.567$ & $-2.80$ & 42 \\ 
      \ion{Fe}{i} & 10818.274 &  3.960 & $-1.948$ & $-2.23$ & 42 \\ 
      \ion{Fe}{i} & 10849.465 &  5.539 & $-1.444$ & $-0.73$ & 41 \\ 
      \ion{Fe}{i} & 10863.518 &  4.733 & $-0.895$ & $-1.06$ & 42 \\ 
      \ion{Fe}{i} & 10881.758 &  2.845 & $-3.604$ & $-3.50$ & 37 \\ 
      \ion{Fe}{i} & 10884.262 &  3.929 & $-1.925$ & $-2.18$ & 40 \\ 
\hline
\end{tabular}
\end{table}

\setcounter{table}{1}
\begin{table}
\centering
   \caption{
      --- continued.
   } 
\begin{tabular}{crrrrr}
\hline
Species & \multicolumn{1}{c}{$\lambda_\mathrm{air}$} & \multicolumn{1}{c}{EP} & \multicolumn{2}{c}{$\log {gf}$ (dex)} & $N_\mathrm{obj}$ \\
 & \multicolumn{1}{c}{(\AA)} & \multicolumn{1}{c}{(eV)} & VALD & MB99 & \\
\hline
      \ion{Fe}{i} & 11607.572 &  2.198 & $-2.009$ & $-2.46$ & 41 \\ 
      \ion{Fe}{i} & 11638.260 &  2.176 & $-2.214$ & $-2.59$ & 41 \\ 
      \ion{Fe}{i} & 11689.972 &  2.223 & $-2.068$ & $-2.67$ & 37 \\ 
      \ion{Fe}{i} & 11783.265 &  2.832 & $-1.574$ & $-1.86$ & 42 \\ 
      \ion{Fe}{i} & 11882.844 &  2.198 & $-1.668$ & $-2.20$ & 42 \\ 
      \ion{Fe}{i} & 11884.083 &  2.223 & $-2.083$ & $-2.45$ & 42 \\ 
      \ion{Fe}{i} & 11973.046 &  2.176 & $-1.483$ & $-2.28$ & 42 \\ 
      \ion{Fe}{i} & 12053.082 &  4.559 & $-1.543$ & $-1.75$ & 41 \\ 
      \ion{Fe}{i} & 12119.494 &  4.593 & $-1.635$ & $-1.88$ & 39 \\ 
      \ion{Fe}{i} & 12190.098 &  3.635 & $-2.330$ & $-2.75$ & 39 \\ 
      \ion{Fe}{i} & 12213.336 &  4.638 & $-1.845$ & $-2.09$ & 38 \\ 
      \ion{Fe}{i} & 12227.112 &  4.607 & $-1.368$ & $-1.60$ & 42 \\ 
      \ion{Fe}{i} & 12283.298 &  6.169 & $-0.537$ & $-0.61$ & 39 \\ 
      \ion{Fe}{i} & 12342.916 &  4.638 & $-1.463$ & $-1.68$ & 28 \\ 
      \ion{Fe}{i} & 12556.996 &  2.279 & $-3.626$ & $-4.07$ & 41 \\ 
      \ion{Fe}{i} & 12615.928 &  4.638 & $-1.517$ & $-1.77$ & 40 \\ 
      \ion{Fe}{i} & 12638.703 &  4.559 & $-0.783$ & $-1.00$ & 41 \\ 
      \ion{Fe}{i} & 12648.741 &  4.607 & $-1.140$ & $-1.32$ & 41 \\ 
      \ion{Fe}{i} & 12789.450 &  5.010 & $-1.514$ & $-1.92$ & 37 \\ 
      \ion{Fe}{i} & 12807.152 &  3.640 & $-2.452$ & $-2.76$ & 35 \\ 
      \ion{Fe}{i} & 12808.243 &  4.988 & $-1.362$ & $-1.87$ & 26 \\ 
      \ion{Fe}{i} & 12824.859 &  3.018 & $-3.835$ & $-3.68$ & 28 \\ 
      \ion{Fe}{i} & 12840.574 &  4.956 & $-1.329$ & $-1.76$ & 36 \\ 
      \ion{Fe}{i} & 12879.766 &  2.279 & $-3.458$ & $-3.61$ & 35 \\ 
      \ion{Fe}{i} & 12896.118 &  4.913 & $-1.424$ & $-1.80$ & 30 \\ 
      \ion{Fe}{i} & 12934.666 &  5.393 & $-0.948$ & $-1.28$ & 41 \\ 
      \ion{Fe}{i} & 13006.684 &  2.990 & $-3.744$ & $-3.49$ & 41 \\ 
      \ion{Fe}{i} & 13014.841 &  5.446 & $-1.693$ & $-1.68$ & 38 \\ 
      \ion{Fe}{i} & 13039.647 &  5.655 & $-0.731$ & $-1.32$ & 27 \\ 
      \ion{Fe}{i} & 13098.876 &  5.010 & $-1.290$ & $-1.73$ & 34 \\ 
      \ion{Fe}{i} & 13147.920 &  5.393 & $-0.814$ & $-0.93$ & 42 \\ 
      \ion{Fe}{ii} & 9997.5980 &  5.484 & $-1.867$ & \multicolumn{1}{c}{---} & 27 \\ 
      \ion{Fe}{ii} & 10173.515 &  5.511 & $-2.736$ & $-2.79$ & 13 \\ 
      \ion{Fe}{ii} & 10366.167 &  6.724 & $-1.825$ & $-1.76$ & 16 \\ 
      \ion{Fe}{ii} & 10501.500 &  5.549 & $-2.086$ & $-2.17$ & 40 \\ 
      \ion{Fe}{ii} & 10862.652 &  5.589 & $-2.199$ & $-2.11$ & 23 \\ 
      \ion{Ca}{i} & 10343.819 &  2.933 & $-0.300$ & $-0.40$ & 42 \\ 
      \ion{Ca}{i} & 10516.156 &  4.744 & $-1.438$ & $-0.52$ & 41 \\ 
      \ion{Ca}{i} & 10838.970 &  4.878 & $0.238$ & $0.03$ & 42 \\ 
      \ion{Ca}{i} & 10846.792 &  4.744 & $-1.318$ & $-0.64$ & 41 \\ 
      \ion{Ca}{i} & 11767.481 &  4.532 & $-0.536$ & $-0.80$ & 32 \\ 
      \ion{Ca}{i} & 11955.955 &  4.131 & $-0.849$ & $-0.91$ & 41 \\ 
      \ion{Ca}{i} & 12105.841 &  4.554 & $-0.305$ & $-0.54$ & 41 \\ 
      \ion{Ca}{i} & 12823.867 &  3.910 & $-0.997$ & $-1.34$ & 25 \\ 
      \ion{Ca}{i} & 12909.070 &  4.430 & $-0.224$ & $-0.50$ & 42 \\ 
      \ion{Ca}{i} & 13033.554 &  4.441 & $-0.064$ & $-0.31$ & 42 \\ 
      \ion{Ca}{i} & 13134.942 &  4.451 & $0.085$ & $-0.14$ & 33 \\ 
      \ion{Ca}{ii} & 9854.7588 &  7.505 & $-0.205$ & \multicolumn{1}{c}{---} & 26 \\ 
      \ion{Ca}{ii} & 9890.6280 &  8.438 &  1.261$^*$ & \multicolumn{1}{c}{---} & 27 \\ 
      \ion{Ca}{ii} & 9931.3741 &  7.515 & $0.092$ & \multicolumn{1}{c}{---} & 28 \\ 
      \ion{Ca}{ii} & 11838.997 &  6.468 & $0.312$ & $0.24$ & 40 \\ 
      \ion{Ca}{ii} & 11949.744 &  6.468 & $0.006$ & $-0.04$ & 39 \\ 
\hline
\end{tabular}
\end{table}

\clearpage

Here presented are individual plots for Figure~\ref{fig:pairs1}
which are available as the online material.

\begin{tabular}{c}
\includegraphics[clip,width=0.98\hsize]{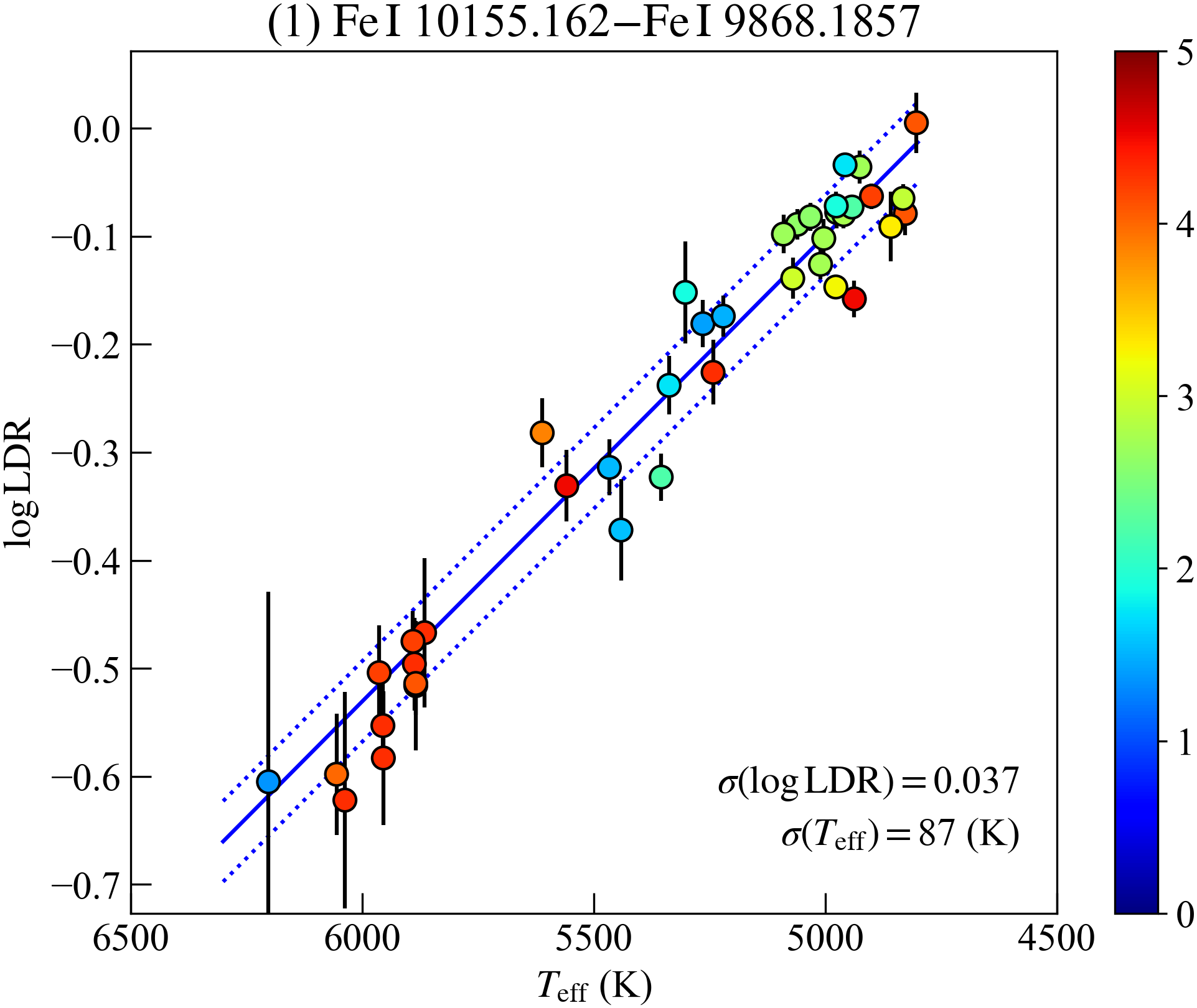}
\includegraphics[clip,width=0.98\hsize]{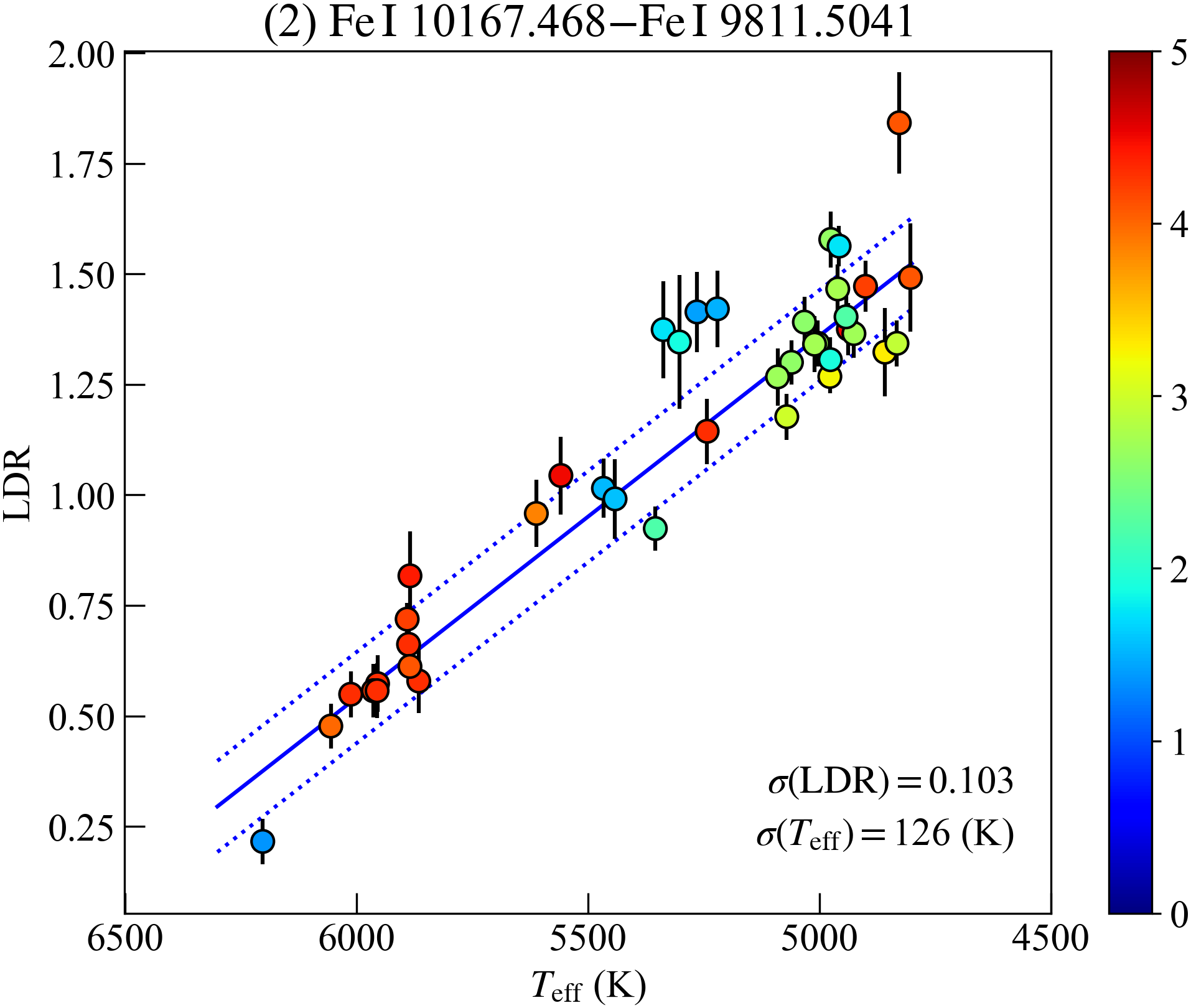}
\\
\includegraphics[clip,width=0.98\hsize]{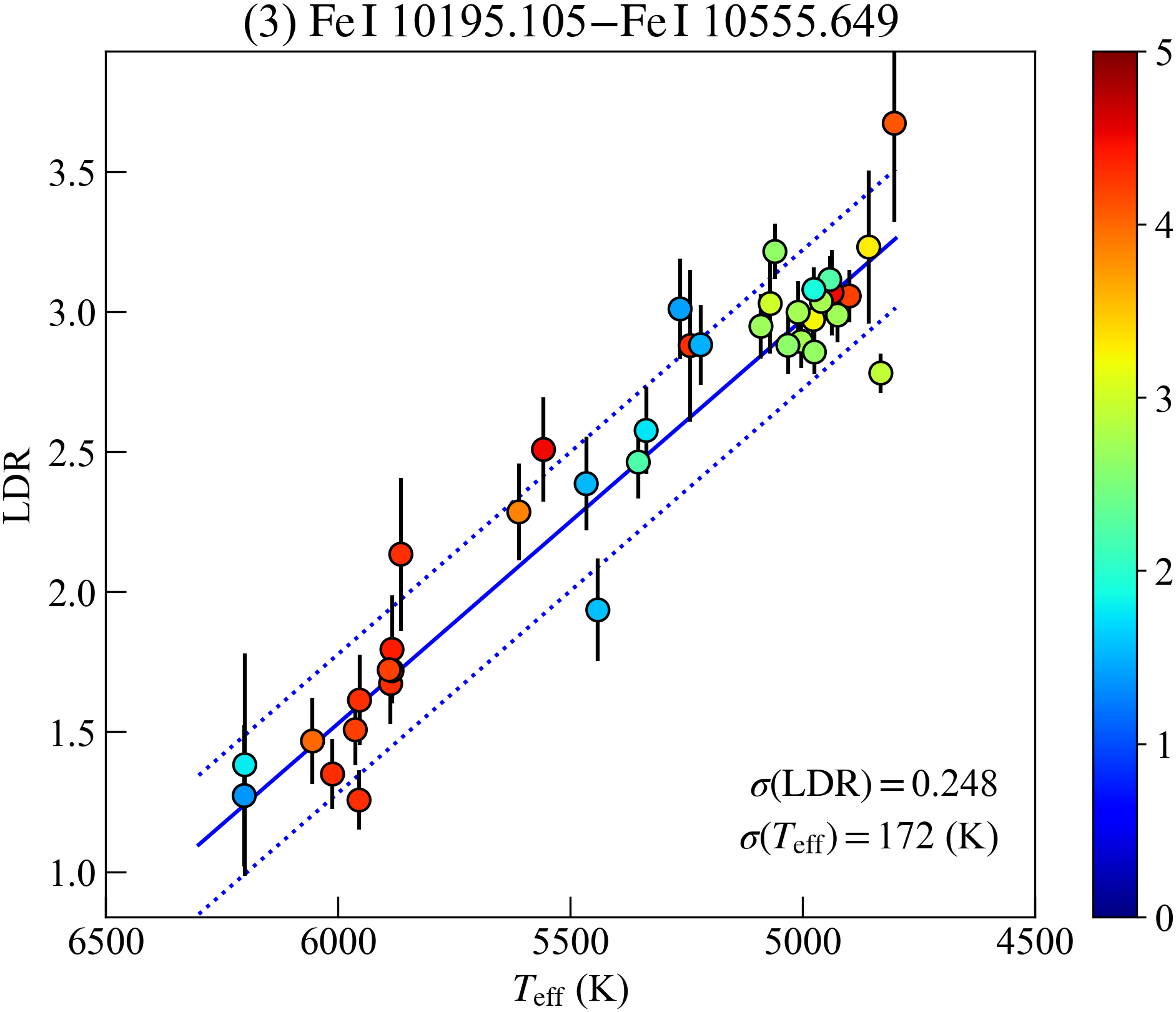}
\includegraphics[clip,width=0.98\hsize]{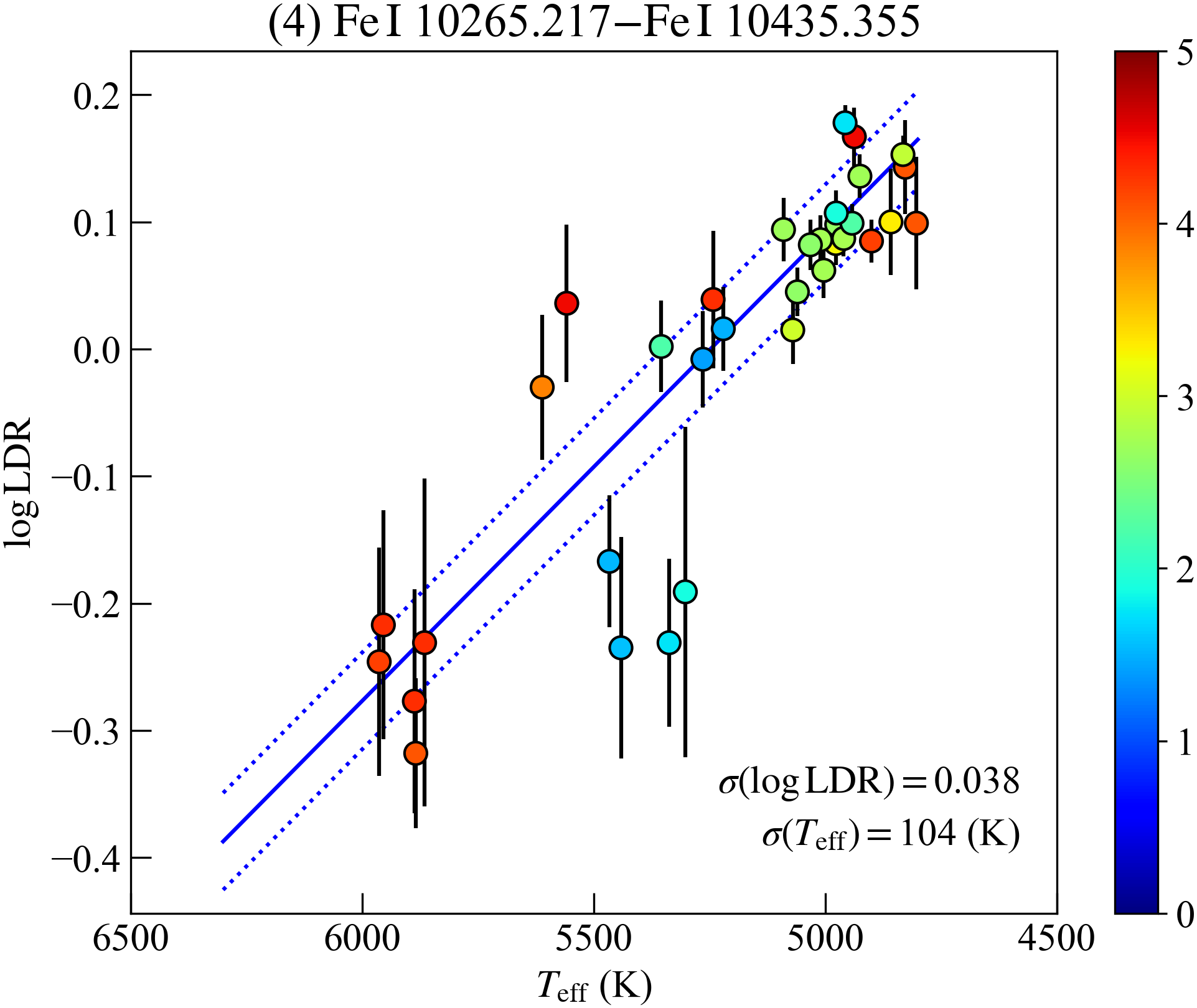}
\\
\includegraphics[clip,width=0.98\hsize]{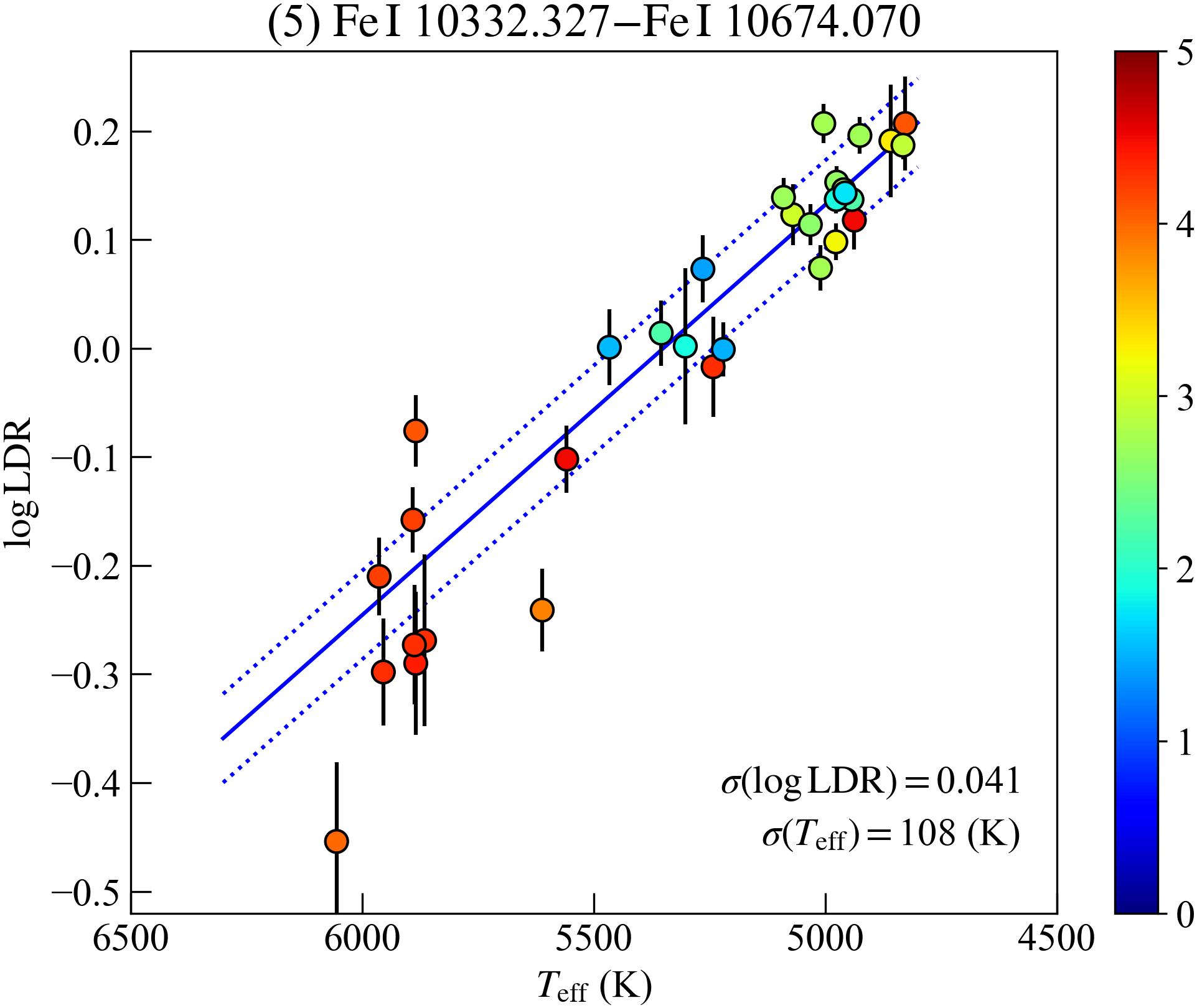}
\includegraphics[clip,width=0.98\hsize]{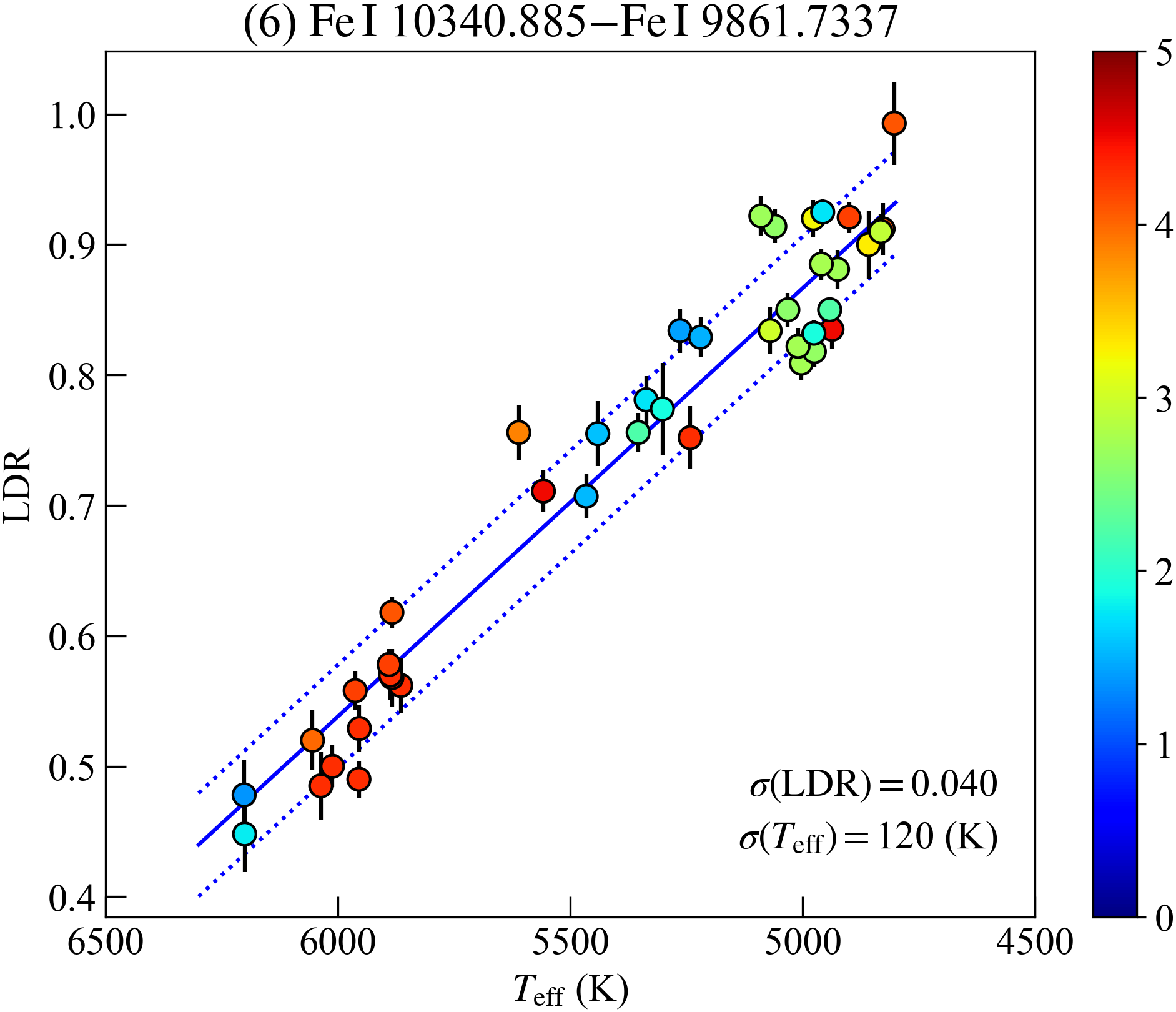}
\end{tabular}

\clearpage

Here presented are individual plots for Figure~\ref{fig:pairs1}
which are available as the online material.

\begin{tabular}{c}
\includegraphics[clip,width=0.98\hsize]{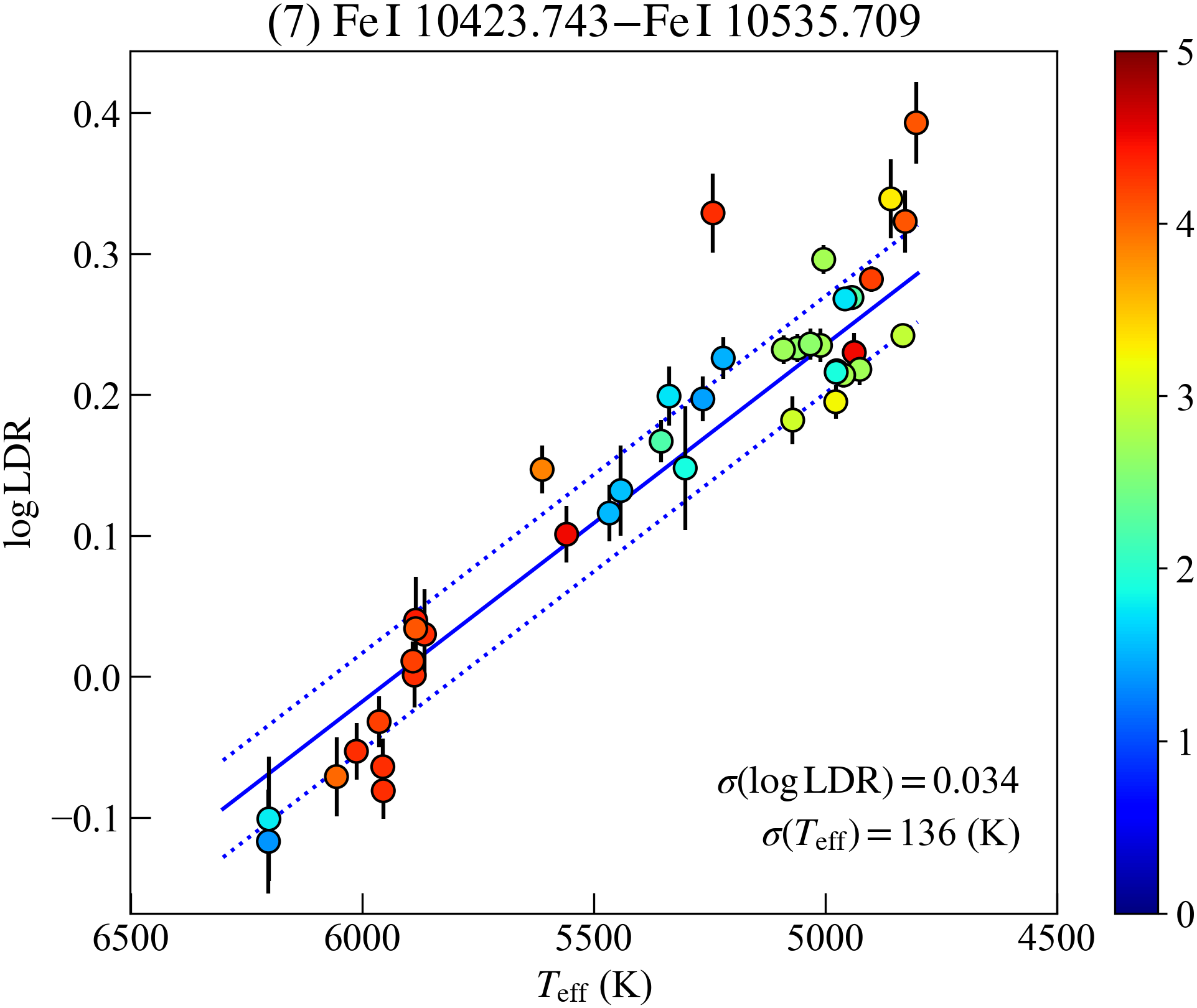}
\includegraphics[clip,width=0.98\hsize]{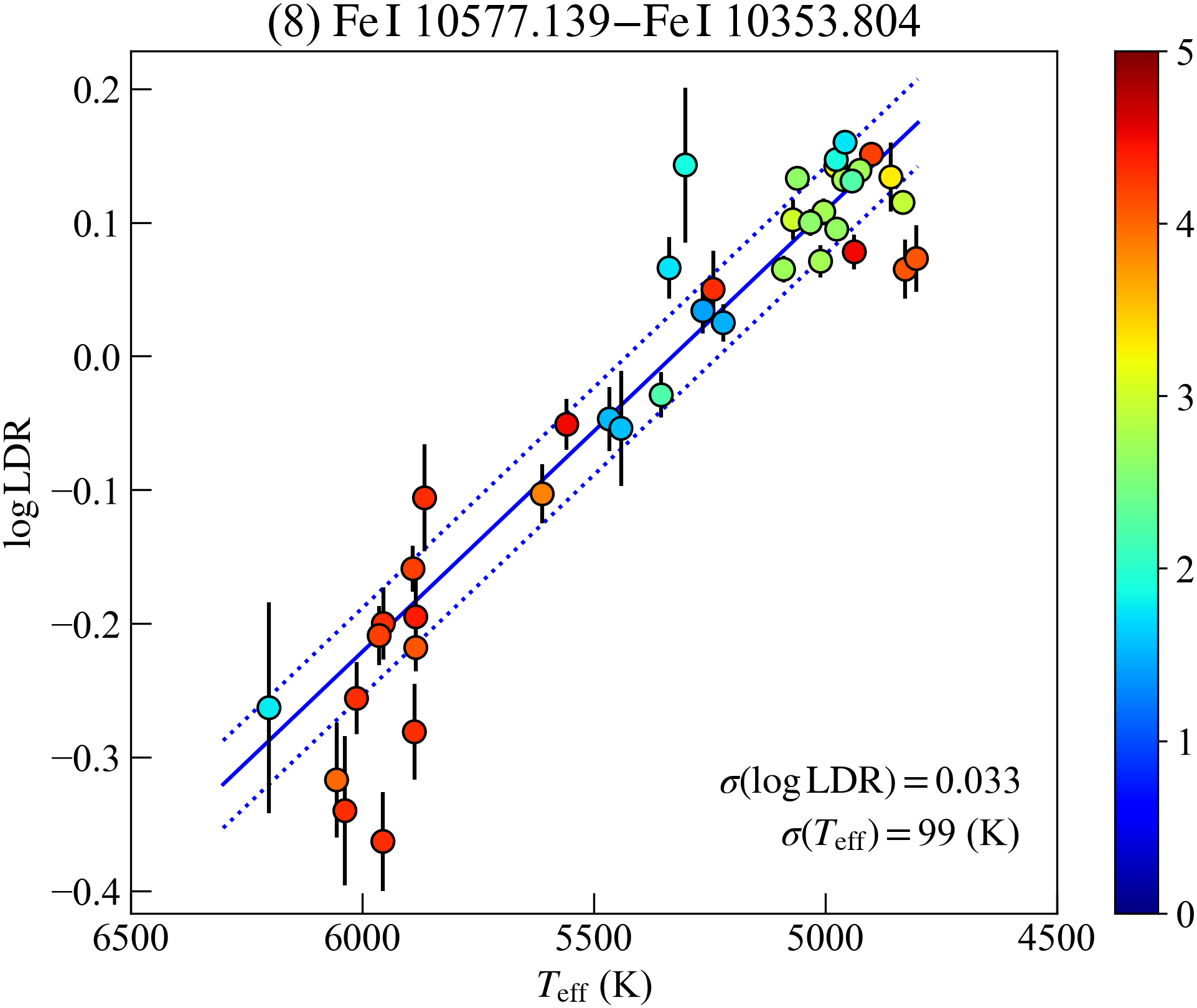}
\\
\includegraphics[clip,width=0.98\hsize]{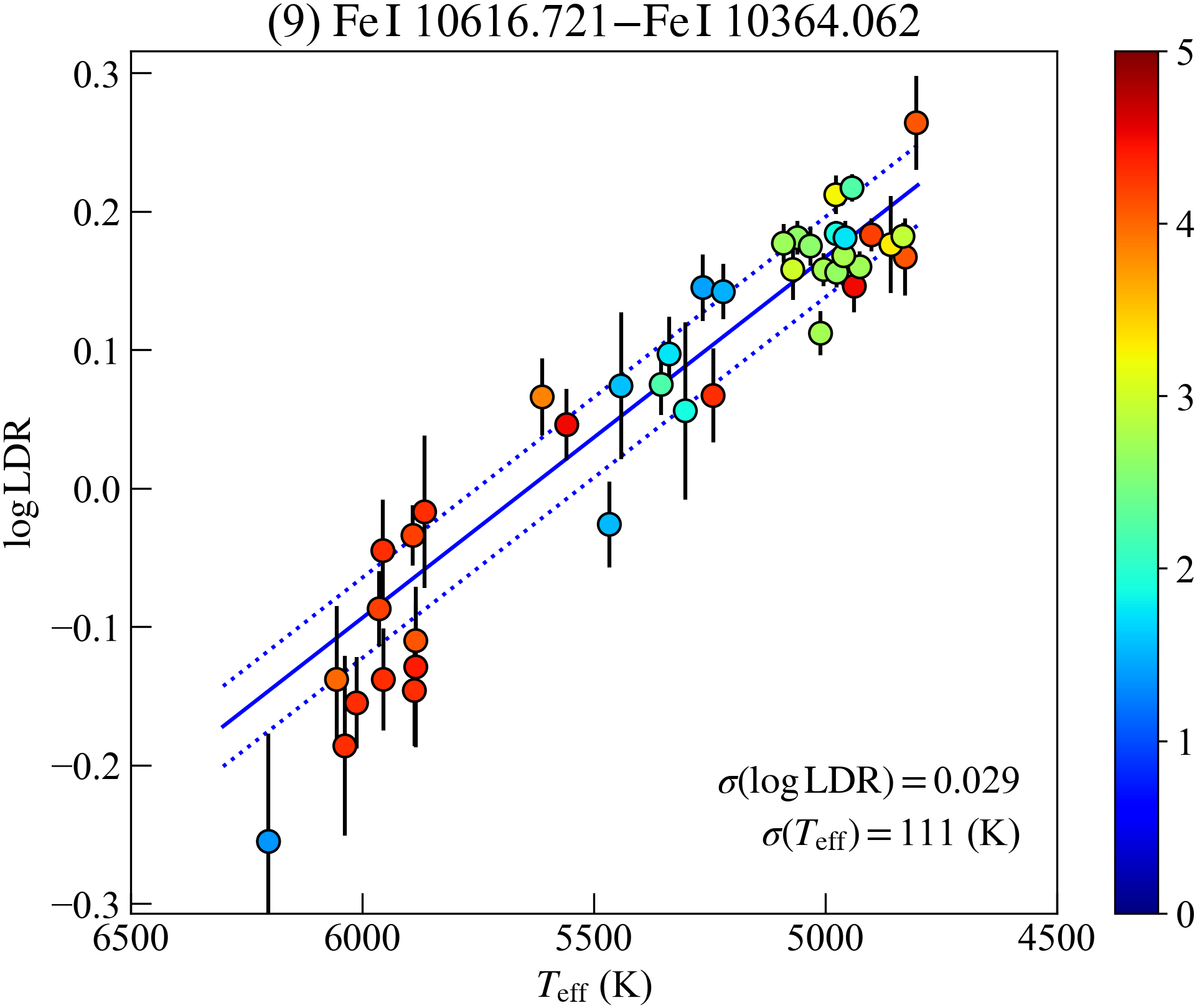}
\includegraphics[clip,width=0.98\hsize]{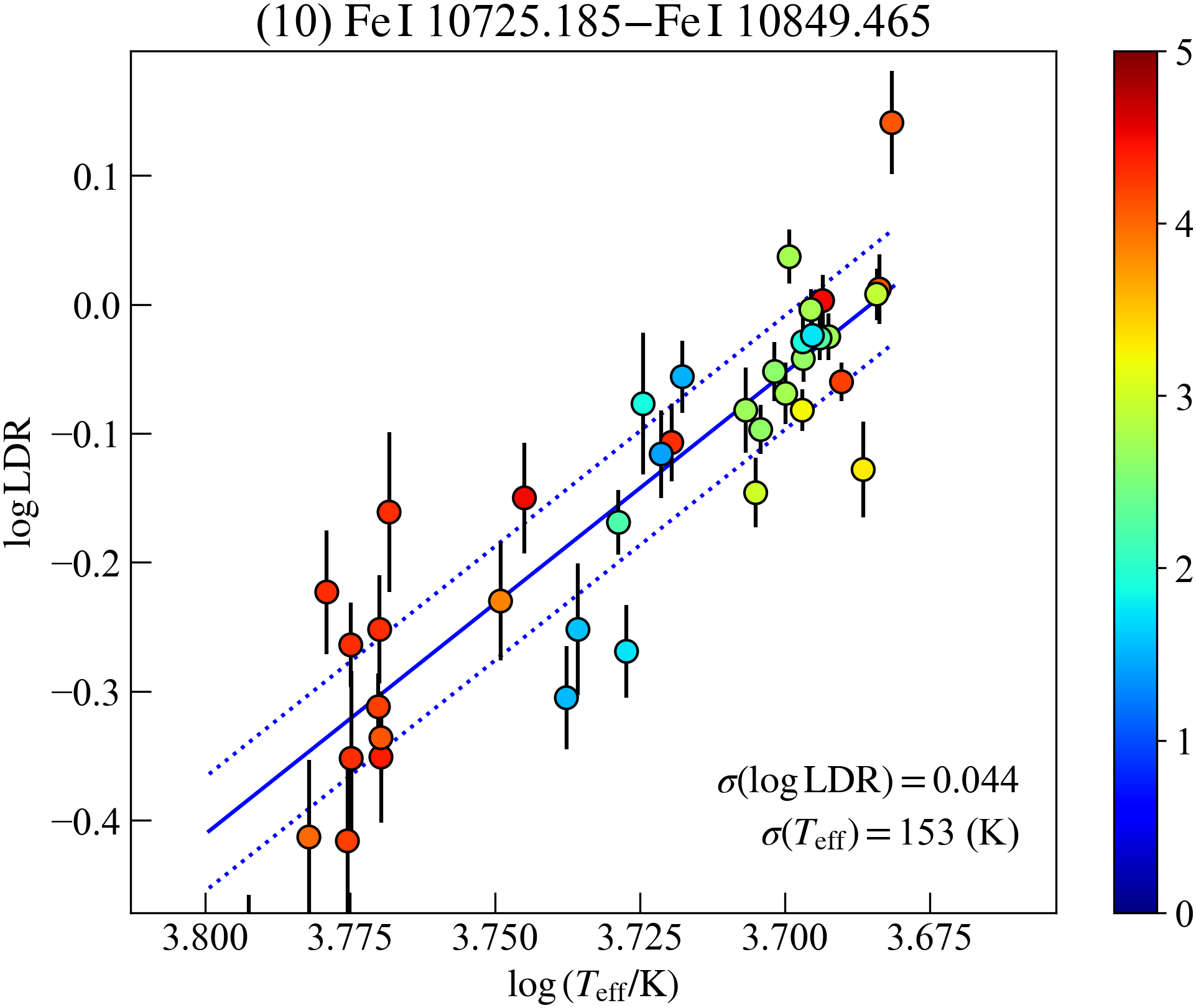}
\\
\includegraphics[clip,width=0.98\hsize]{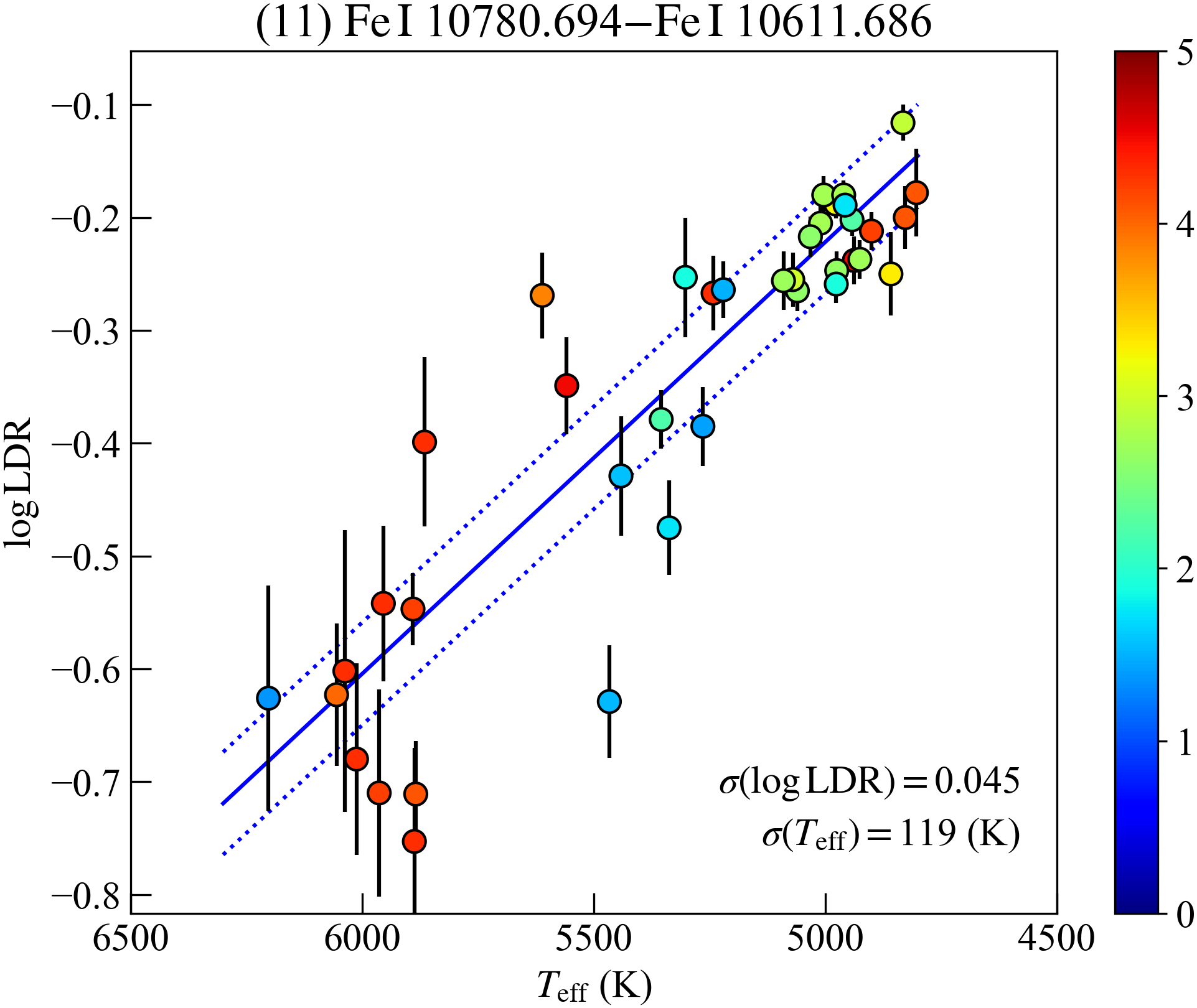}
\includegraphics[clip,width=0.98\hsize]{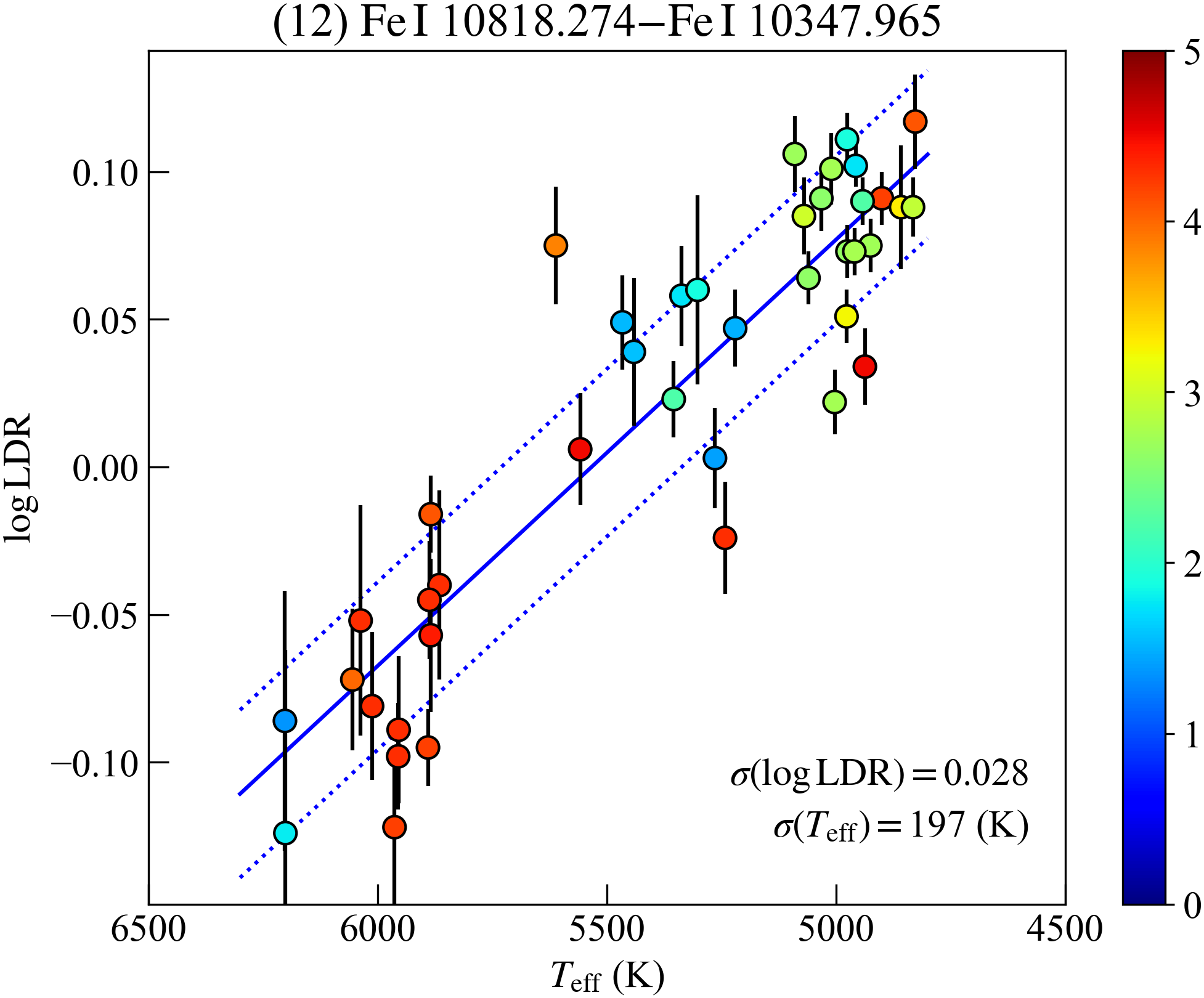}
\end{tabular}

\clearpage

Here presented are individual plots for Figure~\ref{fig:pairs1}
which are available as the online material.

\begin{tabular}{c}
\includegraphics[clip,width=0.98\hsize]{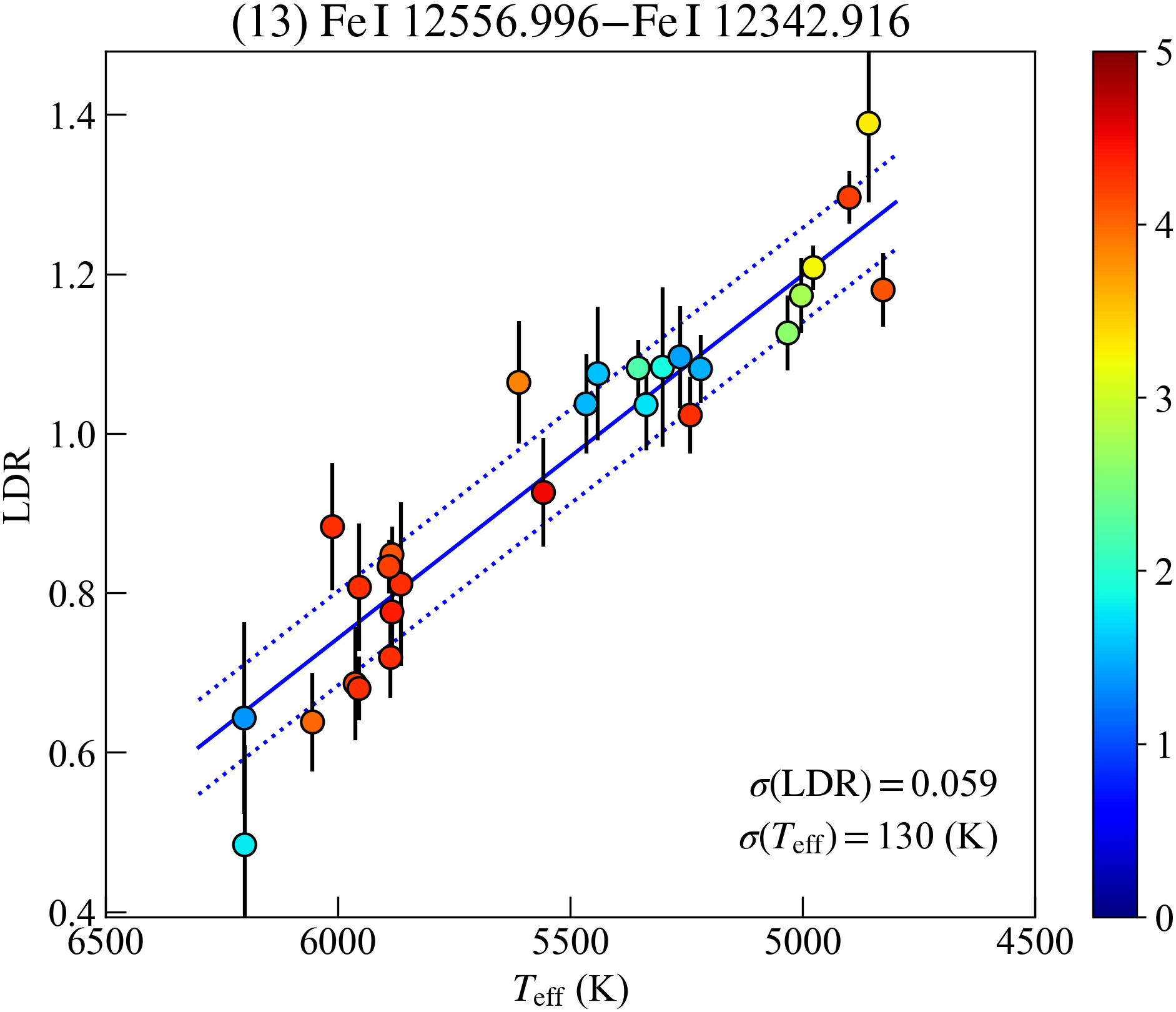}
\end{tabular}

\clearpage

Here presented are individual plots for Figure~\ref{fig:pairs2}
which are available as the online material.

\begin{tabular}{c}
\includegraphics[clip,width=0.98\hsize]{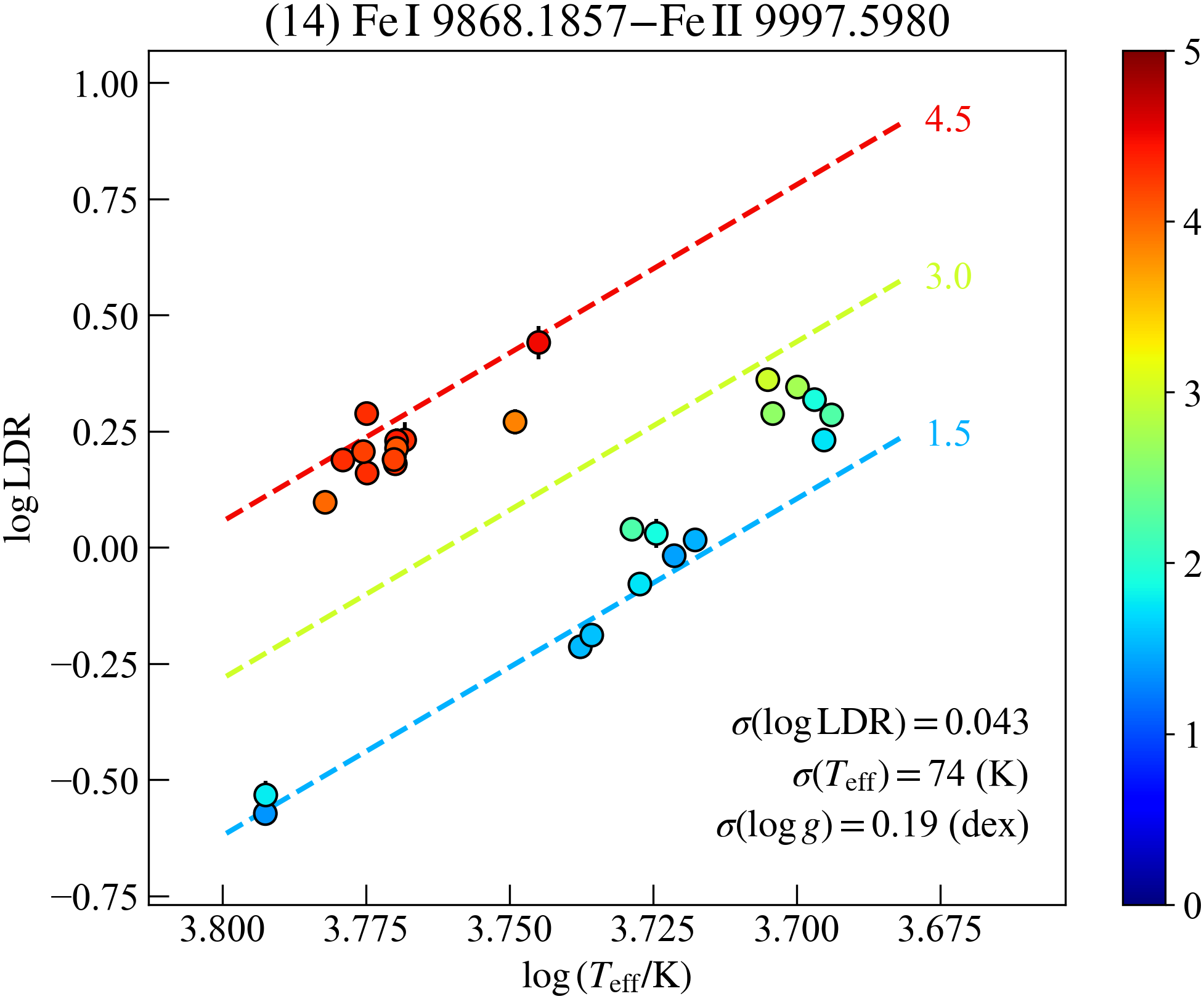}
\includegraphics[clip,width=0.98\hsize]{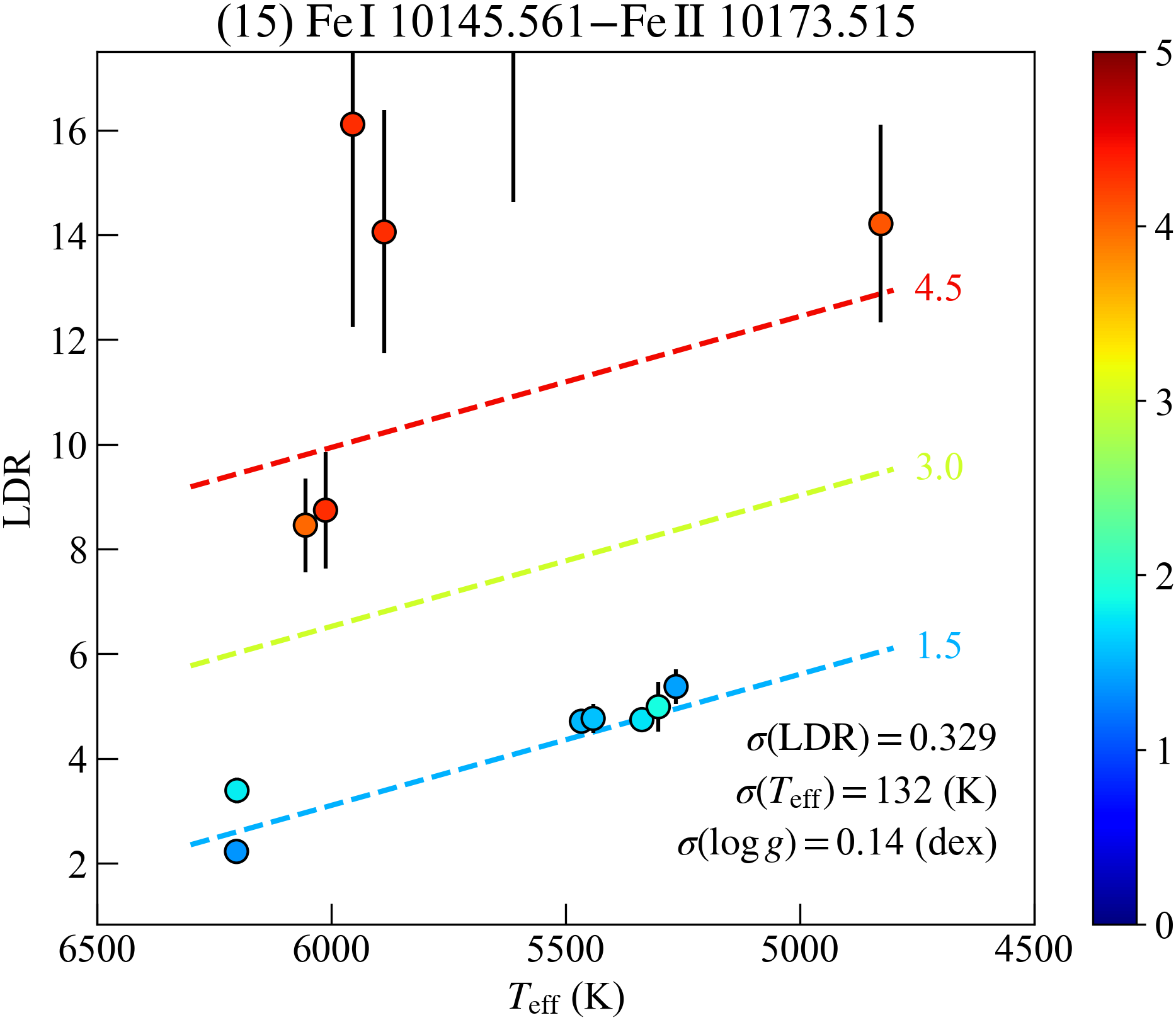}
\\
\includegraphics[clip,width=0.98\hsize]{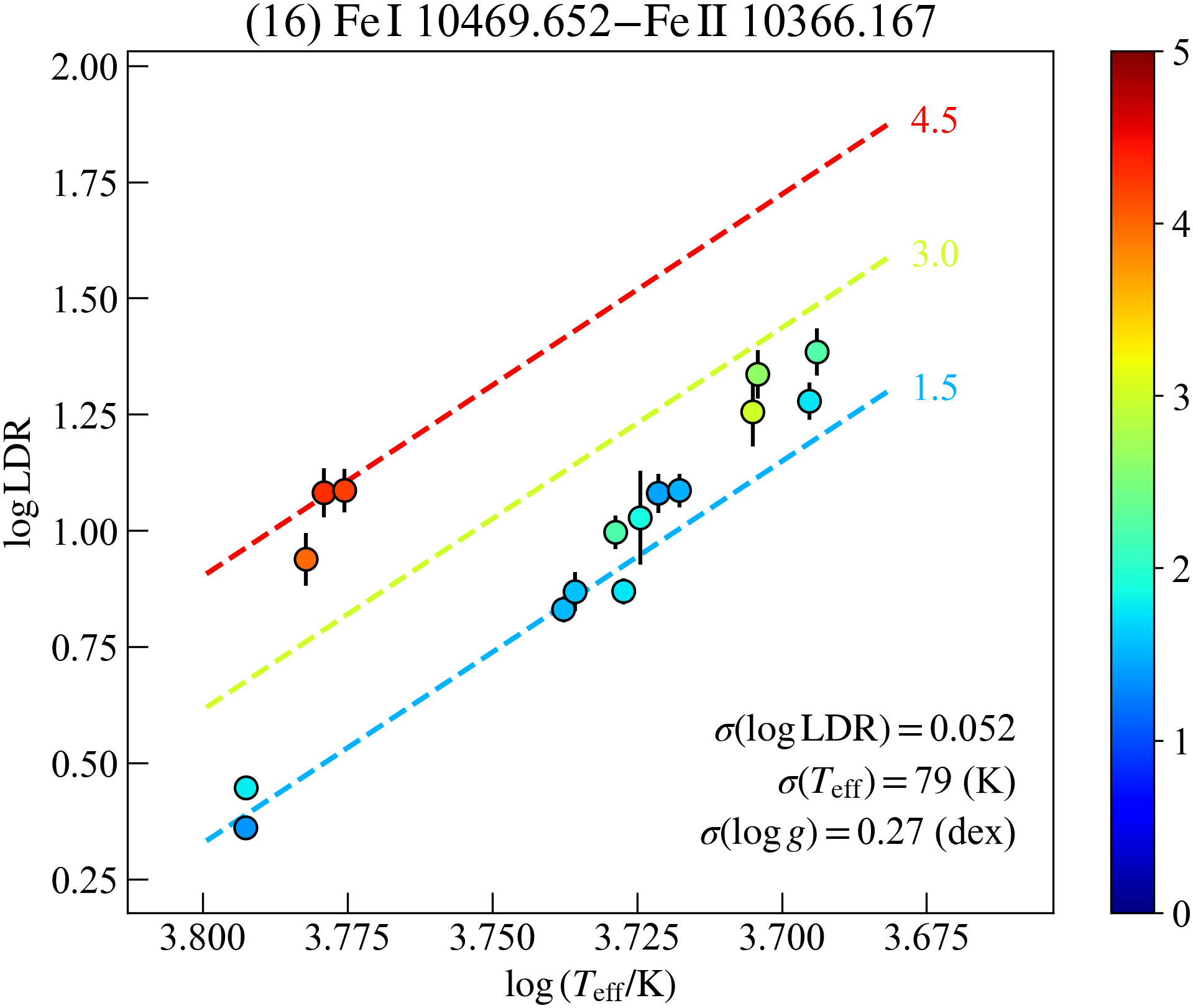}
\includegraphics[clip,width=0.98\hsize]{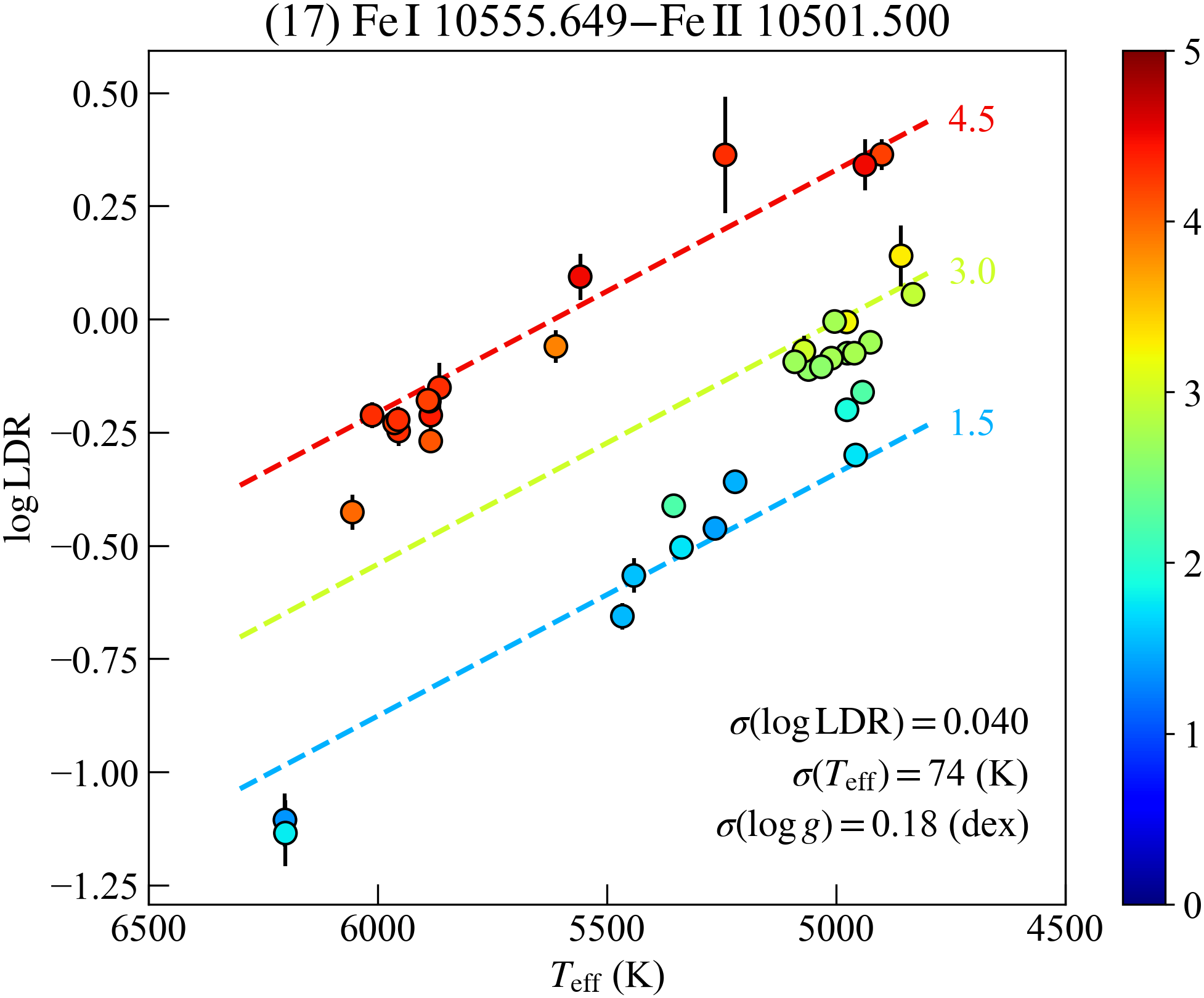}
\\
\includegraphics[clip,width=0.98\hsize]{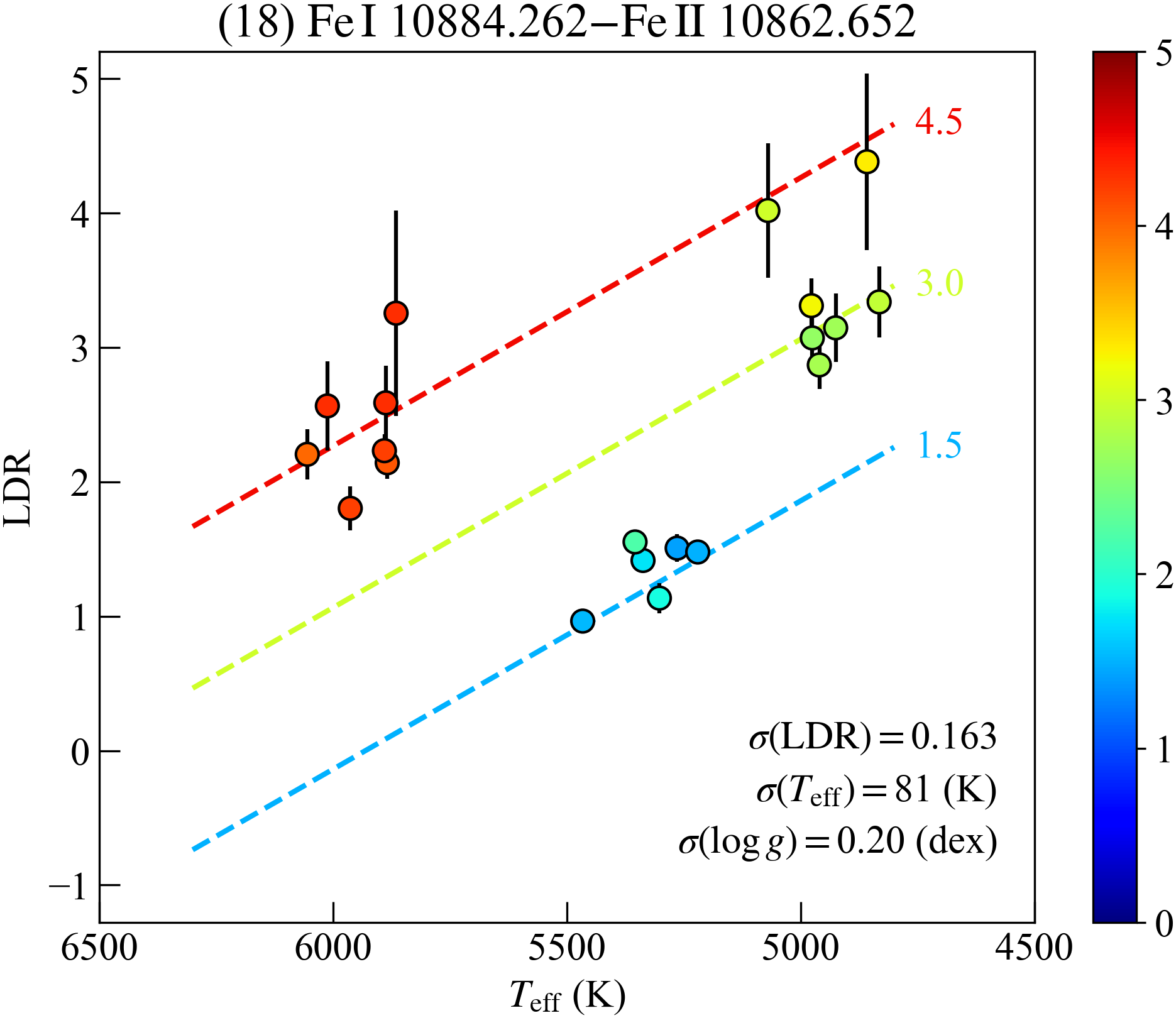}
\includegraphics[clip,width=0.98\hsize]{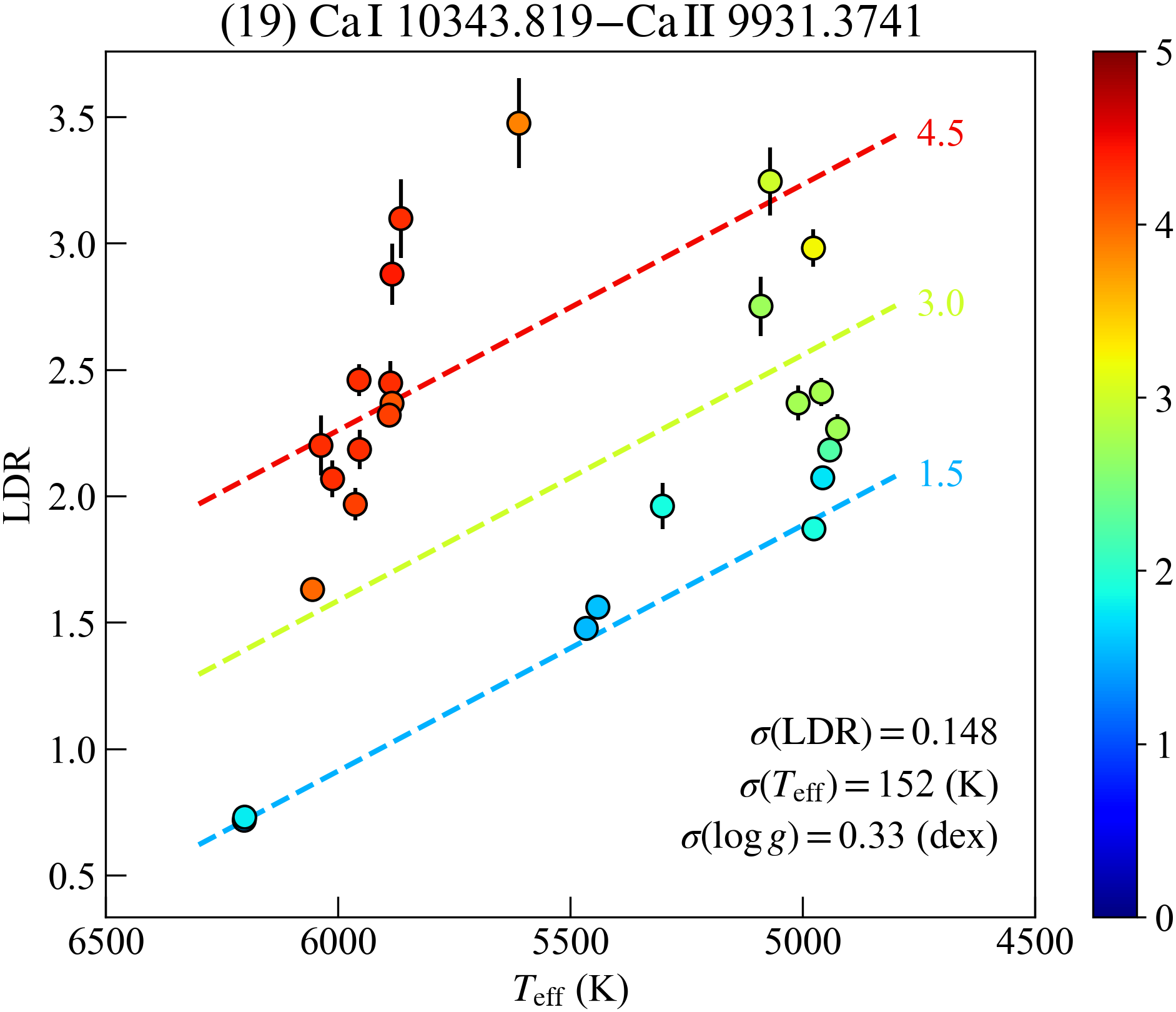}
\end{tabular}

\clearpage

Here presented are individual plots for Figure~\ref{fig:pairs2}
which are available as the online material.

\begin{tabular}{c}
\includegraphics[clip,width=0.98\hsize]{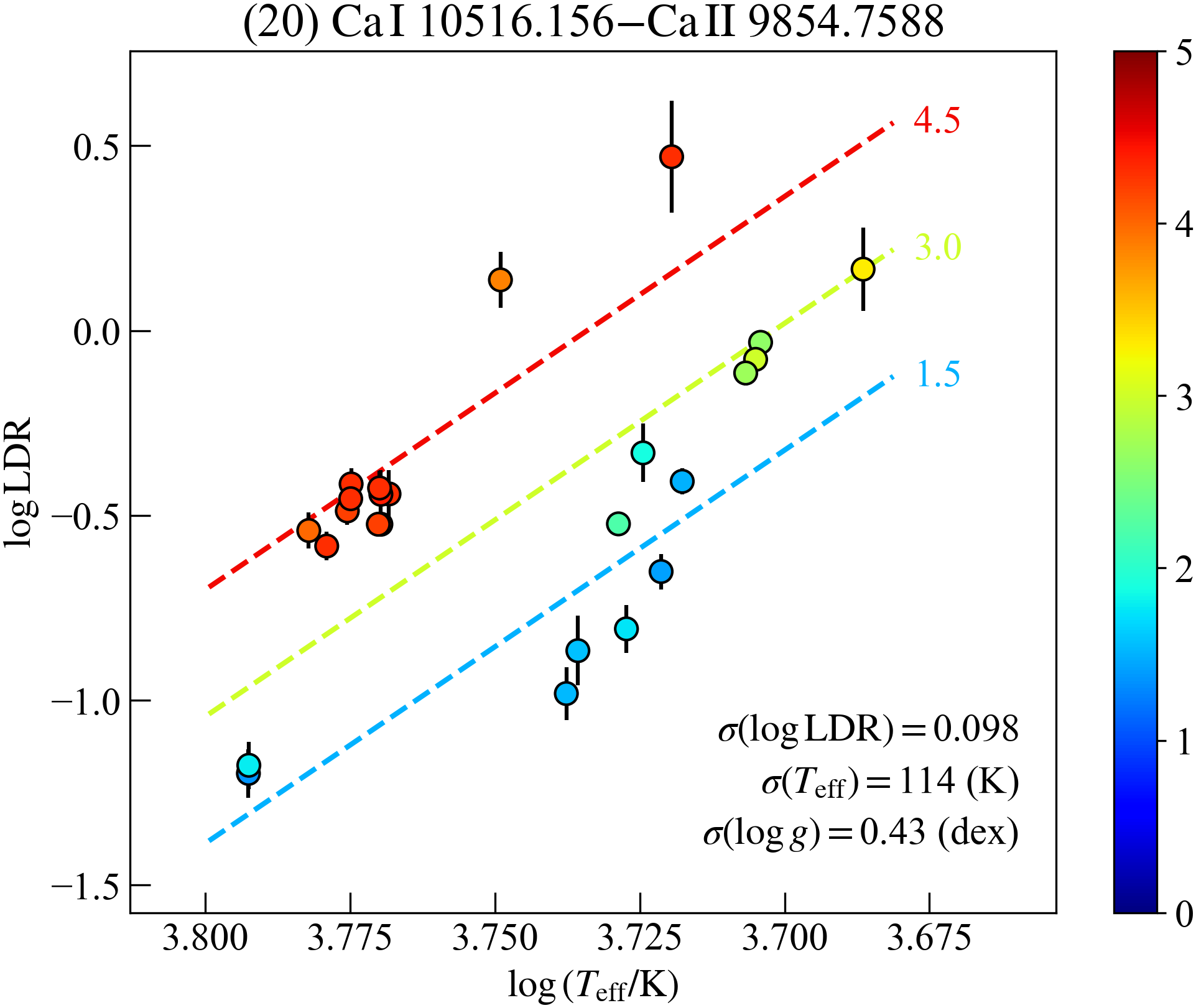}
\includegraphics[clip,width=0.98\hsize]{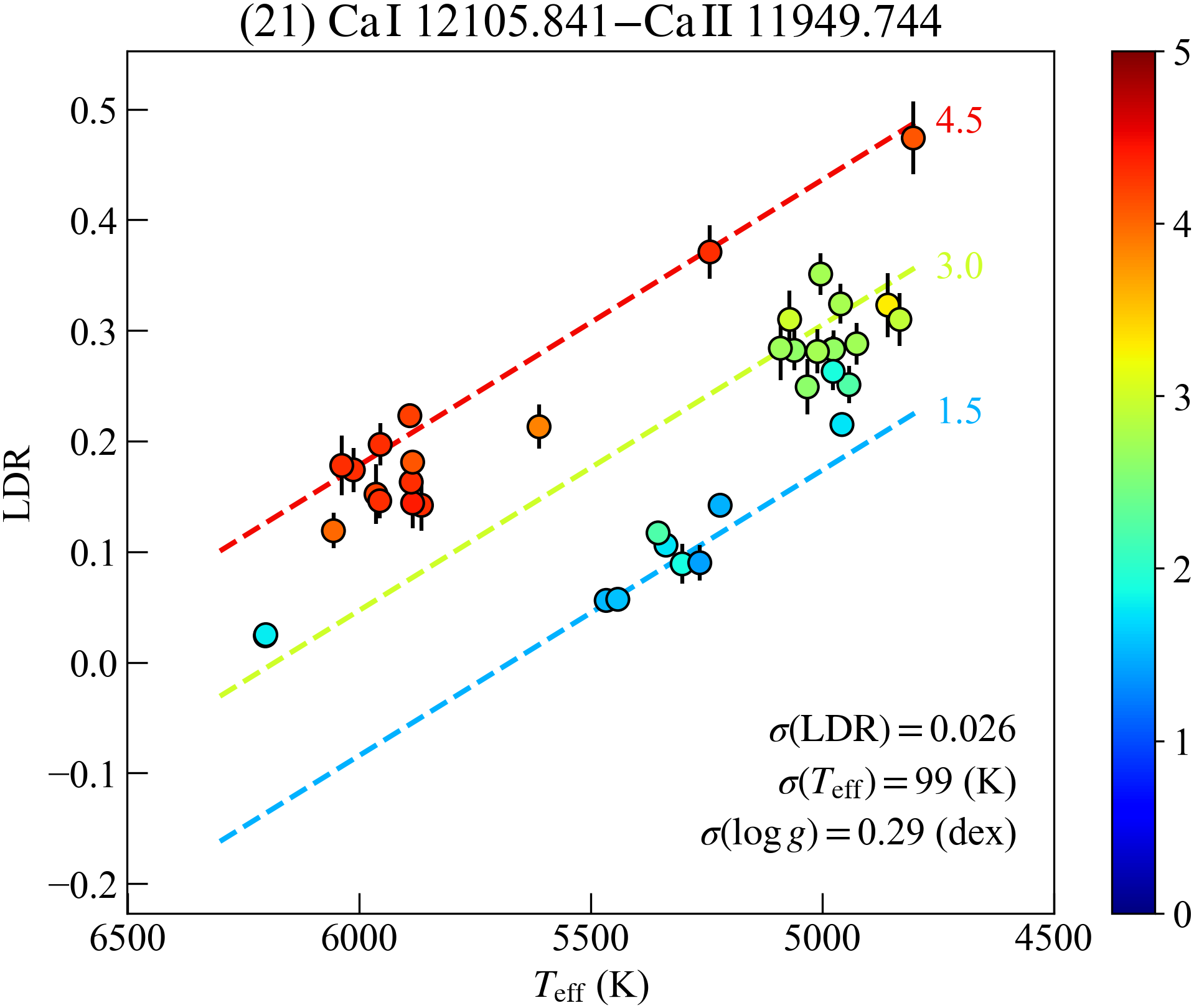}
\\
\includegraphics[clip,width=0.98\hsize]{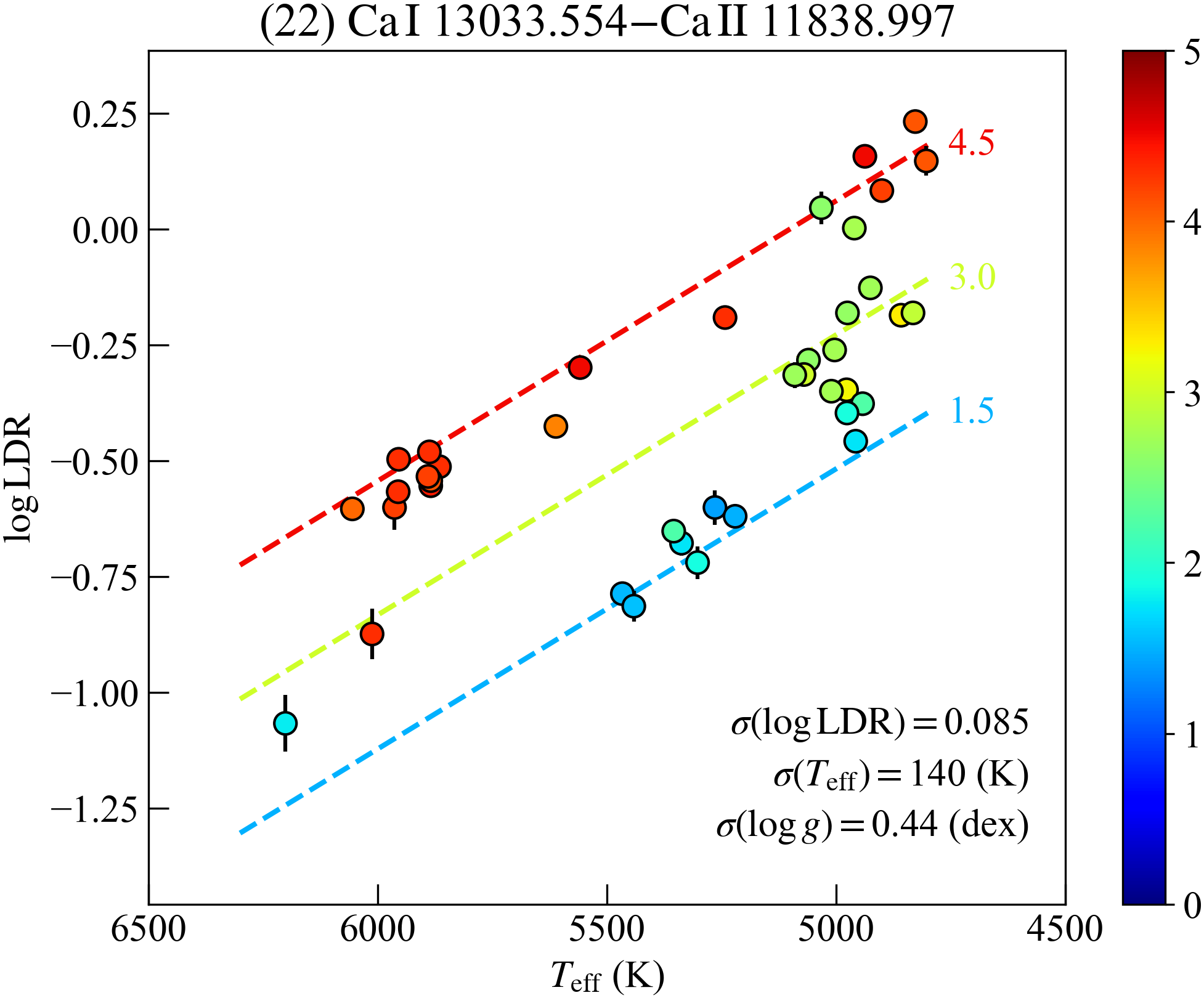}
\end{tabular}

\bsp	
\label{lastpage}
\end{document}

%% file: tab1.tex
\begin{table} 
   \centering 
   \caption{ 
      The objects used for deriving the LDR relations listed in descending order of $\Teff$. The stellar parameters are adopted from \citet{Jian-2020}, and the last column indicates the table in the paper in which each star was listed.
   } \label{tab:targets} 
   \begin{tabular}{lrrrrl} 
      \hline
      \multicolumn{1}{c}{Object} & \multicolumn{1}{c}{$\Teff$} & \multicolumn{1}{c}{$\log g$} & \multicolumn{1}{c}{$\FeH$} & \multicolumn{1}{c}{Obs.~date} & Table \\ 
       & \multicolumn{1}{c}{(K)} & \multicolumn{1}{c}{(dex)} & \multicolumn{1}{c}{(dex)} &  &  \\ 
      \hline
      HD\,194093 & 6202 & 1.35 & $ 0.05$ & 2015.10.26 & Table~3 \\
      HD\,171635 & 6201 & 1.78 & $-0.10$ & 2016.03.11 & Table~3 \\
      HD\,102870 & 6055 & 4.00 & $ 0.18$ & 2016.01.27 & Table~1 \\
      HD\,137107 & 6037 & 4.30 & $ 0.05$ & 2016.01.26 & Table~1 \\
      HD\,115383 & 6012 & 4.30 & $ 0.16$ & 2016.01.26 & Table~1 \\
      HD\,19373 & 5963 & 4.20 & $ 0.11$ & 2016.03.11 & Table~1 \\
      HD\,39587 & 5955 & 4.30 & $ 0.02$ & 2016.03.17 & Table~1 \\
      HD\,114710 & 5954 & 4.30 & $ 0.12$ & 2016.01.26 & Table~1 \\
      HD\,34411 & 5890 & 4.20 & $ 0.15$ & 2016.02.28 & Table~1 \\
      HD\,95128 & 5887 & 4.30 & $ 0.06$ & 2016.02.03 & Table~1 \\
      HD\,141004 & 5884 & 4.10 & $ 0.03$ & 2016.02.27 & Table~1 \\
      HD\,72905 & 5884 & 4.40 & $-0.02$ & 2016.01.27 & Table~1 \\
      HD\,143761 & 5865 & 4.30 & $-0.06$ & 2016.01.26 & Table~1 \\
      HD\,117176 & 5611 & 3.86 & $-0.10$ & 2016.01.26 & Table~1 \\
      HD\,101501 & 5558 & 4.50 & $ 0.02$ & 2016.01.26 & Table~1 \\
      HD\,204867 & 5466 & 1.54 & $ 0.03$ & 2015.10.26 & Table~3 \\
      HD\,31910 & 5441 & 1.57 & $-0.01$ & 2015.10.31 & Table~3 \\
      HD\,92125 & 5354 & 2.22 & $ 0.03$ & 2016.03.21 & Table~3 \\
      HD\,26630 & 5337 & 1.74 & $ 0.09$ & 2015.10.28 & Table~3 \\
      HD\,3421 & 5302 & 1.88 & $-0.20$ & 2015.10.31 & Table~3 \\
      HD\,74395 & 5264 & 1.41 & $ 0.09$ & 2016.01.27 & Table~3 \\
      HD\,10476 & 5242 & 4.30 & $ 0.00$ & 2015.10.31 & Table~1 \\
      HD\,159181 & 5220 & 1.50 & $-0.15$ & 2016.02.03 & Table~3 \\
      HD\,68312 & 5090 & 2.70 & $-0.12$ & 2016.03.15 & Table~2 \\
      HD\,78235 & 5070 & 3.00 & $-0.03$ & 2015.10.28 & Table~2 \\
      HD\,76813 & 5060 & 2.63 & $-0.06$ & 2015.10.28 & Table~2 \\
      HD\,62345 & 5032 & 2.61 & $-0.01$ & 2016.01.27 & Table~2 \\
      HD\,34559 & 5010 & 2.74 & $ 0.00$ & 2016.01.19 & Table~2 \\
      HD\,27348 & 5003 & 2.75 & $ 0.05$ & 2015.10.25 & Table~2 \\
      HD\,11559 & 4977 & 3.23 & $ 0.16$ & 2015.10.23 & Table~2 \\
      HD\,202109 & 4976 & 1.90 & $ 0.10$ & 2016.03.21 & Table~3 \\
      HD\,27697 & 4975 & 2.64 & $ 0.12$ & 2015.10.25 & Table~2 \\
      HD\,27371 & 4960 & 2.76 & $ 0.15$ & 2015.10.28 & Table~2 \\
      HD\,77912 & 4957 & 1.75 & $-0.14$ & 2016.03.21 & Table~3 \\
      HD\,99648 & 4942 & 2.24 & $-0.03$ & 2016.03.12 & Table~3 \\
      HD\,122064 & 4937 & 4.50 & $ 0.12$ & 2016.02.03 & Table~1 \\
      HD\,28305 & 4925 & 2.72 & $ 0.20$ & 2015.10.25 & Table~2 \\
      HD\,219134 & 4900 & 4.20 & $ 0.10$ & 2015.10.25 & Table~1 \\
      HD\,198149 & 4858 & 3.29 & $-0.19$ & 2015.10.26 & Table~2 \\
      HD\,19787 & 4832 & 2.92 & $ 0.12$ & 2016.03.11 & Table~2 \\
      HD\,82106 & 4827 & 4.10 & $-0.06$ & 2016.03.12 & Table~1 \\
      HD\,79555 & 4803 & 4.10 & $-0.14$ & 2016.05.01 & Table~1 \\
      \hline
   \end{tabular}
\end{table}

%% file: tab2.tex
\begin{table} 
   \centering 
   \caption{ 
      List of the 97 confirmed lines that are used to search line-depth ratio pairs (see Section~\ref{subsec:select-line-obs}). The last column lists the number of objects for which the depth measurements were validated. \ion{Fe}{i}, \ion{Fe}{ii}, \ion{Ca}{i} and \ion{Ca}{ii} lines are combined together. The first 10 lines are presented here, and the full table is available in the supporting information section.
   } \label{tab:lines} 
   \begin{tabular}{crrrrr} 
      \hline
      Species & \multicolumn{1}{c}{$\lambda_\mathrm{air}$} & \multicolumn{1}{c}{EP} & \multicolumn{2}{c}{$\log {gf}$ (dex)} & $N_\mathrm{obj}$ \\ 
       & \multicolumn{1}{c}{(\AA)} & \multicolumn{1}{c}{(eV)} & VALD & MB99 & \\ 
      \hline
      \ion{Fe}{i} & 9800.3075 &  5.086 & $-0.453$ & \multicolumn{1}{c}{---} & 37 \\ 
      \ion{Fe}{i} & 9811.5041 &  5.012 & $-1.362$ & \multicolumn{1}{c}{---} & 41 \\ 
      \ion{Fe}{i} & 9861.7337 &  5.064 & $-0.142$ & \multicolumn{1}{c}{---} & 42 \\ 
      \ion{Fe}{i} & 9868.1857 &  5.086 & $-0.979$ & \multicolumn{1}{c}{---} & 42 \\ 
      \ion{Fe}{i} & 9889.0351 &  5.033 & $-0.446$ & \multicolumn{1}{c}{---} & 41 \\ 
      \ion{Fe}{i} & 9944.2065 &  5.012 & $-1.338$ & \multicolumn{1}{c}{---} & 40 \\ 
      \ion{Fe}{i} & 9980.4629 &  5.033 & $-1.379$ & \multicolumn{1}{c}{---} & 42 \\ 
      \ion{Fe}{i} & 10041.472 &  5.012 & $-1.772$ & $-1.84$ & 37 \\ 
      \ion{Fe}{i} & 10065.045 &  4.835 & $-0.289$ & $-0.57$ & 40 \\ 
      \ion{Fe}{i} & 10081.393 &  2.424 & $-4.537$ & $-4.53$ & 33 \\ 
      \hline
   \end{tabular}
\end{table}

%% file: tab3.tex
\begin{table*} 
   \centering 
   \caption{ 
      List of the selected line pairs.
One of the forms (T1)--(T4) was selected for each of the \ion{Fe}{i}--\ion{Fe}{i} pairs, while one of the forms (TG1)--(TG4) was selected for each of the \ion{Fe}{i}--\ion{Fe}{ii} and \ion{Ca}{i}--\ion{Ca}{ii} pairs, and we also remark which of $r$ and $\log r$ and which of $\Teff$ and $\log \Teff$ are included in the given form. According to the form of each line pair, the residual around the relation, $\sigma_{y}$ where $y$ is $r$ or $\log r$, and the corresponding scatter in $\Teff$ (Equation~\ref{eq:sigma-teff}) or $\log g$ (Equation~\ref{eq:sigma-logg}) are listed. 
   } \label{tab:pairs} 
   \begin{tabular}{rllcccccrrr} 
      \hline
     \multicolumn{1}{c}{ID} & \multicolumn{1}{c}{Line 1} & \multicolumn{1}{c}{Line 2} & \multicolumn{2}{c}{Form} & $\alpha$ & $\beta$ & $\gamma$ & \multicolumn{1}{c}{$\sigma_y$} & \multicolumn{1}{c}{$\sigma_p$} & \multicolumn{1}{c}{$N$} \\ 
      \hline
(1) & \ion{Fe}{i}~10155.162 & \ion{Fe}{i}~9868.1857 & T3 & ($\log\,r$,~$\Teff$) & 2.0576 & $-4.314\times 10^{-4}$ & \multicolumn{1}{c}{---} & 0.0373 & 87 & 39 \\
(2) & \ion{Fe}{i}~10167.468 & \ion{Fe}{i}~9811.5041 & T1 & ($r$,~$\Teff$) & 5.4548 & $-8.190\times 10^{-4}$ & \multicolumn{1}{c}{---} & 0.1031 & 126 & 40 \\
(3) & \ion{Fe}{i}~10195.105 & \ion{Fe}{i}~10555.649 & T1 & ($r$,~$\Teff$) & 10.190 & $-14.435\times 10^{-4}$ & \multicolumn{1}{c}{---} & 0.2479 & 172 & 39 \\
(4) & \ion{Fe}{i}~10265.217 & \ion{Fe}{i}~10435.355 & T3 & ($\log\,r$,~$\Teff$) & 1.9317 & $-3.681\times 10^{-4}$ & \multicolumn{1}{c}{---} & 0.0381 & 104 & 32 \\
(5) & \ion{Fe}{i}~10332.327 & \ion{Fe}{i}~10674.070 & T3 & ($\log\,r$,~$\Teff$) & 2.0211 & $-3.778\times 10^{-4}$ & \multicolumn{1}{c}{---} & 0.0408 & 108 & 33 \\
(6) & \ion{Fe}{i}~10340.885 & \ion{Fe}{i}~9861.7337 & T1 & ($r$,~$\Teff$) & 2.5081 & $-3.283\times 10^{-4}$ & \multicolumn{1}{c}{---} & 0.0396 & 120 & 42 \\
(7) & \ion{Fe}{i}~10423.743 & \ion{Fe}{i}~10535.709 & T3 & ($\log\,r$,~$\Teff$) & 1.5019 & $-2.533\times 10^{-4}$ & \multicolumn{1}{c}{---} & 0.0344 & 136 & 41 \\
(8) & \ion{Fe}{i}~10577.139 & \ion{Fe}{i}~10353.804 & T3 & ($\log\,r$,~$\Teff$) & 1.7581 & $-3.299\times 10^{-4}$ & \multicolumn{1}{c}{---} & 0.0327 & 99 & 41 \\
(9) & \ion{Fe}{i}~10616.721 & \ion{Fe}{i}~10364.062 & T3 & ($\log\,r$,~$\Teff$) & 1.4699 & $-2.606\times 10^{-4}$ & \multicolumn{1}{c}{---} & 0.0290 & 111 & 41 \\
(10) & \ion{Fe}{i}~10725.185 & \ion{Fe}{i}~10849.465 & T4 & ($\log\,r$,~$\log\,\Teff$) & 13.182 & $-3.577$ & \multicolumn{1}{c}{---} & 0.0442 & 153 & 39 \\
(11) & \ion{Fe}{i}~10780.694 & \ion{Fe}{i}~10611.686 & T3 & ($\log\,r$,~$\Teff$) & 1.6917 & $-3.827\times 10^{-4}$ & \multicolumn{1}{c}{---} & 0.0455 & 119 & 38 \\
(12) & \ion{Fe}{i}~10818.274 & \ion{Fe}{i}~10347.965 & T3 & ($\log\,r$,~$\Teff$) & 0.79900 & $-1.444\times 10^{-4}$ & \multicolumn{1}{c}{---} & 0.0284 & 197 & 42 \\
(13) & \ion{Fe}{i}~12556.996 & \ion{Fe}{i}~12342.916 & T1 & ($r$,~$\Teff$) & 3.4772 & $-4.557\times 10^{-4}$ & \multicolumn{1}{c}{---} & 0.0591 & 130 & 28 \\
      \hline
(14) & \ion{Fe}{i}~9868.1857 & \ion{Fe}{ii}~9997.5980 & TG4 & ($\log\,r$,~$\log\,\Teff$) & 26.584 & $-7.248$ & 0.2254 & 0.0435 & 0.19 & 27 \\
(15) & \ion{Fe}{i}~10145.561 & \ion{Fe}{ii}~10173.515 & TG1 & ($r$,~$\Teff$) & 14.714 & $-25.042\times 10^{-4}$ & 2.278 & 0.3294 & 0.14 & 12 \\
(16) & \ion{Fe}{i}~10469.652 & \ion{Fe}{ii}~10366.167 & TG4 & ($\log\,r$,~$\log\,\Teff$) & 31.283 & $-8.222$ & 0.1915 & 0.0524 & 0.27 & 16 \\
(17) & \ion{Fe}{i}~10555.649 & \ion{Fe}{ii}~10501.500 & TG3 & ($\log\,r$,~$\Teff$) & 2.0032 & $-5.358\times 10^{-4}$ & 0.2234 & 0.0395 & 0.18 & 37 \\
(18) & \ion{Fe}{i}~10884.262 & \ion{Fe}{ii}~10862.652 & TG1 & ($r$,~$\Teff$) & 10.642 & $-19.971\times 10^{-4}$ & 0.8016 & 0.1627 & 0.20 & 21 \\
      \hline
(19) & \ion{Ca}{i}~10343.819 & \ion{Ca}{ii}~9931.3741 & TG1 & ($r$,~$\Teff$) & 6.0750 & $-9.729\times 10^{-4}$ & 0.4494 & 0.1476 & 0.33 & 27 \\
(20) & \ion{Ca}{i}~10516.156 & \ion{Ca}{ii}~9854.7588 & TG4 & ($\log\,r$,~$\log\,\Teff$) & 38.703 & $-10.640$ & 0.2287 & 0.0978 & 0.43 & 24 \\
(21) & \ion{Ca}{i}~12105.841 & \ion{Ca}{ii}~11949.744 & TG1 & ($r$,~$\Teff$) & 1.3329 & $-2.581\times 10^{-4}$ & 0.0875 & 0.0256 & 0.29 & 36 \\
(22) & \ion{Ca}{i}~13033.554 & \ion{Ca}{ii}~11838.997 & TG3 & ($\log\,r$,~$\Teff$) & 2.2166 & $-6.047\times 10^{-4}$ & 0.1928 & 0.0849 & 0.44 & 40 \\
      \hline
   \end{tabular}
\end{table*}

%% file: tab4.tex
\begin{table*} 
   \centering 
   \caption{ 
      List of the $\Teff$ and $\log g$ derived with the LDR relations. The definitions of the parameters are given in the text.
   } \label{tab:calcTGs} 
   \begin{tabular}{lrrrrrrrrr} 
      \hline
      \multicolumn{1}{c}{Object} & \multicolumn{1}{c}{$\TLDR$} & \multicolumn{1}{c}{$\epsT$} & \multicolumn{1}{c}{$N_{T}$} & \multicolumn{1}{c}{$\Delta_{T}$} & \multicolumn{1}{c}{$\log\,\gLDR$} & \multicolumn{1}{c}{$\epsG$} & \multicolumn{1}{c}{$\delG$} & \multicolumn{1}{c}{$N_{G}$} & \multicolumn{1}{c}{$\Delta_{G}$} \\ 
      \hline
      HD\,194093 & 6291 & 73 & 10 & $89$ & 1.73 & 0.26 & 0.11 & 7 & $0.38$ \\ 
      HD\,171635 & 6283 & 97 & 6 & $82$ & 2.06 & 0.23 & 0.12 & 8 & $0.28$ \\ 
      HD\,102870 & 6163 & 53 & 12 & $108$ & 4.25 & 0.14 & 0.10 & 9 & $0.25$ \\ 
      HD\,137107 & 6201 & 91 & 6 & $164$ & --- & --- & --- & --- & --- \\ 
      HD\,115383 & 6060 & 54 & 10 & $48$ & 4.46 & 0.14 & 0.14 & 9 & $0.16$ \\ 
      HD\,19373 & 6001 & 46 & 13 & $38$ & 4.29 & 0.11 & 0.08 & 8 & $0.09$ \\ 
      HD\,39587 & 6103 & 58 & 10 & $148$ & 4.85 & 0.12 & 0.15 & 6 & $0.55$ \\ 
      HD\,114710 & 6032 & 49 & 13 & $78$ & 4.49 & 0.13 & 0.09 & 6 & $0.19$ \\ 
      HD\,34411 & 5845 & 41 & 12 & $-45$ & 4.16 & 0.10 & 0.06 & 7 & $-0.04$ \\ 
      HD\,95128 & 5984 & 48 & 13 & $97$ & 4.53 & 0.12 & 0.11 & 7 & $0.23$ \\ 
      HD\,72905 & 5910 & 56 & 11 & $26$ & 4.39 & 0.21 & 0.10 & 6 & $-0.01$ \\ 
      HD\,141004 & 5896 & 51 & 13 & $12$ & 4.15 & 0.10 & 0.13 & 7 & $0.05$ \\ 
      HD\,143761 & 5792 & 56 & 13 & $-73$ & 4.22 & 0.26 & 0.13 & 7 & $-0.08$ \\ 
      HD\,117176 & 5451 & 65 & 13 & $-160$ & 3.65 & 0.23 & 0.17 & 5 & $-0.21$ \\ 
      HD\,101501 & 5465 & 43 & 13 & $-93$ & 4.32 & 0.17 & 0.08 & 3 & $-0.18$ \\ 
      HD\,204867 & 5526 & 56 & 13 & $60$ & 1.58 & 0.08 & 0.13 & 9 & $0.04$ \\ 
      HD\,31910 & 5496 & 54 & 12 & $55$ & 1.62 & 0.10 & 0.11 & 8 & $0.05$ \\ 
      HD\,92125 & 5381 & 39 & 13 & $27$ & 2.09 & 0.10 & 0.07 & 7 & $-0.13$ \\ 
      HD\,26630 & 5312 & 67 & 12 & $-25$ & 1.49 & 0.09 & 0.16 & 8 & $-0.25$ \\ 
      HD\,3421 & 5183 & 61 & 12 & $-119$ & 1.43 & 0.13 & 0.13 & 8 & $-0.45$ \\ 
      HD\,74395 & 5185 & 41 & 13 & $-79$ & 1.46 & 0.09 & 0.09 & 8 & $0.05$ \\ 
      HD\,10476 & 5237 & 58 & 13 & $-5$ & 4.57 & 0.36 & 0.17 & 4 & $0.27$ \\ 
      HD\,159181 & 5157 & 38 & 13 & $-63$ & 1.59 & 0.10 & 0.09 & 7 & $0.09$ \\ 
      HD\,68312 & 5011 & 38 & 12 & $-79$ & 2.72 & 0.15 & 0.07 & 5 & $0.02$ \\ 
      HD\,78235 & 5104 & 39 & 12 & $34$ & 3.12 & 0.22 & 0.08 & 8 & $0.12$ \\ 
      HD\,76813 & 5002 & 39 & 11 & $-58$ & 2.49 & 0.11 & 0.09 & 6 & $-0.14$ \\ 
      HD\,62345 & 5009 & 36 & 13 & $-23$ & 2.77 & 0.44 & 0.04 & 3 & $0.16$ \\ 
      HD\,34559 & 5065 & 37 & 12 & $55$ & 2.76 & 0.12 & 0.07 & 5 & $0.02$ \\ 
      HD\,27348 & 4984 & 41 & 13 & $-19$ & 3.04 & 0.16 & 0.11 & 3 & $0.29$ \\ 
      HD\,11559 & 5009 & 35 & 13 & $32$ & 3.10 & 0.21 & 0.08 & 5 & $-0.13$ \\ 
      HD\,202109 & 4976 & 36 & 12 & $0$ & 2.12 & 0.15 & 0.07 & 5 & $0.22$ \\ 
      HD\,27697 & 5003 & 36 & 12 & $28$ & 2.81 & 0.14 & 0.07 & 4 & $0.17$ \\ 
      HD\,27371 & 4955 & 36 & 12 & $-5$ & 2.78 & 0.21 & 0.04 & 5 & $0.02$ \\ 
      HD\,77912 & 4839 & 42 & 12 & $-118$ & 1.45 & 0.11 & 0.12 & 6 & $-0.30$ \\ 
      HD\,99648 & 4938 & 36 & 12 & $-4$ & 2.15 & 0.11 & 0.05 & 6 & $-0.09$ \\ 
      HD\,122064 & 5032 & 38 & 12 & $95$ & --- & --- & --- & --- & --- \\ 
      HD\,28305 & 4938 & 36 & 12 & $13$ & 2.68 & 0.14 & 0.06 & 5 & $-0.04$ \\ 
      HD\,219134 & 4910 & 37 & 12 & $10$ & --- & --- & --- & --- & --- \\ 
      HD\,198149 & 4935 & 46 & 13 & $77$ & 3.32 & 0.21 & 0.12 & 5 & $0.03$ \\ 
      HD\,19787 & 4907 & 36 & 12 & $75$ & 3.01 & 0.15 & 0.07 & 4 & $0.09$ \\ 
      HD\,82106 & 4892 & 50 & 12 & $65$ & --- & --- & --- & --- & --- \\ 
      HD\,79555 & 4714 & 81 & 12 & $-89$ & --- & --- & --- & --- & --- \\ 
      \hline
   \end{tabular}
\end{table*}